\documentclass[aps,prc,twocolumn,showpacs,preprintnumbers,
                          nofootinbib,float,superscriptaddress,longbibliography]{revtex4}
\usepackage{graphicx, fancybox}
\usepackage{amsmath,amssymb}
\usepackage[colorlinks=true, pdfstartview=FitV, linkcolor=red, citecolor=blue, urlcolor=blue]{hyperref}
\usepackage{color}
\usepackage{soul}
\usepackage{url}

\newcommand{\nobracket}{}
\newcommand{\nocomma}{}

\newcommand{\tmop}[1]{\ensuremath{\operatorname{#1}}}

\usepackage[normalem]{ulem}

\begin{document}

\title{Net baryon diffusion in fluid dynamic simulations of relativistic heavy-ion collisions}

\author{Gabriel S. Denicol}
 \affiliation{Instituto de F\'isica, Universidade Federal Fluminense, UFF, Niter\'oi, 24210-346, RJ, Brazil}

 \author{Charles Gale}
 \affiliation{Department of Physics, McGill University, 3600 University
 Street, Montreal,
 QC, H3A 2T8, Canada}

 \author{Sangyong Jeon}
 \affiliation{Department of Physics, McGill University, 3600 University
 Street, Montreal,
 QC, H3A 2T8, Canada}

\author{Akihiko Monnai}
\affiliation{Institut de Physique Th\'eorique, CNRS, CEA/Saclay, F-91191 Gif-sur-Yvette, France}
\affiliation{KEK Theory Center, Institute of Particle and Nuclear Studies, \\
High Energy Accelerator Research Organization (KEK),
1-1, Ooho, Tsukuba, Ibaraki 305-0801, Japan}

\author{Bj\"orn Schenke}
\affiliation{Physics Department, Brookhaven National Laboratory, Upton, NY 11973, USA}

\author{Chun Shen}
\affiliation{Physics Department, Brookhaven National Laboratory, Upton, NY 11973, USA}

\begin{abstract}

A hybrid (hydrodynamics + hadronic transport) theoretical framework is assembled to model the bulk dynamics of relativistic heavy-ion collisions at energies accessible in the Beam Energy Scan (BES) program at the Relativistic Heavy-Ion Collider (RHIC) and the NA61/SHINE experiment at CERN. The system's energy-momentum tensor and net baryon current are evolved according to relativistic hydrodynamics with finite shear viscosity and non-zero net baryon diffusion. Our hydrodynamic description is matched to a hadronic transport model in the dilute region. With this fully integrated theoretical framework, we present a pilot study of the hadronic chemistry, particle spectra, and anisotropic flow. Phenomenological effects of a non-zero net-baryon current and its diffusion on hadronic observables are presented for the first time. The importance of the hadronic transport phase is also investigated. 
\end{abstract}

\maketitle
\date{\today}

\section{Introduction}
The beam energy scan program (BES and BESII) at the Relativistic Heavy Ion Collider (RHIC) at Brookhaven National Laboratory \cite{Adamczyk:2013dal,Adamczyk:2014fia,Adare:2015aqk,Adamczyk:2017iwn} and the NA61/SHINE experiment at the CERN Super Proton Synchrotron (SPS) \cite{Mackowiak-Pawlowska:2017rcx} aim to fully explore the phase diagram of strongly interacting matter.
In the baryon-rich region, the transition from a hadron gas to a quark-gluon-plasma phase (QGP) is expected to be first order, whereas it is a rapid crossover at the low baryon density. Therefore, the existence of a critical point in this phase diagram is widely speculated. 
The existence of such a critical point has yet to be confirmed, and its location in the phase diagram determined.

Measurements sensitive to the presence of a critical point are those of fluctuations of conserved charges, most commonly those of net-baryon number. In the vicinity of a critical point, the correlation length grows, leading to increased fluctuations \cite{Stephanov:1998dy,Stephanov:1999zu,Stephanov:2004wx}. Varying the collision energy should then allow the trajectory of the system to explore the plane spanned by temperature and baryon chemical potential, and to locate the position of the critical point using the fluctuation measurements.

To accomplish this, one needs to know precisely what to expect for the relevant observables in the case that there is no critical point (and fluctuations are entirely non-critical) and how they are modified if a critical point is present. This requires complex simulations of the entire system starting from fluctuating initial states, hydrodynamic evolution at finite net-baryon density (and possible hydrodynamic fluctuations \cite{Kapusta:2011gt,Young:2014pka,Kapusta:2014dja}), as well as microscopic hadronic cascades for the low temperature stage.

Such simulations can then be used to study the effects of an equation of state with a critical point \cite{Nonaka:2004pg}. They can be coupled to evolution equations for the sigma field and Polyakov loop in the so called chiral fluid dynamics \cite{Paech:2005cx,Herold:2013bi}, and can provide important information required for calculations of the non-equilibrium evolution of cumulants of critical fluctuations \cite{Mukherjee:2015swa}.

Calculations that include the viscous relativistic hydrodynamic evolution of the QGP and hadron gas, combined with models for fluctuating initial states and hadronic afterburners, have been very successful in describing the soft observables measured in heavy ion collisions at top RHIC and Large Hadron Collider (LHC) energies. However, at these high collision energies the net-baryon density is typically assumed to be negligible, which is valid at least near mid-rapidity \cite{Shen:2017bsr}. Furthermore, the initial state description is somewhat simplified because an instantaneous interaction of two highly contracted nuclei can be assumed. At lower energies, neither assumption holds. For reviews on relativistic hydrodynamics and hybrid models of heavy ion collisions we refer the reader to \cite{Heinz:2013th,Gale:2013da} and \cite{Petersen:2014yqa}, respectively.

To make progress towards a simulation framework valid at all collision energies, fluctuating initial conditions for lower energy collisions have been addressed recently \cite{Karpenko:2015xea,Okai:2017ofp,Shen:2017bsr}. The simulation \textsc{Music}\footnote{The numerical package can be downloaded from \url{http://www.physics.mcgill.ca/music}.} \cite{Schenke:2010nt} has included the evolution of conserved baryon currents from the beginning, but baryon diffusion has so far been neglected. However, when studying observables that are sensitive to the precise baryon distributions and their fluctuations \cite{Kapusta:2017hfi}, we need to take great care in including all relevant physics in the simulation. In this work we present results of an extended version of \textsc{Music} that includes the most basic effects of baryon diffusion.

Apart from this extension of the hydrodynamic simulation itself, we need to consider an equation of state at finite baryon density. We present a construction of such an equation of state using Lattice QCD results with Taylor expansion in baryon chemical potential coupled to a hadron resonance gas, and use it in all shown calculations. Current lattice QCD simulations have not shown evidence of a critical point and, hence, our results at this stage do not probe any effects from a critical point -- hence,  in this aspect, they can be considered as baseline calculations.

Besides providing a necessary tool for simulating heavy ion collisions over a wide range of energies relevant to the critical point search, the new developments presented in this work also establish a path to the extraction of the heat conductivity of the quark gluon plasma by detailed comparison with experimental measurements. We identify observables that are most sensitive to the effect of baryon diffusion and thus the heat conductivity of the QGP. 

The paper is organized as follows. Sec.~\ref{sec:model} gives a detailed model description of our hybrid framework. The phenomenological impact of net baryon diffusion and hadronic transport on experimental observables are studied in Sec.~\ref{sec:results}. The focus of our studies are Au+Au collisions at 19.6 GeV. Sec.~\ref{sec:conclusion} summarizes the main findings of this work. Additional detailed derivations of net baryon diffusion corrections and numerical validation of the hydrodynamic simulation are presented in the appendices.

\section{The Hybrid framework}\label{sec:model}

\subsection{Initialization of hydrodynamics}
For very high center of mass energies, like the top RHIC energy or LHC energies, the Lorentz contraction of the incoming nuclei is so strong that it is a good approximation to consider them as sheets of negligible width in the longitudinal (beam-) direction. This means that the time of the collision is given precisely by the time the two sheets pass through each other.
In contrast, the collision energies scanned in the RHIC BES program and the NA61/SHINE experiment are not high enough to neglect the finite thickness of the colliding nuclei along the longitudinal direction. The time the two nuclei spend passing through one another for a given collision energy $\sqrt{s_\mathrm{NN}}$ can be estimated as
\begin{equation}
\tau_\mathrm{overlap} = \frac{2 R}{\gamma_L v_L} = \frac{2 R}{\sqrt{\gamma_L^2  - 1}},
\label{eq2.1}
\end{equation}
where the Lorentz factor in the longitudinal direction is $\gamma_L = \frac{\sqrt{s_\mathrm{NN}}/2}{m_N}$ with $m_N = 0.938$ GeV, and $R$ is the radius of the colliding nuclei. For gold nuclei $R_\mathrm{Au} \simeq 7.0$\,fm. 
At the lowest BES collision energy of $\sqrt{s_\mathrm{NN}} = 7.7$ GeV, this overlapping time is $3\,{\rm fm}$, comparable to the lifetime of the QGP created in the system. 

In Ref.\,\cite{Shen:2017bsr} two of the authors have presented a new initial state model that treats the early stage of the evolution dynamically by starting hydrodynamic evolution before that time and taking care of additional deposited entropy and baryon densities via source terms.

Because the focus of this work is the effect of baryon diffusion, we employ a simpler initial state description, where the initial entropy and baryon densities are assumed to be smooth average quantities and the hydrodynamic simulations are started at $\tau_0 = \tau_\mathrm{overlap}$.
The smooth initial conditions are generated by averaging over 10,000 fluctuating Monte Carlo (MC)-Glauber events in the given centrality bin, which is determined using the configurations' total entropy. When averaging the spatial structure, events within the same centrality bin are aligned using their second-order participant plane angles, $\Psi^\mathrm{PP}_2$, defined as
\begin{equation}
\varepsilon_2 e^{i 2 \Psi^\mathrm{PP}_2} = - \frac{\int d^2 {\bf r}\,r^2 s(r, \phi) e^{i2\phi}}{\int d^2 {\bf r}\,r^2 s(r, \phi)}.
\label{eq2.2}
\end{equation}
Here $s(r, \phi)$ is the transverse plane entropy density profile at mid-rapidity. 

To construct the entropy density as a function of the transverse coordinates and of the space-time rapidity, 
we first define the contributions from the right moving ($+$) and left moving ($-$) nuclei as
\begin{equation}
  s_\pm(x, y) = \sum_{j = 1}^{N^\pm_\mathrm{part}} \frac{1}{2\pi \sigma^2 }\exp\left(-\frac{({\bf r-r}^\pm_j)^2}{2\sigma^2} \right)\,,
\end{equation}
where $\mathbf{r}=(x,y)$ and $\mathbf{r}^\pm_j$ are the positions of the partici\-pant nucleons in the two nuclei.
The Gaussian width parameter is set to $\sigma = 0.5$ fm. 

The full initial 3D density profiles follow from folding $s_\pm (x, y)$ with envelope functions along the rapidity direction,
\begin{equation}
s(x, y, \eta; \tau_0) = \frac{s_0}{\tau_0} \sum_{i=\pm} f^{s}_i (\eta) s_i (x, y).
\label{eq:entropydensity}
\end{equation}
Here $s_0$ is the peak entropy density which is adjusted to reproduce the experimentally observed charged hadron multiplicities.

\begin{figure*}[ht!]
  \centering
  \begin{tabular}{cc}
  \includegraphics[width=0.48\linewidth]{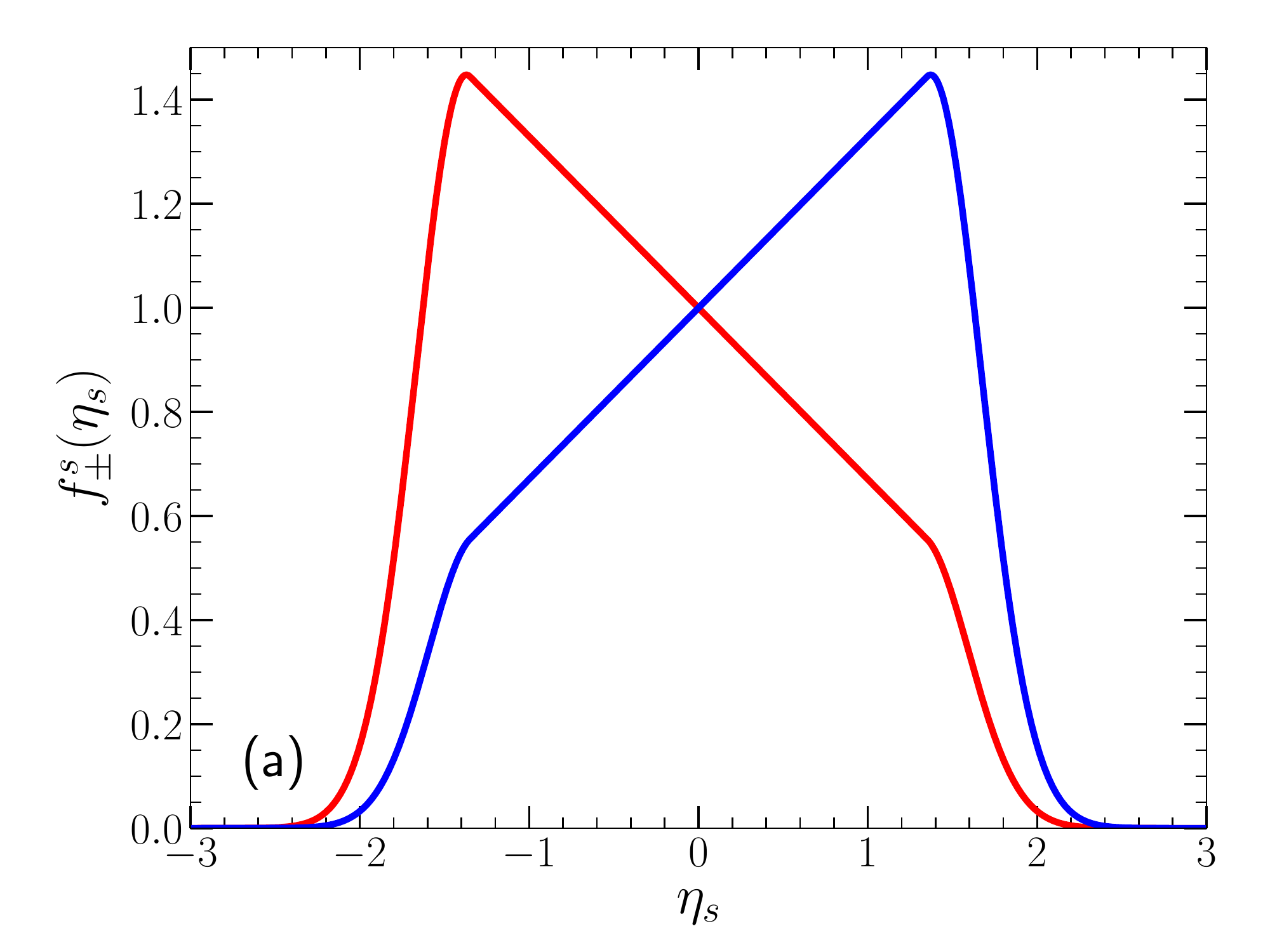} &
  \includegraphics[width=0.48\linewidth]{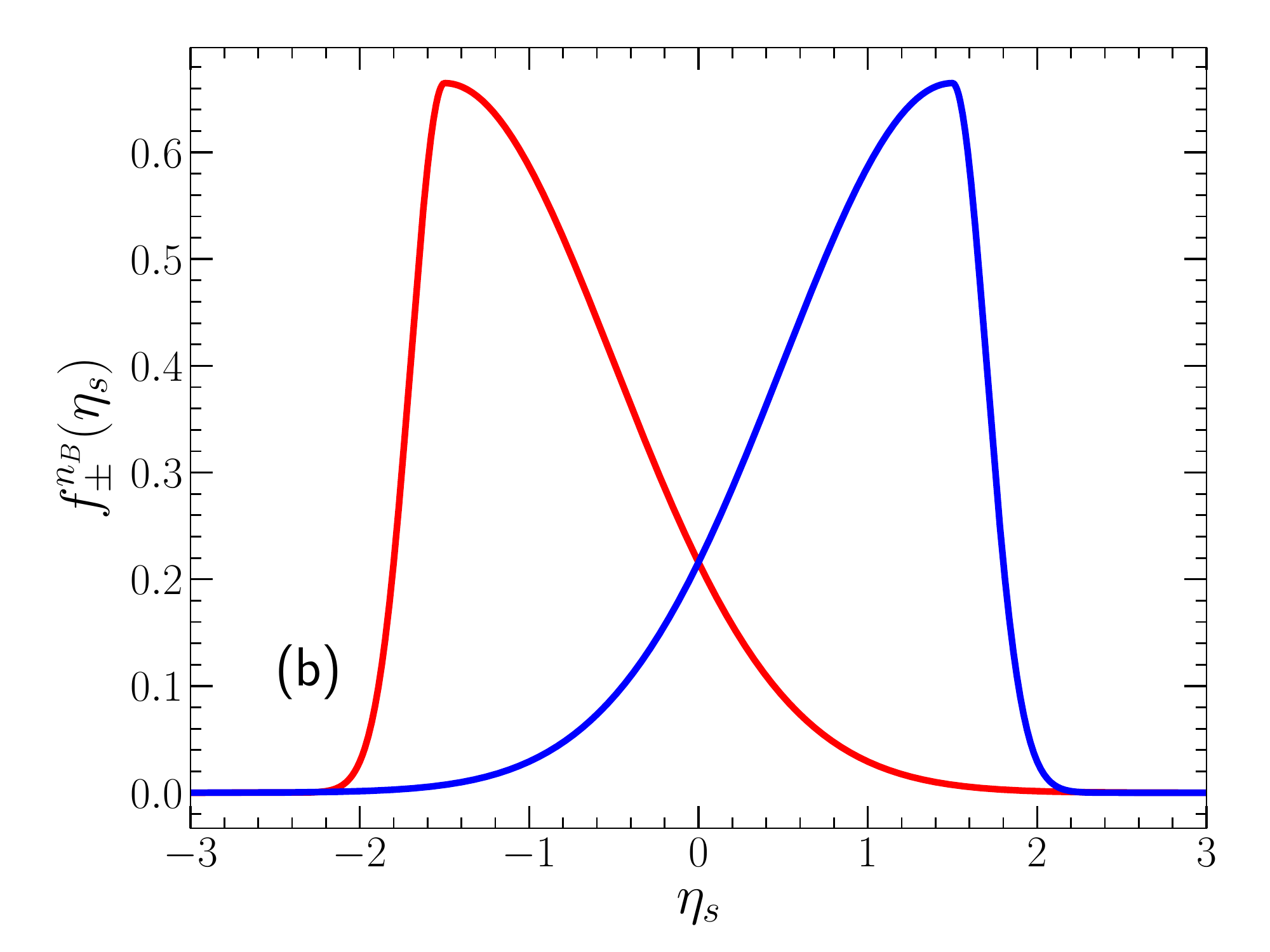}
  \end{tabular}
  \caption{Example of the envelope functions for entropy density and net baryon density $f^s_\pm(\eta_s)$ and $f^{n_B}_\pm(\eta_s)$ in Au+Au collisions at $\sqrt{s_\mathrm{NN}} = 19.6$\,GeV.}
  \label{fig2}
\end{figure*}
%

Similarly, the net baryon density profile can be constructed as
\begin{equation}
n_B(x, y, \eta; \tau_0) = \frac{1}{\tau_0} \sum_{i=\pm} f^{n_B}_i (\eta) s_i (x, y).
\label{eq:baryondensity}
\end{equation}
There is no additional normalization factor for the net baryon density because it is constrained by the total number of participant nucleons $N_\mathrm{part}$, $\int \tau_0 dx dy d\eta n_B(x, y, \eta; \tau_0) = N_\mathrm{part}$. The envelope functions in Eqs. (\ref{eq:entropydensity}) and (\ref{eq:baryondensity}) are chosen as,
\begin{eqnarray}
&&f^s_\pm(\eta) = \theta(\eta_\mathrm{max} - \vert \eta \vert) \left(1 \pm \frac{\eta}{\eta_\mathrm{max}} \right) \notag \\
&& \,\times \left[\theta(\vert \eta \vert - \eta^s_0) \exp\left(-\frac{(|\eta| - \eta^s_0)^2}{2\sigma_{\eta,s}^2} \right) + \theta(\eta^s_0 - \vert \eta \vert) \right]
\label{eq:entropyEnvelope}
\end{eqnarray}
where the maximum extension in space-time pseudo-rapidity $\eta_\mathrm{max}$ is chosen to be equal to the beam rapidity $y_\mathrm{beam} = \mathrm{arctanh}\left( \frac{\sqrt{\gamma_L^2 - 1}}{\gamma_L} \right)$ of incoming nucleons. The parameters $\eta^s_0$ and $\sigma_{\eta, s}$ are determined to reproduce the pseudo-rapidity distribution of charged hadrons $dN^\mathrm{ch}/d\eta$.  

For the net baryon density envelope profile,
\begin{eqnarray}
f^{n_B}_\pm (\eta) &=& \frac{1}{\mathcal{N}} \left[\theta(\eta - \eta^{n_B, \pm}_0) \exp\left(-\frac{(\eta - \eta^{n_B, \pm}_0)^2}{2\sigma_{\eta,\pm}^2} \right) \right. \notag \\
&& \left. + \theta(\eta^{n_B, \pm}_0 - \eta) \exp\left(-\frac{(\eta - \eta^{n_B, \pm}_0)^2}{2\sigma_{\eta, \mp}^2} \right) \right]
\label{eq:baryonEnvelope}
\end{eqnarray}
where $\mathcal{N}$ is the normalization of the envelope profile which ensures
\begin{equation}
\int d\eta f^{n_B}_\pm (\eta) = 1.
\end{equation}
The peak position $\eta^{n_B,\pm}_0$ is determined by the measured rapidity loss in the net proton distribution and the width parameters $\sigma_{\eta, \pm}$ determine the shape of the final $dN^{p - \bar{p}}/dy$. 
Figure \ref{fig2} shows an example of the $\eta_s$ envelope functions for entropy density and net baryon density. The parameters in Eqs.\ (\ref{eq:entropyEnvelope}) and (\ref{eq:baryonEnvelope}) are determined for Au+Au collisions and shown in Table \ref{table1} for different collision energies.

\begin{table}[ht!]
  \centering
  \begin{tabular}{c|c|c|c|c|c|c|c|c}
  \hline \hline
   $\sqrt{s_\mathrm{NN}}$ (GeV) & $y_\mathrm{beam}$ & $\tau_0$ (fm) & $s_0$ & $\eta^s_0$ & $\sigma_{\eta,s}$ &$\eta^{n_B}_0$  & $\sigma_{\eta,+}$ & $\sigma_{\eta,-}$ \\ \hline
   19.6 & 3.04& 1.5 & 6.3 & 2.7 & 0.3 & 1.5 & 0.2 & 1.0 \\ \hline
   \hline
   \end{tabular}
  \caption{A list of parameters for MC-Glauber initial conditions for Au+Au collisions at different collision energies.}
  \label{table1}
\end{table}
%

\begin{figure*}[ht!]
  \centering
  \begin{tabular}{cc}
  \includegraphics[width=0.48\linewidth]{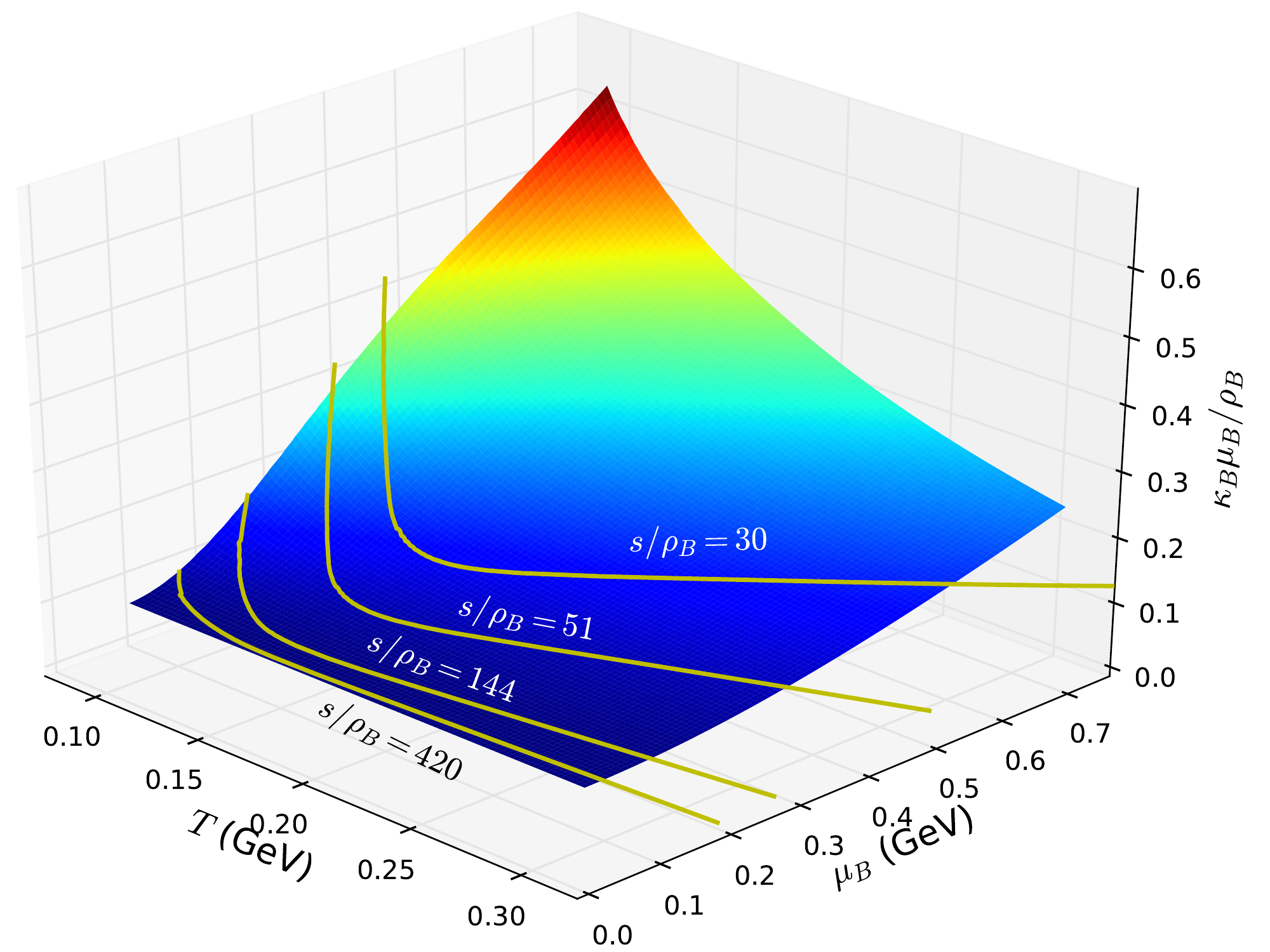} &
  \includegraphics[width=0.48\linewidth]{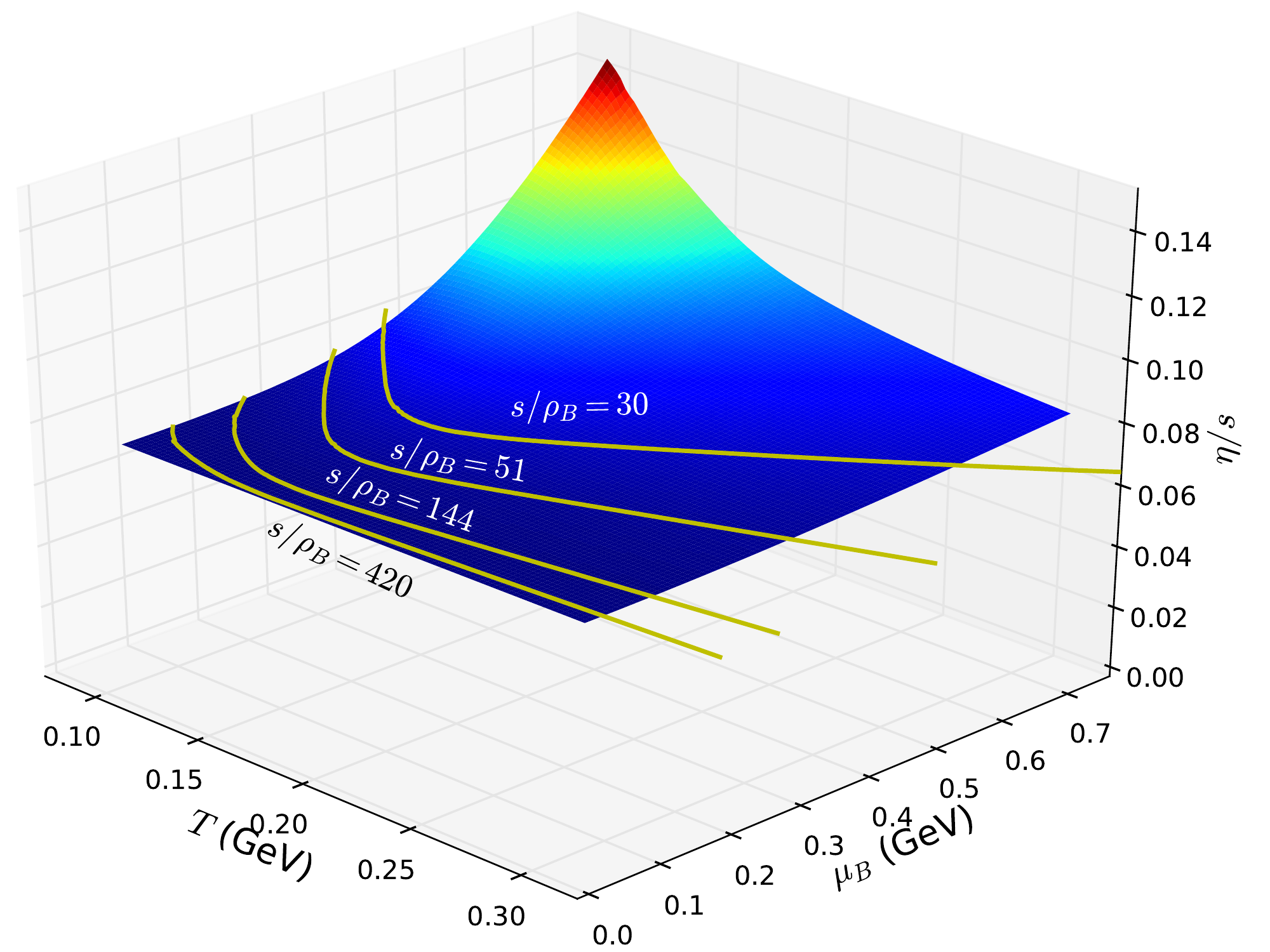}
  \end{tabular}
   \caption{The temperature and net baryon chemical potential dependence of the net baryon diffusion constant and specific shear viscosity for $C_B = 0.4$ and $C_\eta = 008$.}
  \label{fig2B.0}
\end{figure*}
%

\subsection{Hydrodynamics at finite baryon density}

The hydrodynamical equation of motion at finite net baryon density can be written as,
\begin{equation}
\partial_\mu T^{\mu\nu} = 0,
\label{eq2.10}
\end{equation}
\begin{equation}
\partial_\mu J_B^{\mu} = 0,
\label{eq2.11}
\end{equation}
where the system's energy momentum tensor can be decomposed as
\begin{equation}
T^{\mu\nu} = eu^\mu u^\nu - (P + \Pi) \Delta^{\mu\nu} + \pi^{\mu\nu},
\end{equation}
and
\begin{equation}
J_B^{\mu} = n_B u^\mu + q^\mu.
\end{equation}
Here $\Delta^{\mu\nu} = g^{\mu\nu}-u^\mu u^\nu$ is a projection operator, $u^\mu$ is the flow velocity, and $g^{\mu\nu}={\rm diag}(1,-1,-1,-1)$ is the space-time metric.
The dissipative currents in the system are the bulk viscous pressure $\Pi$, the net baryon diffusion current $q^\mu$, and shear stress tensor $\pi^{\mu\nu}$. In this work, we consider only the effects of the shear stress tensor and net baryon diffusion. These two currents are described by the Israel-Stewart-like equations,
\begin{eqnarray}
\Delta^{\mu\nu} D q_\nu &=& - \frac{1}{\tau_q} \left(q^\mu - \kappa_B \nabla^\mu \frac{\mu_B}{T} \right) - \frac{\delta_{qq}}{\tau_q} q^\mu \theta - \frac{\lambda_{qq}}{\tau_q} q_\nu \sigma^{\mu\nu} \notag \\
&& + \frac{l_{q\pi}}{\tau_q} \Delta^{\mu\nu} \partial_{\lambda} \pi^{\lambda}\,_\nu - \frac{\lambda_{q\pi}}{\tau_q} \pi^{\mu\nu} \nabla_\nu \frac{\mu_B}{T},
\label{eq2.14}
\end{eqnarray}
and
\begin{eqnarray}
\Delta^{\mu\nu}_{\alpha \beta} D \pi^{\alpha \beta} &=& - \frac{1}{\tau_\pi} (\pi^{\mu\nu} - 2 \eta \sigma^{\mu\nu}) \notag \\
&& - \frac{\delta_{\pi\pi}}{\tau_{\pi}} \pi^{\mu\nu} \theta - \frac{\tau_{\pi\pi}}{\tau_\pi}\pi^{\lambda \langle} \sigma^{\nu \rangle}\,_\lambda
     + \frac{\phi_7}{\tau_\pi} \pi^{\langle \mu}\,_\alpha \pi^{\nu \rangle \alpha} \notag \\
&&  + \frac{l_{\pi q}}{\tau_\pi} \nabla^{\langle \mu} q^{\nu \rangle} + \frac{\lambda_{\pi q}}{\tau_\pi} q^{\langle \mu} \nabla^{\nu \rangle} \frac{\mu_B}{T}\,.
\label{eq2.15}
\end{eqnarray}
Here the evolution of the diffusion current is driven by the gradient of the net baryon chemical potential $\mu_B$ divided by temperature $T$. The thermodynamic force for the shear viscous pressure is the velocity shear tensor $\sigma^{\mu\nu} = \nabla^{\langle \mu} u^{\nu\rangle}$, and $A^{\langle \mu\nu\rangle} = \Delta^{\mu\nu}_{\alpha\beta} A^{\alpha\beta}$ projects out the part that is traceless and transverse to the flow velocity $u_\mu$ using the double, symmetric, and traceless projection operator, $\Delta^{\mu\nu}_{\alpha\beta} = \frac{1}{2}\left[\Delta^\mu_\alpha \Delta^\nu_\beta + \Delta^\nu_\alpha \Delta^\mu_\beta -\frac{2}{3}\Delta^{\mu\nu}_{\alpha\beta}\right]$. The system's expansion rate is $\theta = \partial_\mu u^\mu + u^\tau/\tau$.

The transport coefficients $\eta$ and the baryon diffusion constant $\kappa_B$ are chosen as
\begin{equation}
\frac{\eta T}{e + \mathcal{P}} = C_\eta
\end{equation}
and
\begin{equation}
\kappa_B = \frac{C_B}{T} n_B \left(\frac{1}{3} \coth\left(\frac{\mu_B}{T}\right) - \frac{n_B T}{e + \mathcal{P}} \right). 
\label{eq2.17}
\end{equation}
The specific shear viscosity is chosen to be $C_\eta = 0.08$. 
The constant coefficient $C_B$ will be varied to study the effect of the net baryon diffusion. The $T$ and $\mu_B$ dependence of $\kappa_B$ in Eq.~(\ref{eq2.17}) is derived from the Boltzmann equation in the relaxation time approximation (see Appendix \ref{appendix_A}). We show the dimensionless quantity $\kappa_B \mu_B/n_B$ along with $\eta/s$ as functions of $T$ and $\mu_B$ in Fig.~\ref{fig2B.0}. Four lines of constant $s/n_B$, that reflect the averaged values realized at the four different collision energies we consider, demonstrate what values of the transport parameters typically contribute.

\begin{table}[ht!]
  \def\arraystretch{1.5}
  \centering
  \begin{tabular}{c|c|c|c|c|c}
  \hline \hline
   $\tau_q$ & $\delta_{qq}$ & $\lambda_{qq}$ & $l_{q\pi}$ & $\lambda_{q\pi}$ & \\ \hline
   $\frac{C_B}{T} $ & $\tau_q$ & $\frac{3}{5}\tau_q$ & 0 & 0 & \\ \hline \hline
   $\tau_\pi$ & $\delta_{\pi\pi}$ & $\tau_{\pi\pi}$ & $\phi_7$ & $l_{\pi q}$ & $\lambda_{\pi q}$ \\ \hline
   $\frac{5C_\eta}{T}$ & $\frac{4}{3}\tau_\pi$ & $\frac{10}{7}\tau_\pi$ & $\frac{9}{70} \frac{4}{\varepsilon + \mathcal{P}}$ & 0 & 0 \\ \hline \hline
   \end{tabular}
  \caption{A list for the second order transport coefficients used in the evolution equations for the net baryon diffusion current $q^\mu$ and the shear stress tensor $\pi^{\mu\nu}$.}
  \label{table2}
\end{table}
%
Table~\ref{table2} summarizes the choice of the second order transport coefficients used in Eqs.~(\ref{eq2.14}) and (\ref{eq2.15}). 
The expression for the baryon diffusion relaxation time, $\tau_{q}$ is chosen to be proportional to $1/T$ (as it is exactly in a conformal system), with the proportionality constant $C_B$ a free parameter. The remaining transport coefficients listed in the table are from calculations assuming kinetic theory in the massless limit \cite{Denicol:2010xn,Denicol:2012cn,Molnar:2013lta,Denicol:2014vaa}. Recent calculations of transport coefficients taking into account a finite (and thermal) mass, were performed in Ref.\,\cite{Czajka:2017wdo}.

%
\begin{figure*}[ht!]
  \centering
  \begin{tabular}{cc}
  \includegraphics[width=0.48\linewidth]{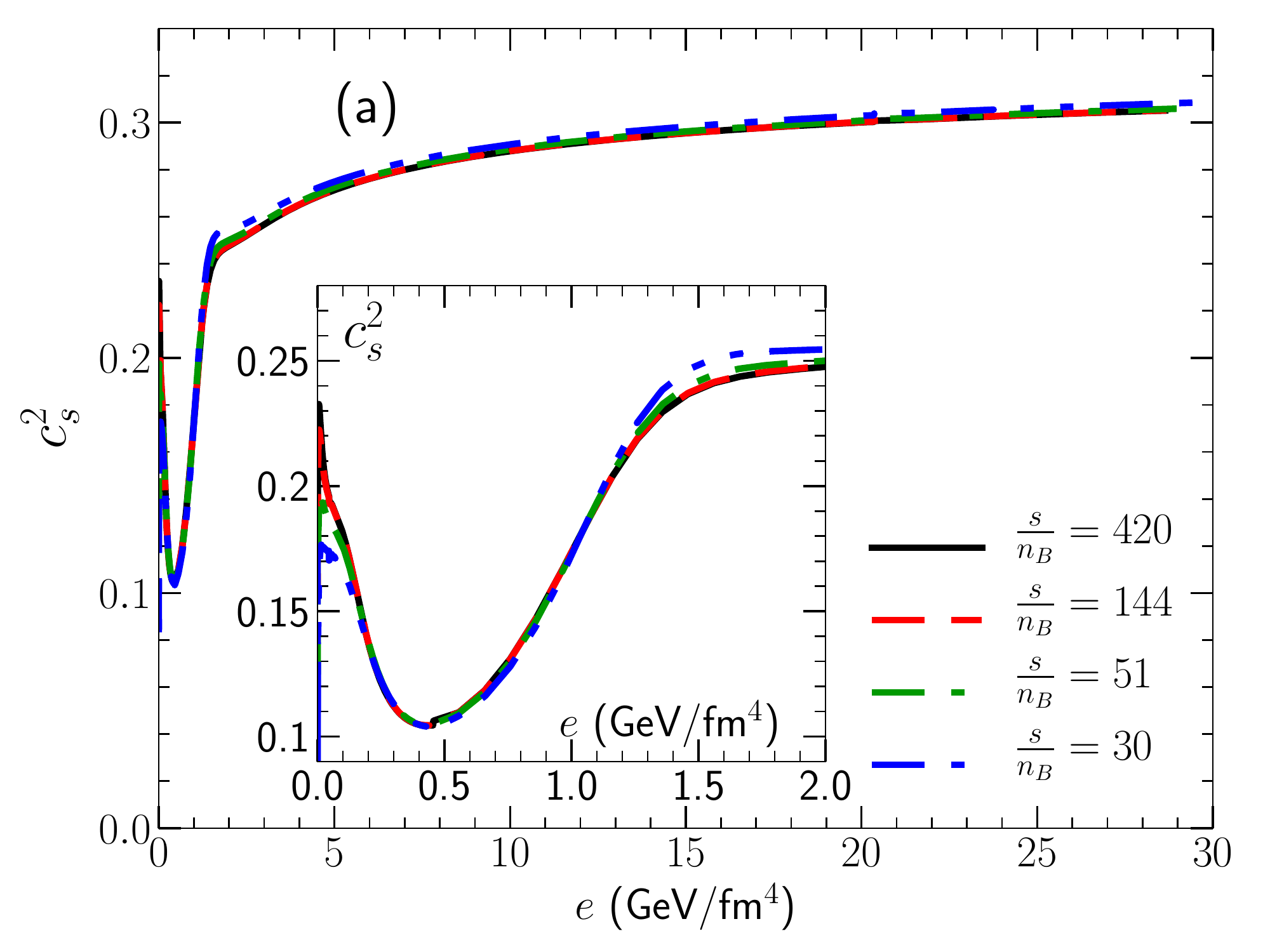} &
  \includegraphics[width=0.48\linewidth]{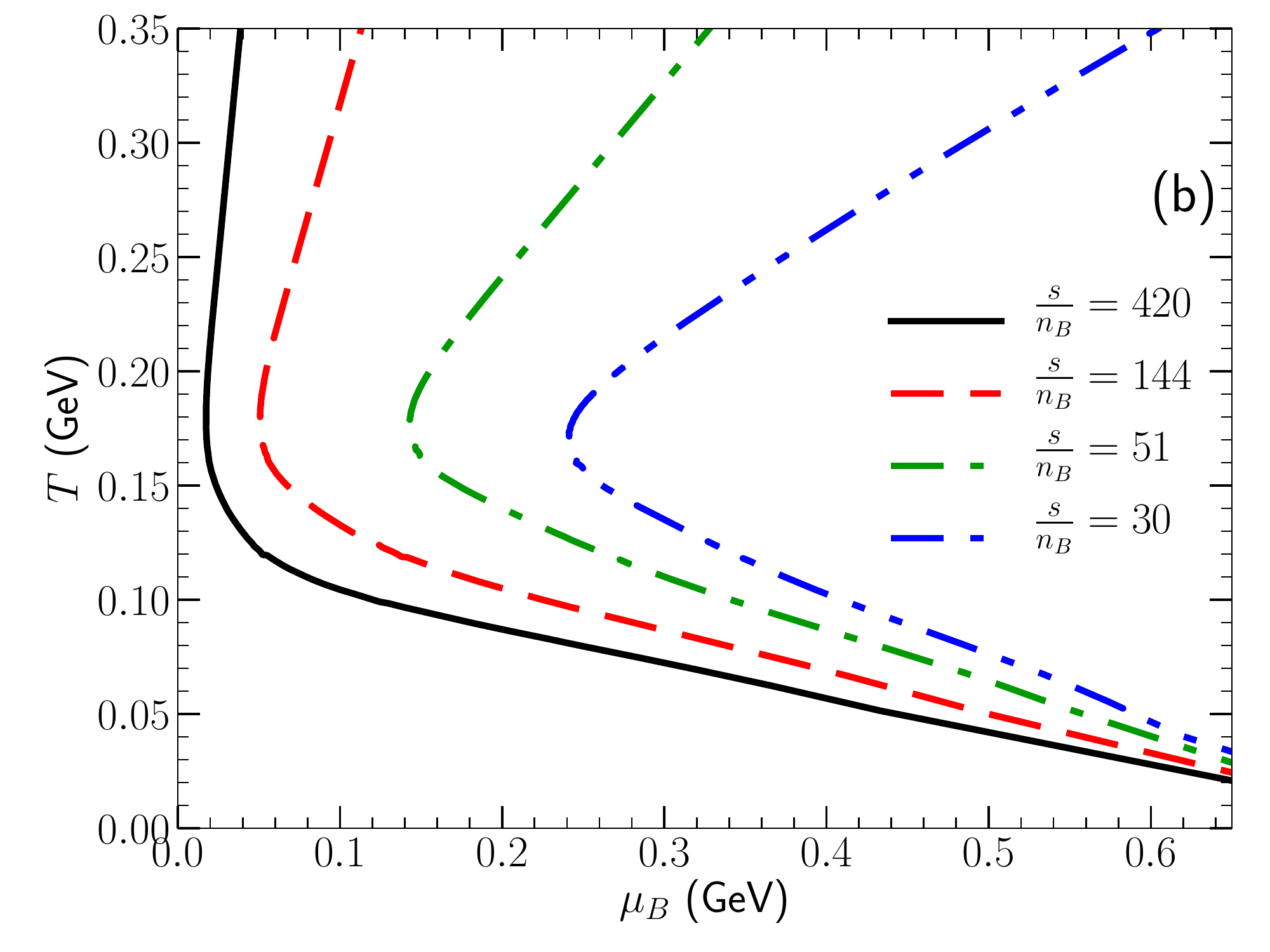}
  \end{tabular}
   \caption{{\it Panel (a)}: The square of the speed of sound as a function of the local energy density along constant $s/n_B$ lines. {\it Panel (b)}: Temperature as a function of net baryon chemical potential along constant $s/n_B$ lines. The collision energies correspond to these $s/n_B$ lines (from top down in the legend) are $\sqrt{s_\mathrm{NN}} = 200, 62.4, 19.6$, and $14.5$\,GeV according to Ref.~\cite{Gunther:2016vcp}.}
  \label{fig2B.1}
\end{figure*}
%

The system of hydrodynamic equations (\ref{eq2.10}) and (\ref{eq2.11}) needs to be closed with the equation of state (EoS) of the fluid. In this work, the EoS of the QCD matter is constructed using lattice QCD calculations \cite{Borsanyi:2013bia,Borsanyi:2011sw}. We consider a crossover-type EoS and leave implementation of the QCD critical point for future study. At zero baryon chemical potential, the pressure of the system is computed as a function of the local temperature via \cite{Huovinen:2009yb},
\begin{equation}
\frac{\mathcal{P}(T)}{T^4} = \frac{\mathcal{P}(T_\mathrm{low})}{T^4_\mathrm{low}} + \int_{T_\mathrm{low}}^T \frac{d T^\prime}{T^\prime} \frac{e - 3\mathcal{P}}{T^{\prime4}},
\end{equation}
where the trace anomaly $e - 3\mathcal{P}$ is computed from lattice QCD as a function of temperature. The lower integration limit $T_\mathrm{low}$ is chosen to be sufficiently small such that $\mathcal{P}(T_\mathrm{low})$ can be neglected because of the exponential suppression. Since of the sign problem, it is not possible to directly calculate the EoS at finite baryon density using lattice QCD. Instead, the $\mu_B$ dependence of the EoS is constructed using the following Taylor expansion,
\begin{eqnarray}
\frac{\mathcal{P}(T, \mu_B)}{T^4} &=& \frac{\mathcal{P}(T)}{T^4} \bigg\vert_{\mu_B = 0} + c_2(T) \left(\frac{\mu_B}{T}\right)^2 + c_4(T) \left(\frac{\mu_B}{T}\right)^4 \notag \\
&& + \mathcal{O}\left( \left(\frac{\mu_B}{T}\right)^6\right) ,
\end{eqnarray}
where $c_2(T)$ and $c_4(T)$ are the expansion coefficients. The former coefficient is extracted using lattice QCD susceptibility calculations \cite{Borsanyi:2011sw} while the latter follows from ratios of the second and fourth order susceptibilities, computed hadron resonance and parton gas pictures. It is noteworthy that the lattice QCD EoS and baryon susceptibilities are known to agree with those of the resonance gas slightly below the crossover. For temperatures below the transition temperature $T_\mathrm{trans}(\mu_B)$, the lattice QCD EoS is smoothly matched to the hadron resonance gas EoS because the Taylor expansion is not well defined at lower $T$ and energy, momentum, and net baryon number need to be conserved at the Cooper-Frye freeze-out \cite{Monnai:2015sca}:
\begin{eqnarray}
\frac{\mathcal{P}}{T^4} &=& \frac{1}{2}\left(1- \tanh \frac{T-T_\mathrm{trans}}{\Delta T_\mathrm{trans}} \right) \frac{\mathcal{P}_{\mathrm{HRS}}(T,\mu_B)}{T^4} \notag \\
&+& \frac{1}{2}\left(1+ \tanh \frac{T-T_\mathrm{trans}}{\Delta T_\mathrm{trans}} \right) \frac{\mathcal{P}_{\mathrm{lat}}(T_s,\mu_B)}{T_s^4} .
\end{eqnarray} 
For the transition temperature of the two EoS, we use the ansatz $T_\mathrm{trans}(\mu_B) = 0.166~\mathrm{GeV} - 0.4 (0.139~\mathrm{GeV}^{-1} \mu_B^2 + 0.053~\mathrm{GeV}^{-3} \mu_B^4)$ motivated by a chemical freeze-out curve \cite{Cleymans:2005xv}. The shift $T_s = T + 0.4[T_\mathrm{trans}(0) - T_\mathrm{trans}(\mu_B)]$ is introduced to ensure that thermodynamic variables are increasing functions of $T$ and $\mu_B$ at very large baryon chemical potential, and should not affect much the bulk dynamics. The entropy density, the net baryon number, and the energy density can be obtained from the thermodynamic relations $s = \partial \mathcal{P}/\partial T \vert_{\mu_B}$, $n_B = \partial \mathcal{P}/\partial \mu_B \vert_{T}$, and $e = Ts - \mathcal{P} + \mu_B n_B$.
The speed of sound squared at the finite $\mu_B$ is computed as
\begin{equation}
c_s^2 (e, n_B) = \frac{\partial \mathcal{P}}{\partial e} \bigg\vert_{n_B} + \frac{n_B}{(e + \mathcal{P})} \frac{\partial \mathcal{P}}{\partial n_B} \bigg\vert_{e}.
\end{equation}
To see whether we should expect a large effect on the collision dynamics from the finite $\mu_B$ values present in smaller energy collisions, in Fig.~\ref{fig2B.1}a we plot $c_s^2$ as a function of local energy density for several constant $s/n_B$ values. Again, the shown values of $s/n_B$ correspond to the considered collision energies. From $\sqrt{s_\mathrm{NN}} = 14.5$ to $\sqrt{s_\mathrm{NN}} = 200$ GeV, the square of the speed of sound does not change significantly. The constant $s/n_B$ trajectories are shown in the $T-\mu_B$ plane in Fig.~\ref{fig2B.1}b.

In Appendix~\ref{appendix_C} we present several validation studies of our 3+1D numerical hydrodynamic implementation at finite baryon density.

\subsection{Particlization and hadronic cascade}
\label{sec:particlization}

\begin{figure}[ht!]
  \centering
  \includegraphics[width=1.0\linewidth]{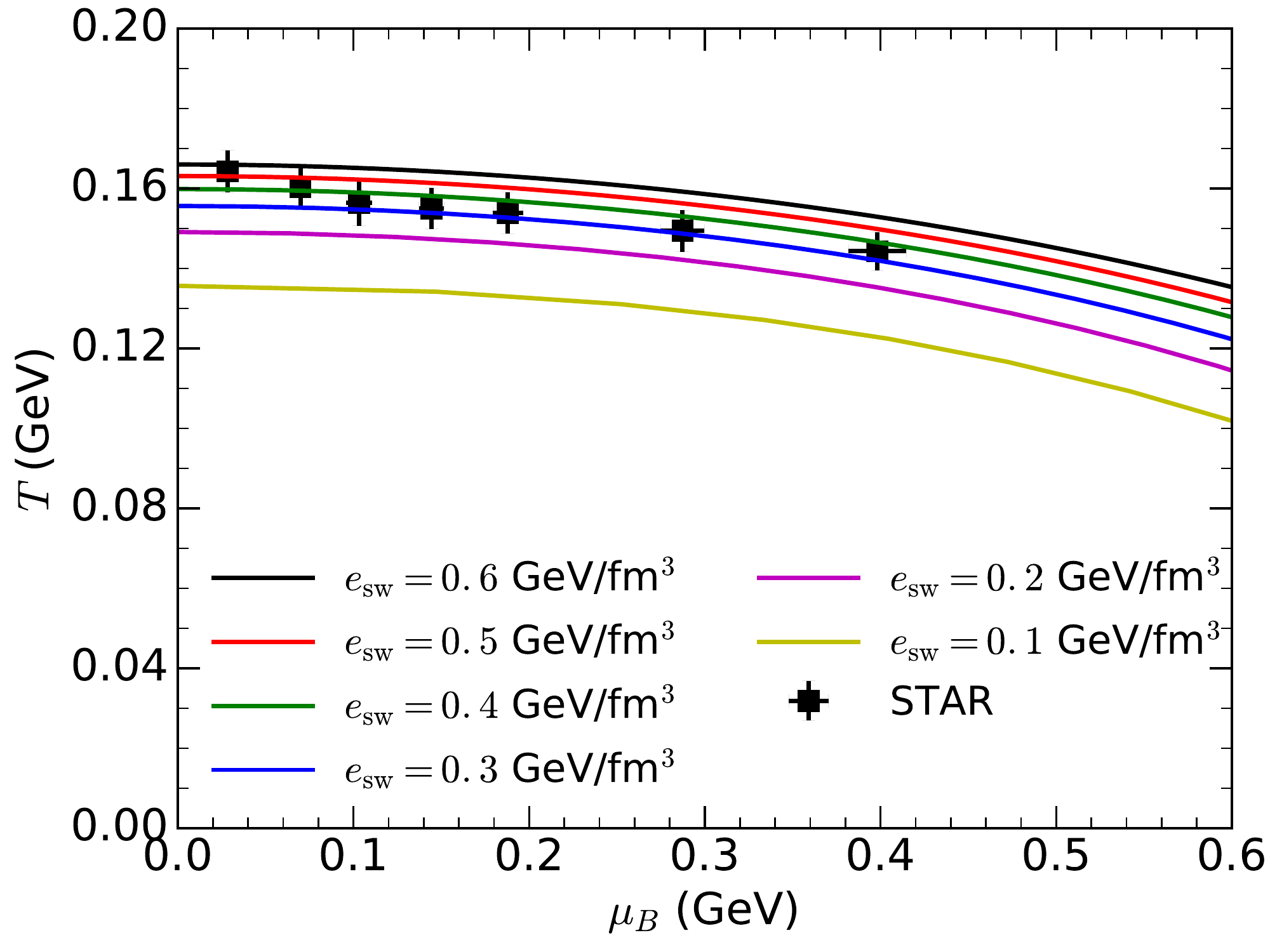}
   \caption{Constant energy density freeze-out lines compared with the extracted chemical freeze-out points from the STAR collaborations \cite{Adamczyk:2017iwn}. }
  \label{fig2B.2}
\end{figure}

As the temperature drops in the hadronic phase, we convert the macroscopic fluid cells into particle samples via the Cooper-Frye procedure \cite{Cooper:1974mv}. At finite $\mu_B$, we choose to perform the Cooper-Frye conversion on a constant switching energy density hyper-surface, $e_\mathrm{sw} = 0.4$ GeV/fm$^3$. This is because the chosen constant energy density line in the $T-\mu_B$ plane follows very well the chemical freeze-out points extracted from the thermal fits done by the STAR Collaboration \cite{Adamczyk:2017iwn}. This is demonstrated in Fig.~\ref{fig2B.2}, where we vary $e_\mathrm{sw}$.

%
%

Because of the long overlapping time at low collision energies, one would have expected that non-negligible amount of matter had already flown out of the switching hyper-surface before the hydrodynamic simulation starts. This is usually referred to as ``corona''. We define the corona as those cells whose local energy densities are between 0.05 GeV/fm$^3$ and $e_{sw} = 0.4$ GeV/fm$^3$. We use the Cooper-Frye formula to convert these corona fluid cells to particles at the first time step of the hydrodynamic evolution and then feed them into the hadronic transport simulation. Because there is no transverse flow velocity at the starting time of the hydrodynamic simulations, these corona particles are emitted isotropically according to their local thermal equilibrium distributions. The effect of the corona on hadronic flow observables will be discussed in the next section.

The momentum distribution of thermally emitted particles from one fluid cell is,
\begin{eqnarray}
E \frac{d^3 N_i}{d^3 p} &=& \frac{g_i}{(2\pi)^3}p^\mu \Delta^3 \sigma_\mu \bigg(f^\mathrm{eq}_i(E, T, \mu_B) \notag \\
&& + \delta f^\mathrm{shear}_i(E, T, \mu_B, \pi^{\mu\nu}) \notag \\
&& + \delta f^\mathrm{diffusion}_i(E, T, \mu_B, q^\mu) \bigg) \bigg\vert_{E = p \cdot u},
\label{eq.CooperFrye}
\end{eqnarray}
where the $\delta f^\mathrm{shear}$ and $\delta f^\mathrm{diffusion}$ are the out-of-equilibrium corrections from shear viscosity and net baryon diffusion. 
As in previous work \cite{Ryu:2015vwa} we employ
\begin{eqnarray}
 \delta f^{\rm shear}_i= f^\mathrm{eq}_i(x, p) (1 \pm f^\mathrm{eq}_i(x, p)) \frac{p^\mu p^\nu \pi_{\mu\nu}}{2 T^2 (e + P)}.
\end{eqnarray}
In the relaxation time approximation, the net baryon diffusion $\delta f^\mathrm{diffusion}$ for a single species of particle $i$ is \cite{Albright:2015fpa,Jaiswal:2015mxa}
\begin{eqnarray}
\delta f^\mathrm{diffusion}_i (x, p) &=& f^\mathrm{eq}_i(x, p) (1 \pm f^\mathrm{eq}_i(x, p)) \notag \\ 
&& \times \left(\frac{n_B}{e + \mathcal{P}} - \frac{b_i}{E} \right) \frac{p^{\langle \mu \rangle} q_\mu}{\hat{\kappa}_B},
\label{eq.diffusion_deltaf}
\end{eqnarray}
where $b_i$ is the baryon number of particle species $i$, $p^{\langle \mu \rangle}=\Delta^{\mu\nu}p_\nu$, and the transport coefficient $\hat{\kappa}_B$ is defined in Appendix~\ref{appendix_A}. An alternative form of diffusion out-of-equilibrium correction was recently derived using the 14-moment method \cite{Monnai:2018rgs}.
We note that $\delta f_i^\mathrm{diffusion}$ is non-zero even for mesons (that have zero baryon number). This is because the changes in the baryon chemical potential can lead to variations in the thermal pressure,  which will change the momentum distributions of mesons.
Using Eq.\,(\ref{eq.CooperFrye}) the system's total net baryon number can be computed as
\begin{eqnarray}
N^B - N^{\bar{B}} &=& \int d^3 \sigma_\mu \sum_i g_i b_i \notag \\
&& \times \int_p p^\mu (f^\mathrm{eq}_i + \delta f^\mathrm{shear}_i + \delta f^\mathrm{diffusion}_i) \notag \\
&=& \int d^3 \sigma_\mu (n_B u^\mu + q^\mu)\,,\label{eq:baryonNumber}
\end{eqnarray}
where $\int_p= \int \frac{d^3p}{E (2\pi)^3}$.
Because the hydrodynamic equation solves $\partial_\mu (n_B u^\mu + q^\mu) = 0$, the net baryon number is conserved during the hydrodynamic evolution as well as on the conversion surface before and after the conversion. 
The inclusion of $\delta f^\mathrm{diffusion}_i$ in the Cooper-Frye formula takes into account contributions from the diffusion current $q^\mu$ in Eq.\,(\ref{eq:baryonNumber}) and is essential to ensure the conservation of net baryon number during the conversion from fluid cells to particles. 

In this work, we generalized the publicly available numerical code \textsc{iSS}\footnote{The latest version of the code package can be downloaded from \url{https://github.com/chunshen1987/iSS}.} to perform the particlization simulations. Detailed implementation and cross checks for the numerical procedure are discussed in Appendix~\ref{appendix_B}.

After the particle conversion, we feed particles into hadronic cascade models, such as UrQMD \cite{Bass:1998ca,Bleicher:1999xi} and JAM\footnote{The latest version of JAM can be downloaded from \url{http://www.aiu.ac.jp/~ynara/jam/}} \cite{Nara:1999dz}, to simulate the transport dynamics in the dilute hadronic phase.

\section{Collectivity in Au+Au collisions at RHIC BES energies}\label{sec:results}

We will focus our study of hadronic flow observables on central and semi-peripheral Au+Au collisions at 19.6 GeV. At this collision energy, the baryon chemical potential can reach up to $\sim$200 MeV in the mid-rapidity region. Consequently, we expect the net baryon current and its diffusion to have sizeable effects on the hadronic flow observables near the mid-rapidity region which can be measured by the STAR experiments.

\subsection{Hydrodynamical evolution with net baryon diffusion}\label{sec:evolution}

Based on the hydrodynamic equations of motion, the net baryon diffusion current only directly affects the evolution of the net baryon density. Nevertheless, it modifies the system's energy density and flow velocity evolution indirectly, via the modification of the pressure $\mathcal{P}(e, n_B)$, given by the equation of state. Thus we expect this dissipative current to have less influence on the system's evolution compared to the usual dissipative effects due to shear and bulk viscosities.

To understand the effect of net baryon diffusion on hadronic flow observables, it is instructive to study the time evolution of $\mu_B/T$, whose spatial gradients are the thermal dynamic force of the net baryon diffusion current, $q^\mu$. 

\begin{figure}[ht!]
  \centering
  \includegraphics[width=0.95\linewidth]{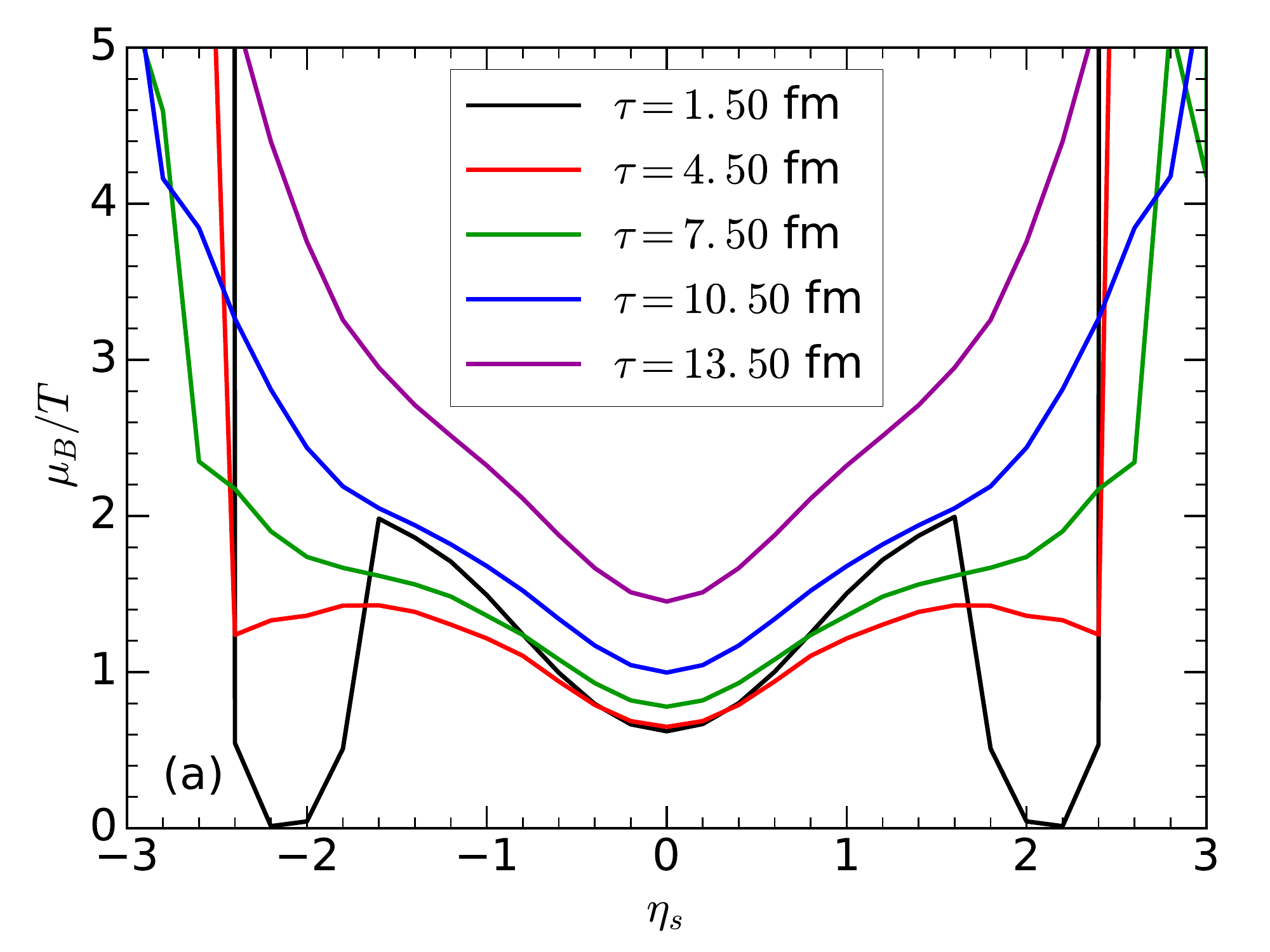}
  \includegraphics[width=0.95\linewidth]{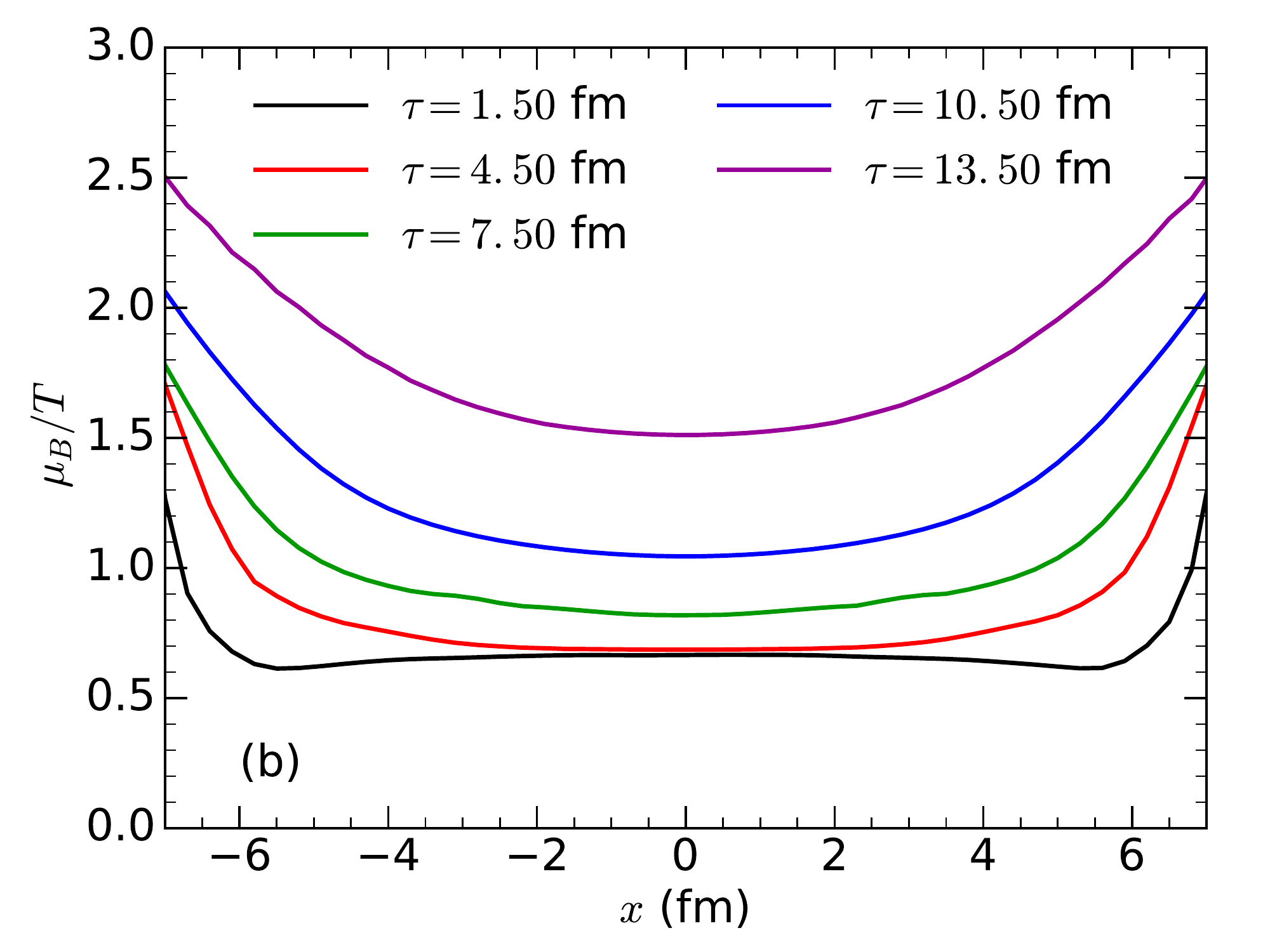}
   \caption{Time evolution of the $\mu_B/T$ along longitudinal direction for points at $x = 0$ and $y = 0$ (panel (a)) and transverse plane along $y = 0$ and $\eta_s = 0$ (panel (b)).}
  \label{fig3A.0}
\end{figure}
%
Figure \ref{fig3A.0} shows the time evolution of $\mu_B/T$ along the longitudinal and transverse directions. At the starting time of the hydrodynamic simulation, the ratio $\mu_B/T$ peaks around $\eta_s = \pm 1.5$. The gradients of $\mu_B/T$ dominantly point to the mid-rapidity region. Thus the baryon diffusion current will transport more baryons from forward rapidities to the central rapidity region. We also find that the value of $\mu_B/T$ increases in dilute energy density regions in both very forward rapidity and in towards the edges in the transverse plane. Such a distribution leads to the spatial gradients of $\mu_B/T$ pointing opposite to the pressure gradients. From these observations in the longitudinal and transverse directions, we expect that the net baryon diffusion current $q^\mu$ will act against the hydrodynamic flow, and will reduce the net baryon flow coefficients. 

\subsection{Effects of baryon diffusion on observables}

In this section, we study how baryon diffusion in the hydrodynamic simulations affects various experimental observables. We vary the amount of diffusion by tuning the value of the pre-factor $C_B$ in Eq.~(\ref{eq2.17}).

\begin{figure}[ht!]
  \centering
  \includegraphics[width=0.95\linewidth]{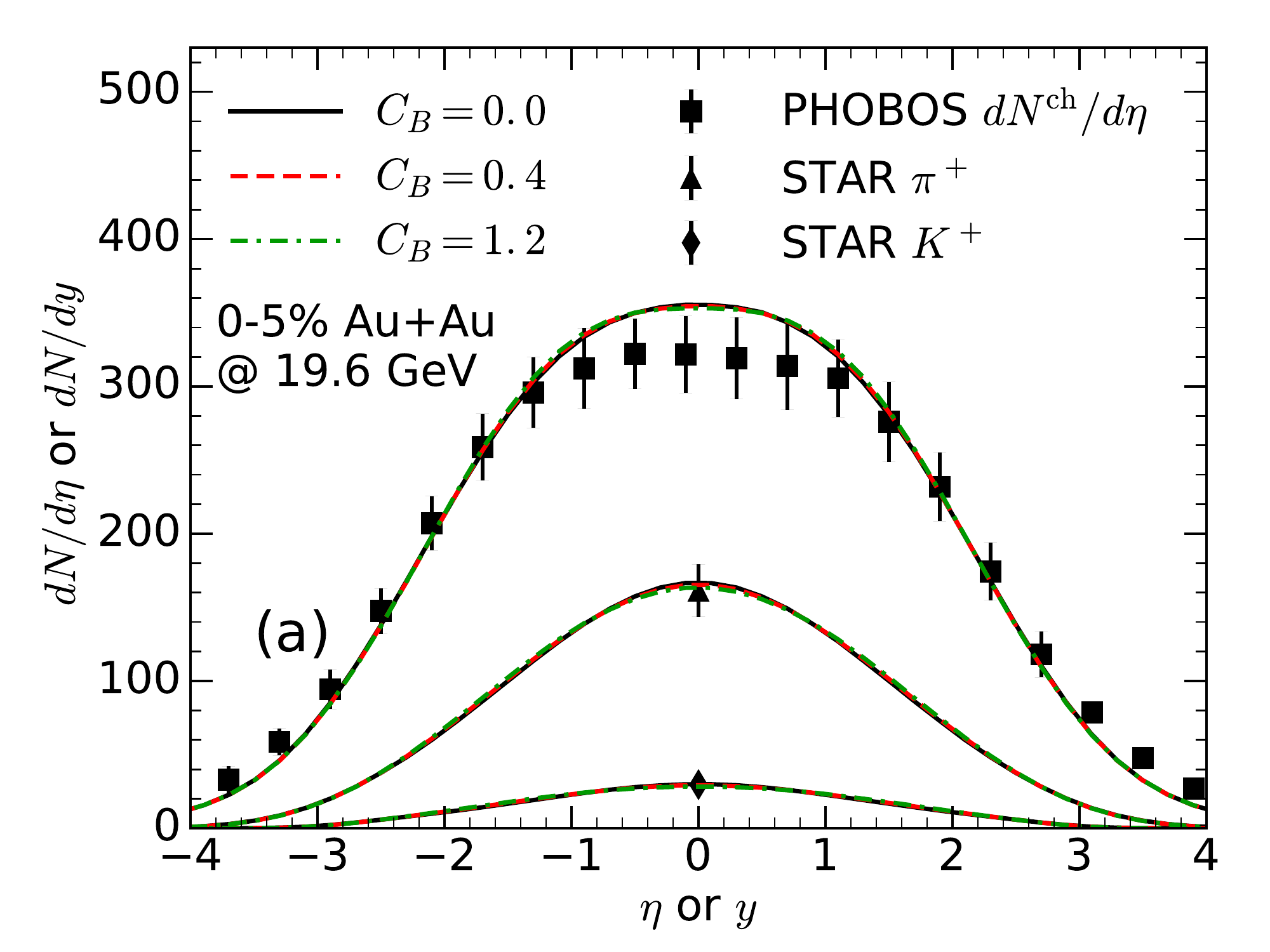}
  \includegraphics[width=0.95\linewidth]{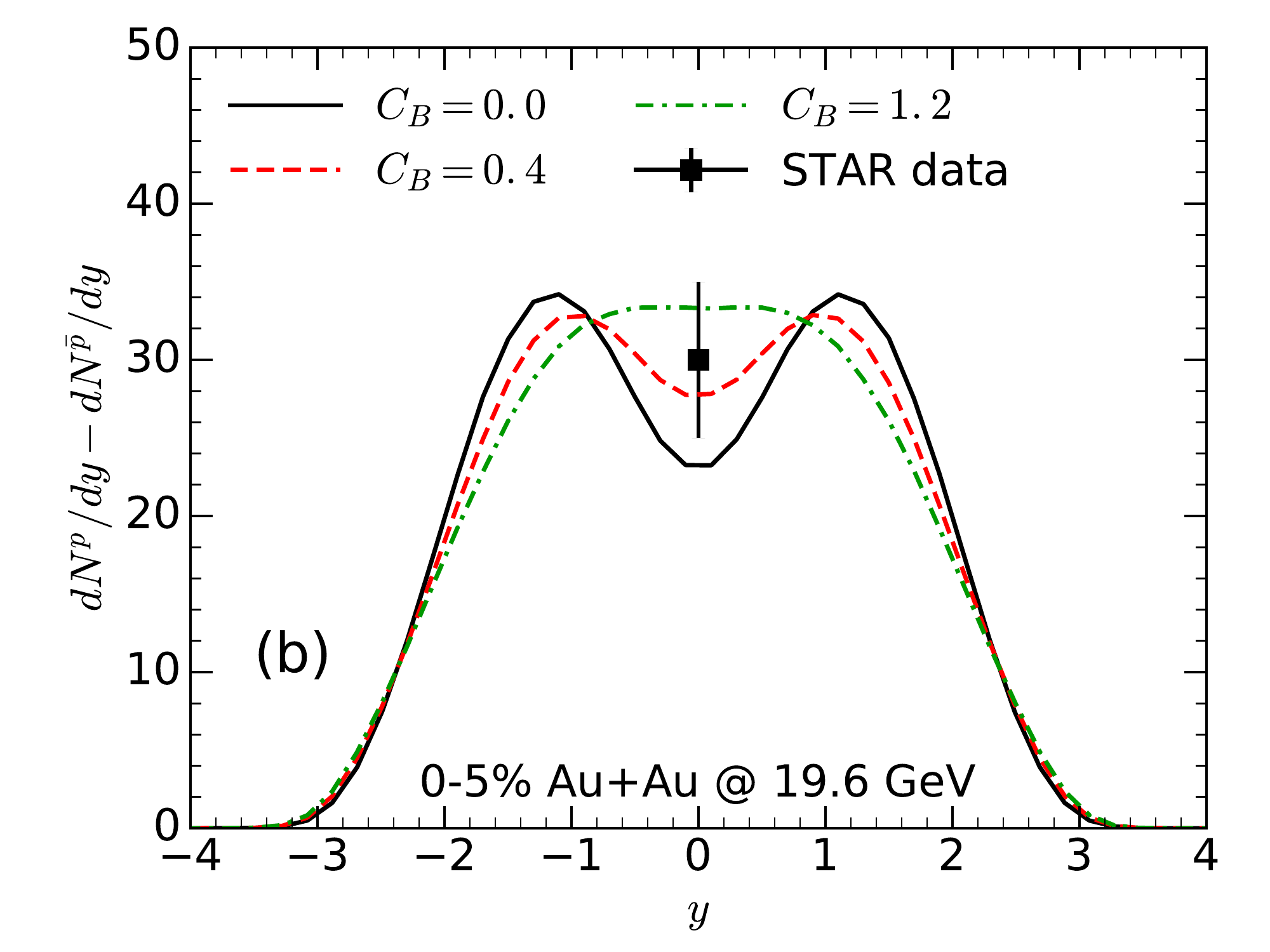}
   \caption{{\it Panel (a)}: The pseudo-rapidity and rapidity distributions of the charged hadron, identified $\pi^+$, and $K^+$ compared to the PHOBOS and STAR measurements in 0-5\% Au+Au collisions at 19.6 GeV\cite{Back:2005hs,Adamczyk:2017iwn}. {\it Panel (b)}: The net proton rapidity distribution with different choices of the net baryon diffusion constant compared with the STAR measurements.}
  \label{fig3A.1}
\end{figure}
%

Figure \ref{fig3A.1}a shows the rapidity distribution of produced hadrons in the top 0-5\% central Au+Au collisions at $\sqrt{s}=$19.6 GeV. The system's total entropy is tuned to reproduce the positive pion yield at mid-rapidity, measured by the STAR collaboration \cite{Adamczyk:2017iwn}. The rapidity envelope profile is tuned to reproduce the rapidity dependence measured by the PHOBOS collaboration \cite{Back:2005hs}. The charged hadron multiplicity is slightly overestimated mainly because the PHOBOS measurement is for the 0-6\% centrality. The net baryon diffusion has negligible effect on mesons and charged hadrons.

In Fig.~\ref{fig3A.1}b we demonstrate that the rapidity dependence of net protons is sensitive to the magnitude of the baryon diffusion coefficient. As discussed in Sec.~\ref{sec:evolution}
, the baryon diffusion current is driven by gradients of $\mu_B/T$, which transports net baryons from forward rapidity to the mid-rapidity region. This effect is visibly stronger for larger $C_B$.

\begin{figure}[ht!]
  \centering
  \begin{tabular}{cc}
  \includegraphics[width=0.95\linewidth]{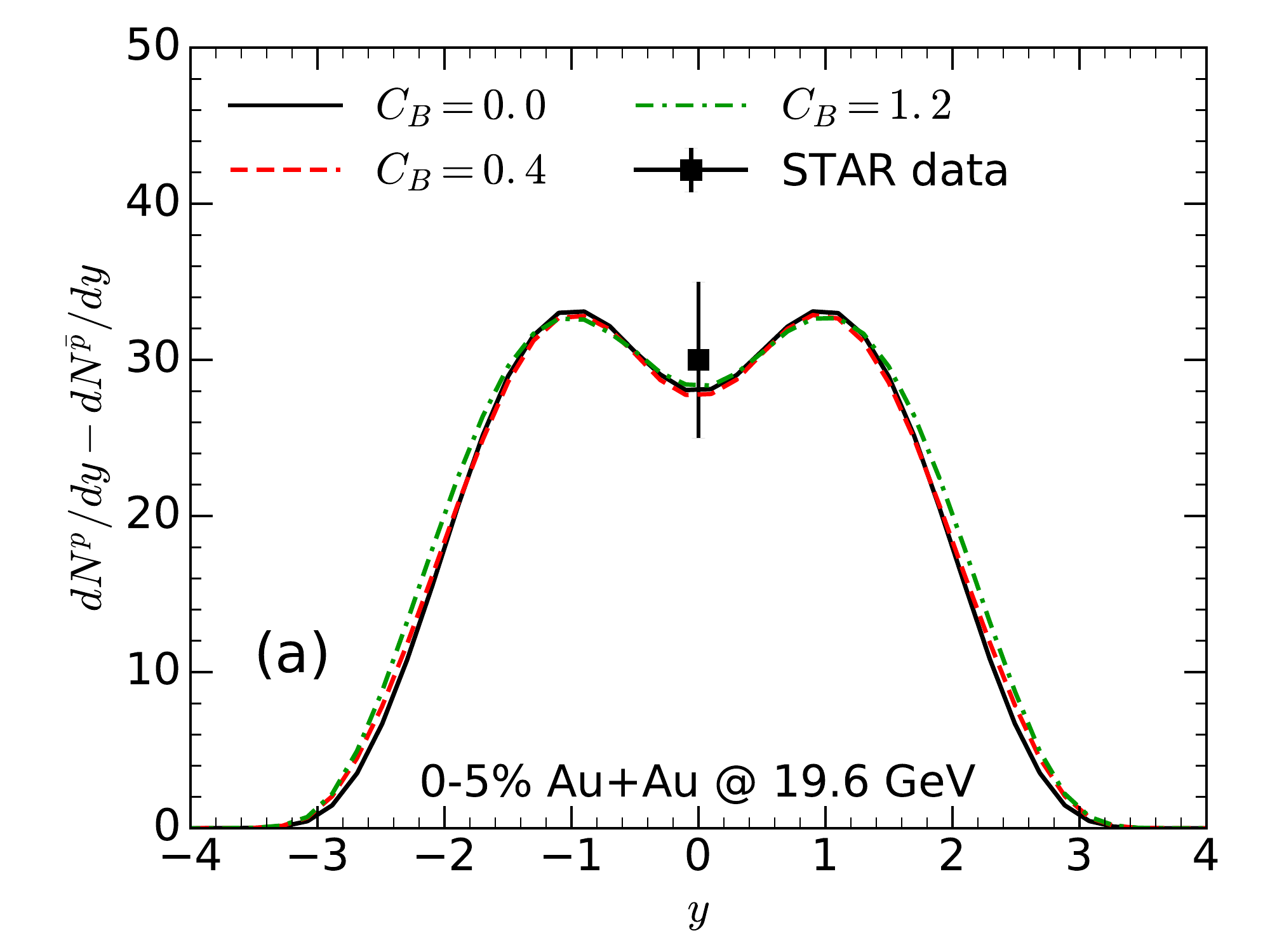} \\
  \includegraphics[width=0.95\linewidth]{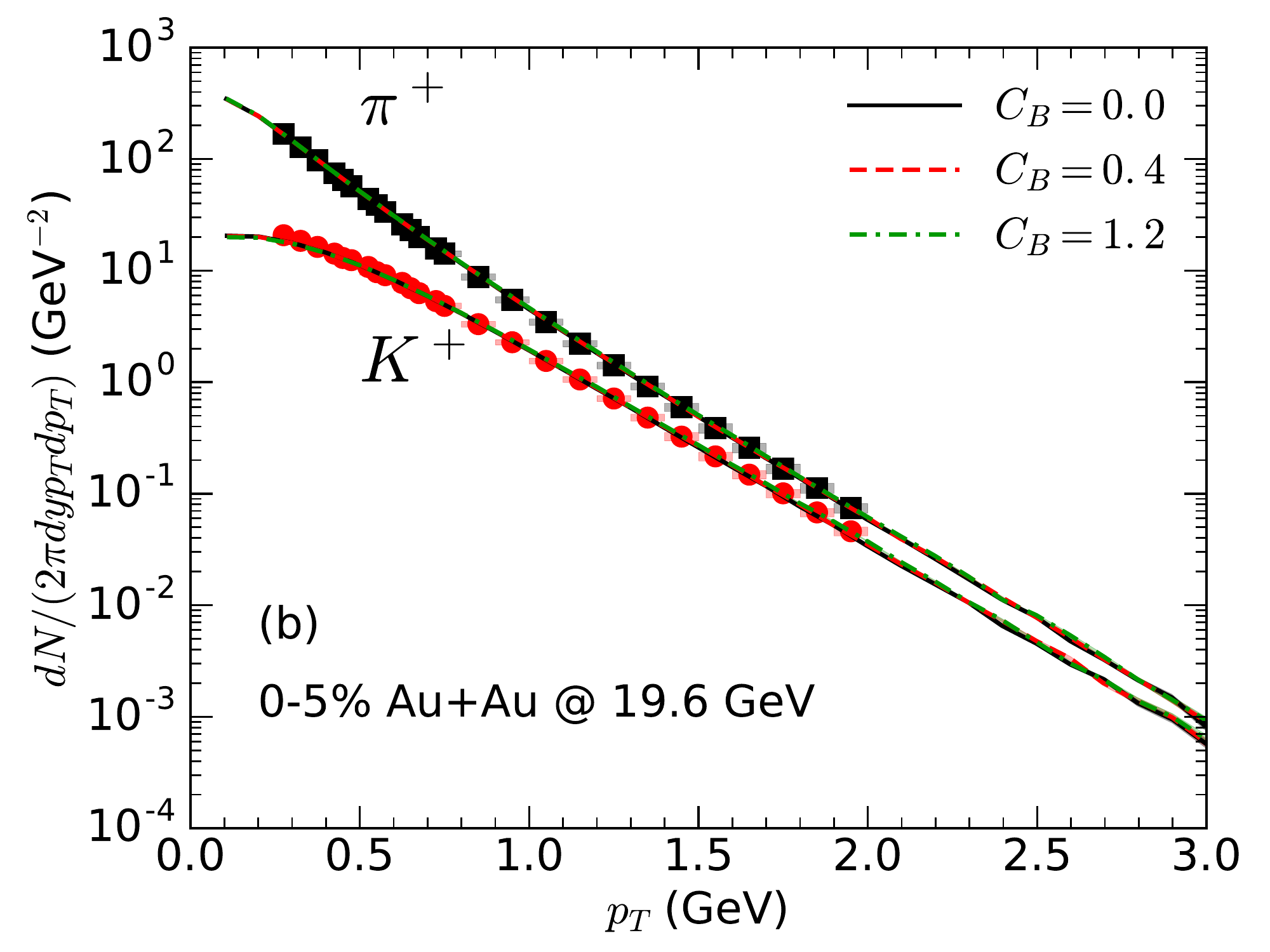} \\
  \includegraphics[width=0.95\linewidth]{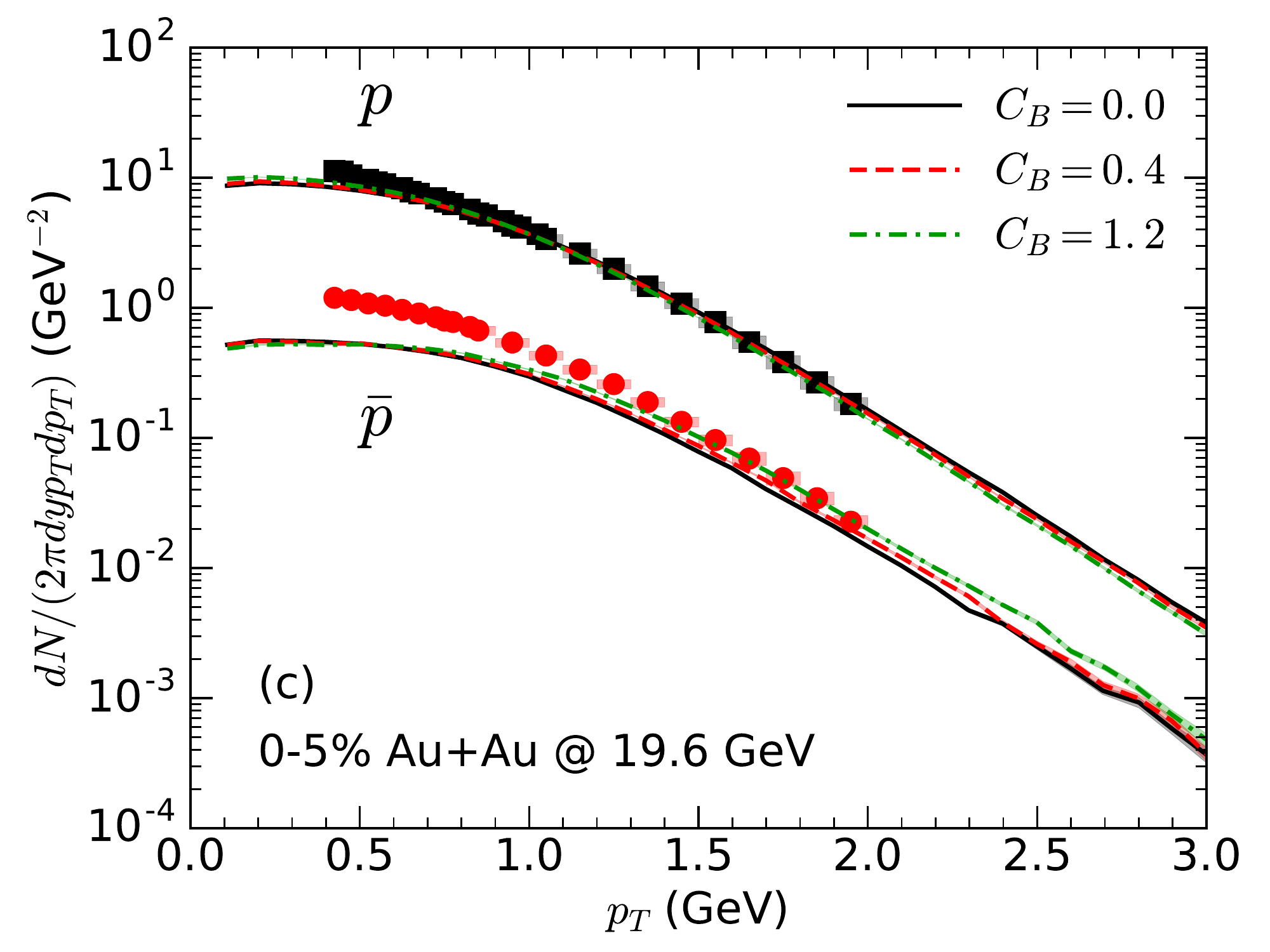}
  \end{tabular}
   \caption{{\it Panel (a)}: The fit of the net proton rapidity distribution with different choices of the net baryon diffusion constant in the simulations. {\it Panel (b)}: The single particle spectra of $\pi^+$ and $K^+$ with different choices of the net baryon diffusion constant. {\it Panel (c)}: The single particle spectra of proton and anti-proton with different choices of the net baryon diffusion constant. }
  \label{fig3A.2}
\end{figure}

Unfortunately, the measured shape of the net proton rapidity distribution cannot be used to constrain the amount of net baryon diffusion, because of the theoretical uncertainties in determining the initial baryon stopping. This is explicitly demonstrated in Fig.~\ref{fig3A.2}a, where we have adjusted the initial baryon rapidity distribution for given values of $C_B$. This shows that approximately the same final distribution is obtained for largely different baryon diffusion currents. 

%

It is possible to find further constraints by considering both experimental data sensitive to longitudinal and transverse dynamics simultaneously. For a given $C_B$, the initial condition can be constrained by the net proton rapidity distribution as above, and studying the transverse dynamics of the collision system could then be used to distinguish different $C_B$ values. 

Figures \ref{fig3A.2}b and \ref{fig3A.2}c show transverse momentum spectra of identified particles. The $p_T$-spectra of light mesons, such as $\pi^+$ and $K^+$, are insensitive to the net baryon diffusion as expected. Proton and anti-proton spectra obtained using different degrees of net baryon diffusion are compared in Fig.~\ref{fig3A.2}c. The effect of the net baryon diffusion constant $C_B$ looks small in the plot. To better quantify the effect, we compare the difference in the average transverse momentum of protons and anti-protons. The result is shown in Table~\ref{table3} for different choices of $C_B$. 

\begin{table}[ht!]
  \centering
  \begin{tabular}{c|c|c|c}
  \hline \hline
    & $C_B = 0.0$ & $C_B = 0.4$ & $C_B = 1.2$ \\ \hline
   $\langle p_T \rangle(\bar{p}) - \langle p_T \rangle(p)$ (GeV) & 0.049 & 0.079 & 0.134  \\ \hline
   $\langle p_T \rangle(\bar{p}) - \langle p_T \rangle(p)$ (GeV) &  & &  \\
   (no diffusion $\delta f$) & 0.049 & 0.050 & 0.056  \\
   \hline \hline
   \end{tabular}
  \caption{The difference of the averaged transverse momentum between anti-protons and protons at different values of the net baryon diffusion constant $C_B$.}
  \label{table3}
\end{table}
%

The hydrodynamic simulation produces a slightly larger mean-$p_T$ for anti-protons than for protons. This difference grows with increasing net baryon diffusion. Part of this effect is caused by the diffusive evolution, because the $\mu_B/T$ gradient in the transverse plane tends to diffuse net baryon number into the central region where the radial flow is relatively smaller. An even larger contribution to the mean-$p_T$ difference is due to the baryon diffusion $\delta f$ corrections to the baryon spectra. We will discuss this effect in more detail in the next section.

\begin{figure}[ht!]
  \centering
  \includegraphics[width=0.9\linewidth]{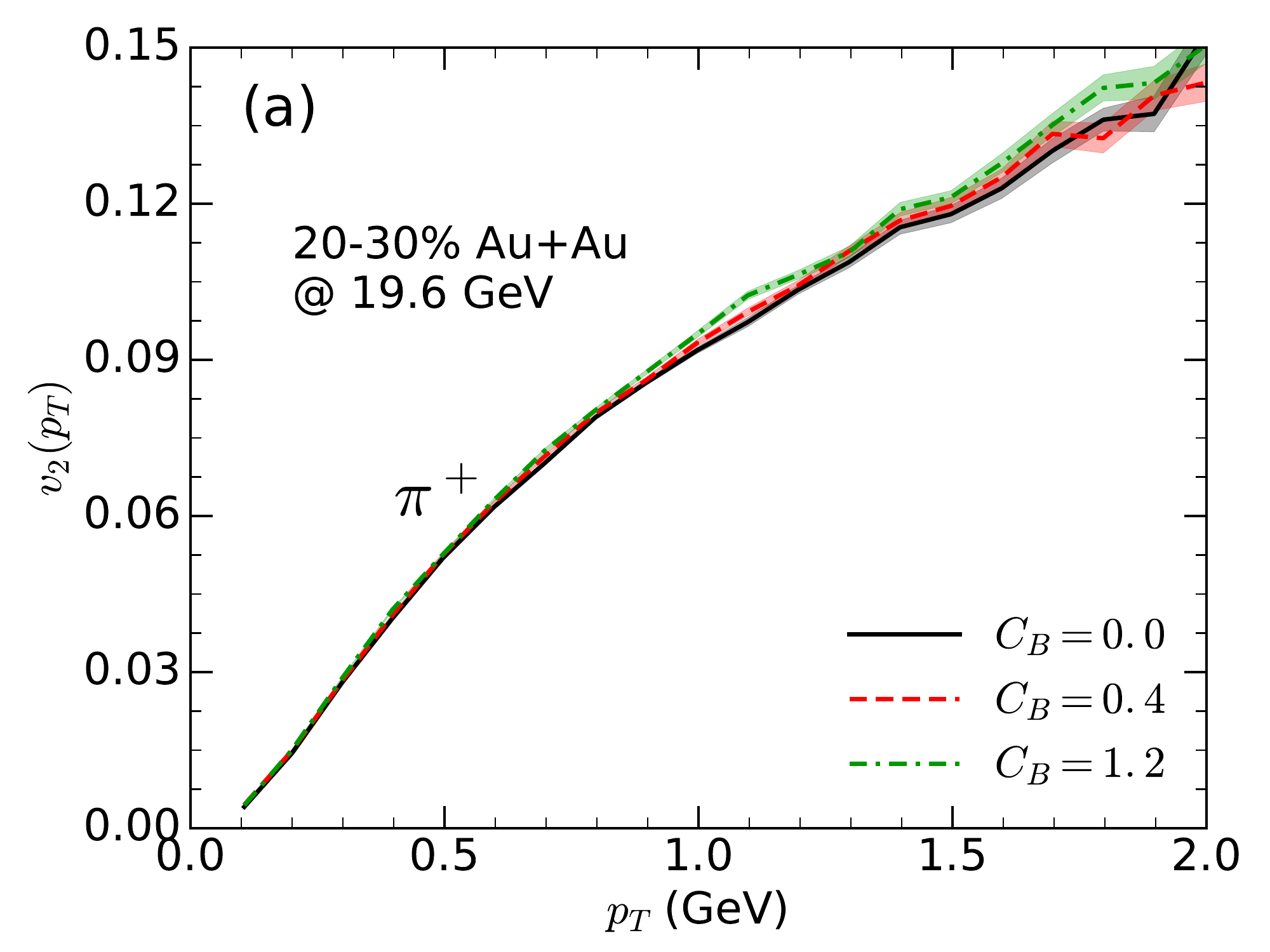}
  \includegraphics[width=0.9\linewidth]{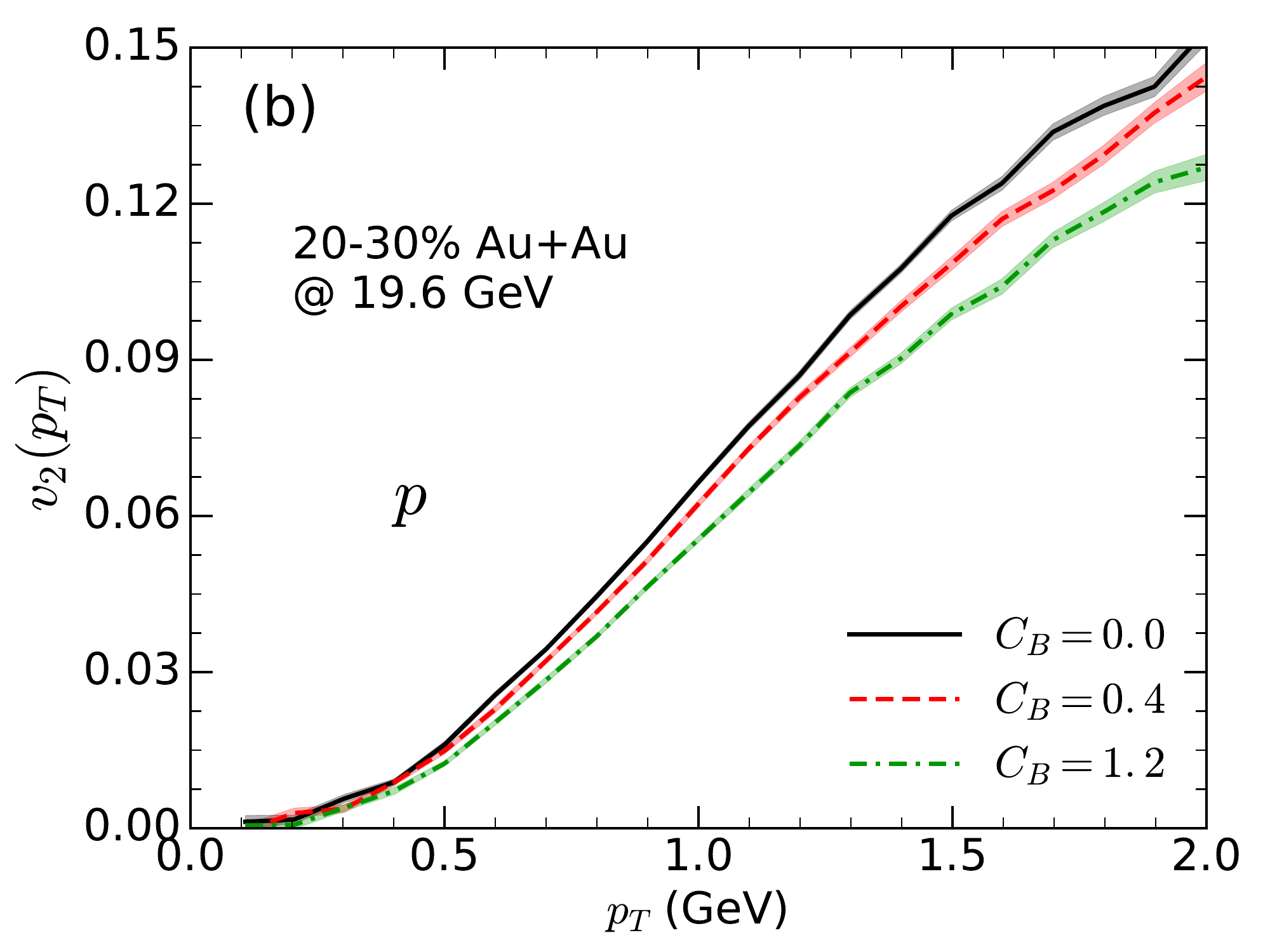}
  \includegraphics[width=0.9\linewidth]{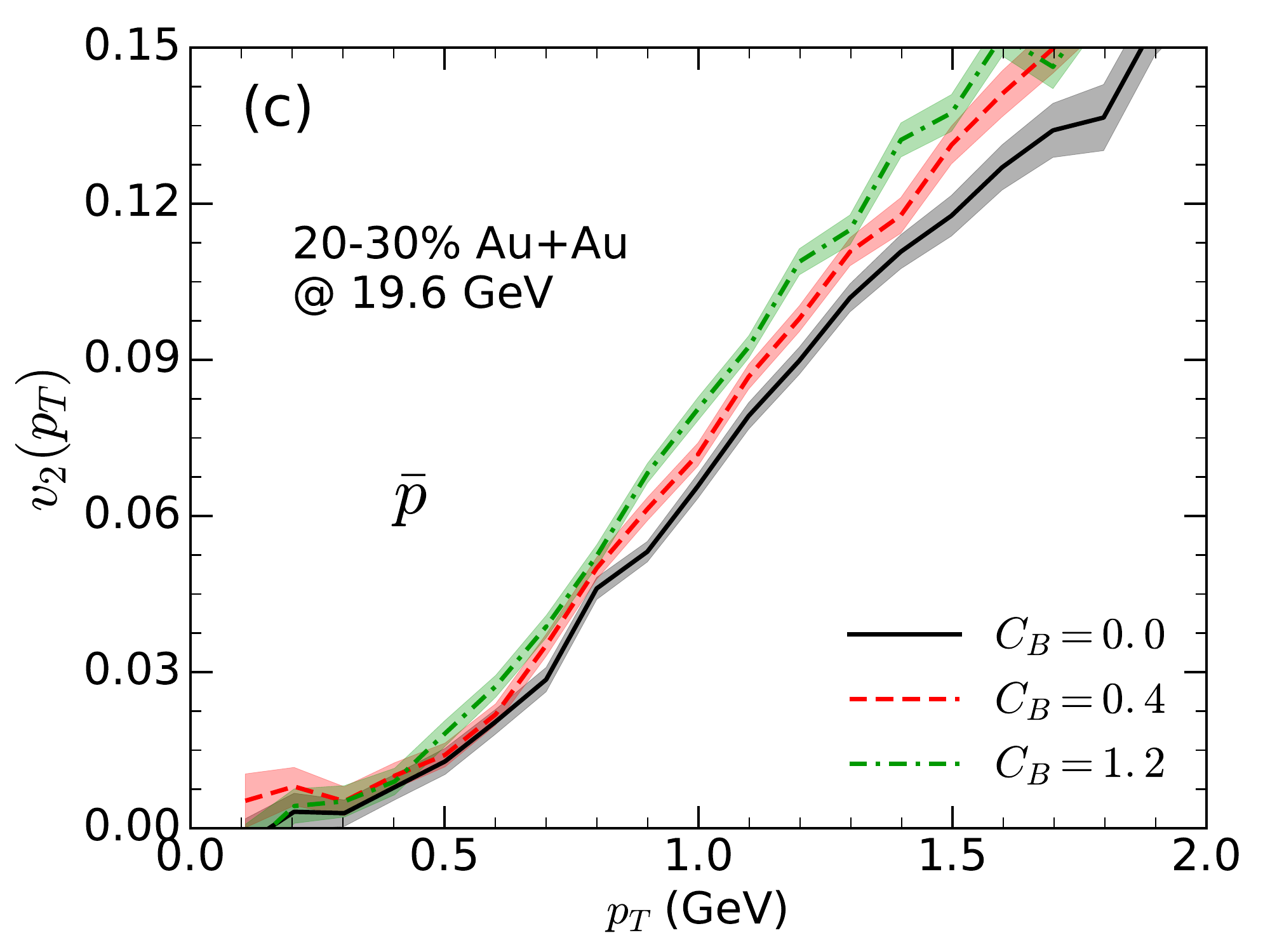}
   \caption{The $p_T$-differential elliptic flow coefficients of identified $\pi^+$ (panel (a)), $p$ (panel (b)), and $\bar{p}$ (panel (c)) with different net baryon diffusion constants in the hydrodynamic simulations for 20-30\% Au+Au collisions at 19.6 GeV. The shaded bands indicate statistical errors.}
  \label{fig:v2}
\end{figure}
%

In Fig.\,\ref{fig:v2} we show the transverse momentum elliptic flow coefficient $v_2$ for positive pions (panel a), protons (b), and anti-protons (c). As for the transverse momentum spectra, pion $v_2$ does not change within statistical errors with the value of $C_B$. For protons and anti-protons we find a sizeable and opposite effect: Proton $v_2$ decreases with increasing $C_B$, while anti-proton $v_2$ increases. We will show in the following section that the differences are to a large part generated by off-equilibrium corrections to the distribution functions. 

We find that the net baryon diffusion has a small influence on the system's transverse dynamics, such as the hydrodynamic flow pattern. The major effects to the baryonic observables are coming from the off-equilibrium corrections at the freeze-out stage. Because of this, baryon diffusion cannot be constrained quantitatively from experimental data in our current analyses.

It should be noted that unlike the other particle species, $\bar{p}$ is not well described for any $C_B$. We expect that it should be possible to improve the agreement by fine tuning both the initial entropy and net-baryon distributions, as well as the switching energy density.

\subsection{The effects of the out-of-equilibrium corrections from net baryon diffusion}

\begin{figure}[ht!]
  \centering
  \includegraphics[width=0.95\linewidth]{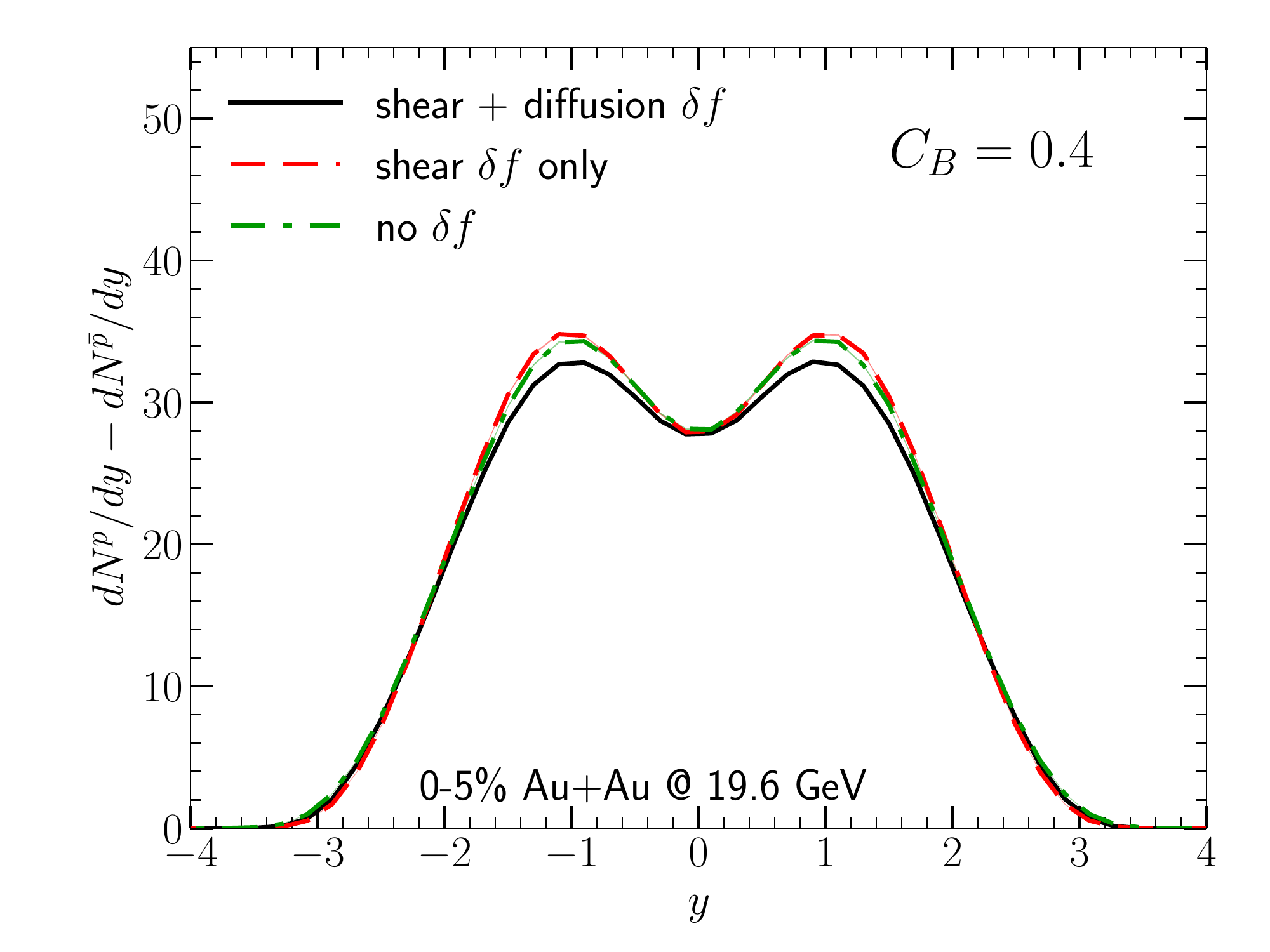}
   \caption{The effects of the out-of-equilibrium corrections $\delta f$ from shear viscosity and net baryon number diffusion on the net proton rapidity distribution.}
  \label{fig3B.1}
\end{figure}

Figure \ref{fig3B.1} shows the effect of the out-of-equilibrium corrections to the particle distributions on net proton rapidity spectra. We show separately the effect from shear viscous corrections and baryon diffusion corrections. The $\delta f$ correction from the net baryon diffusion current reduces the net proton yield. As we discussed in Sec. \ref{sec:particlization}, this $\delta f$ correction is essential to conserve the total net baryon number during the Cooper-Frye conversion procedure.

%
Please note that a non-zero net baryon diffusion current on the freeze-out surface modifies identified particle yields as shown in Eq.\ (\ref{B10}) in Appendix \ref{appendix_B}. Thus, the non-equilibrium evolution of the baryon diffusion current will give corrections the chemical freeze-out parameters determined in thermal model fits \cite{Adamczyk:2017iwn}. Because the baryon diffusion $\delta f$ reduces the net proton yield, the averaged chemical potential on the freeze-out surface is about 30 MeV larger with baryon diffusion compared to the simulations without diffusion.

\begin{figure}[ht!]
  \centering
  \includegraphics[width=0.9\linewidth]{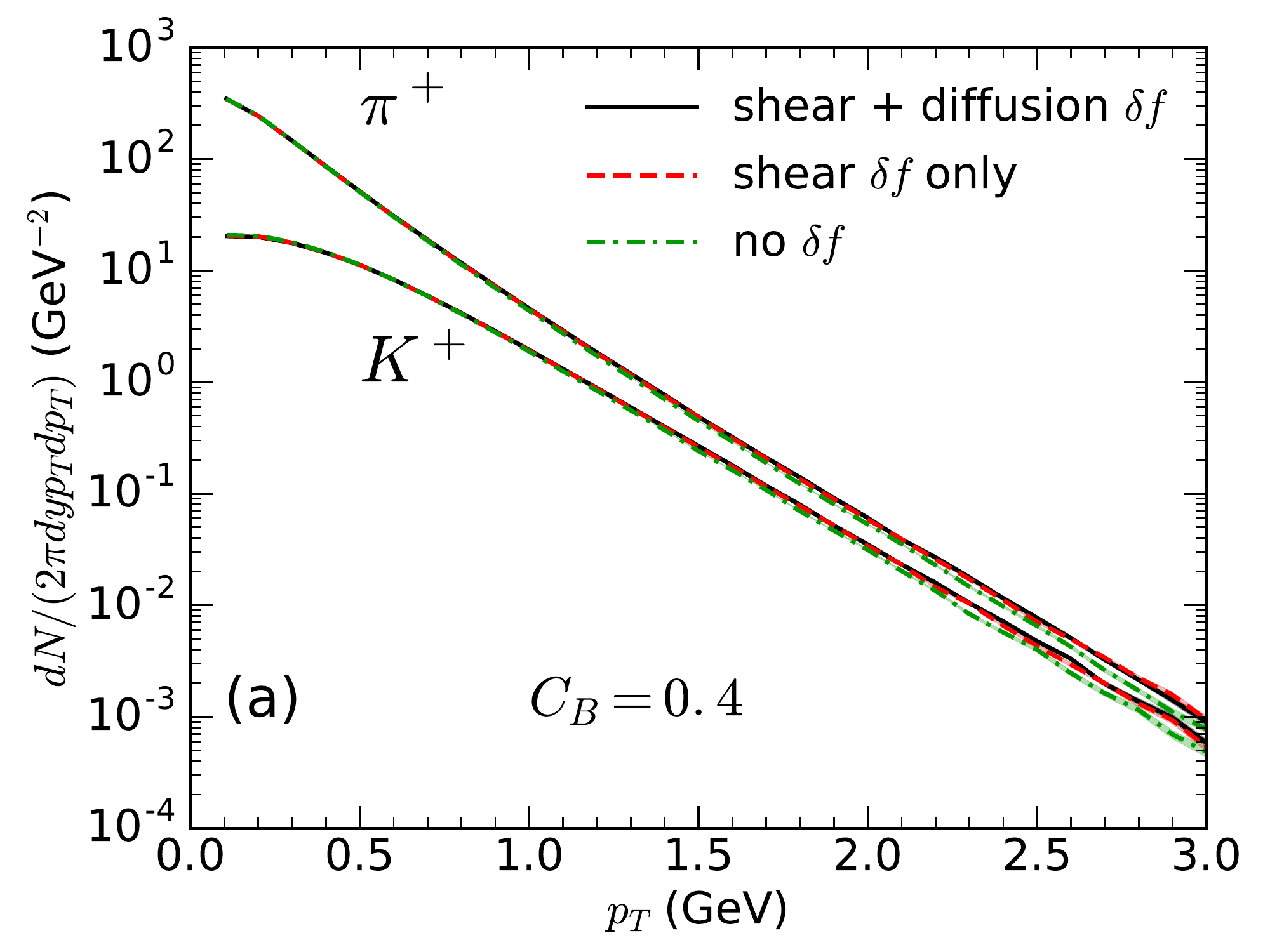}
  \includegraphics[width=0.9\linewidth]{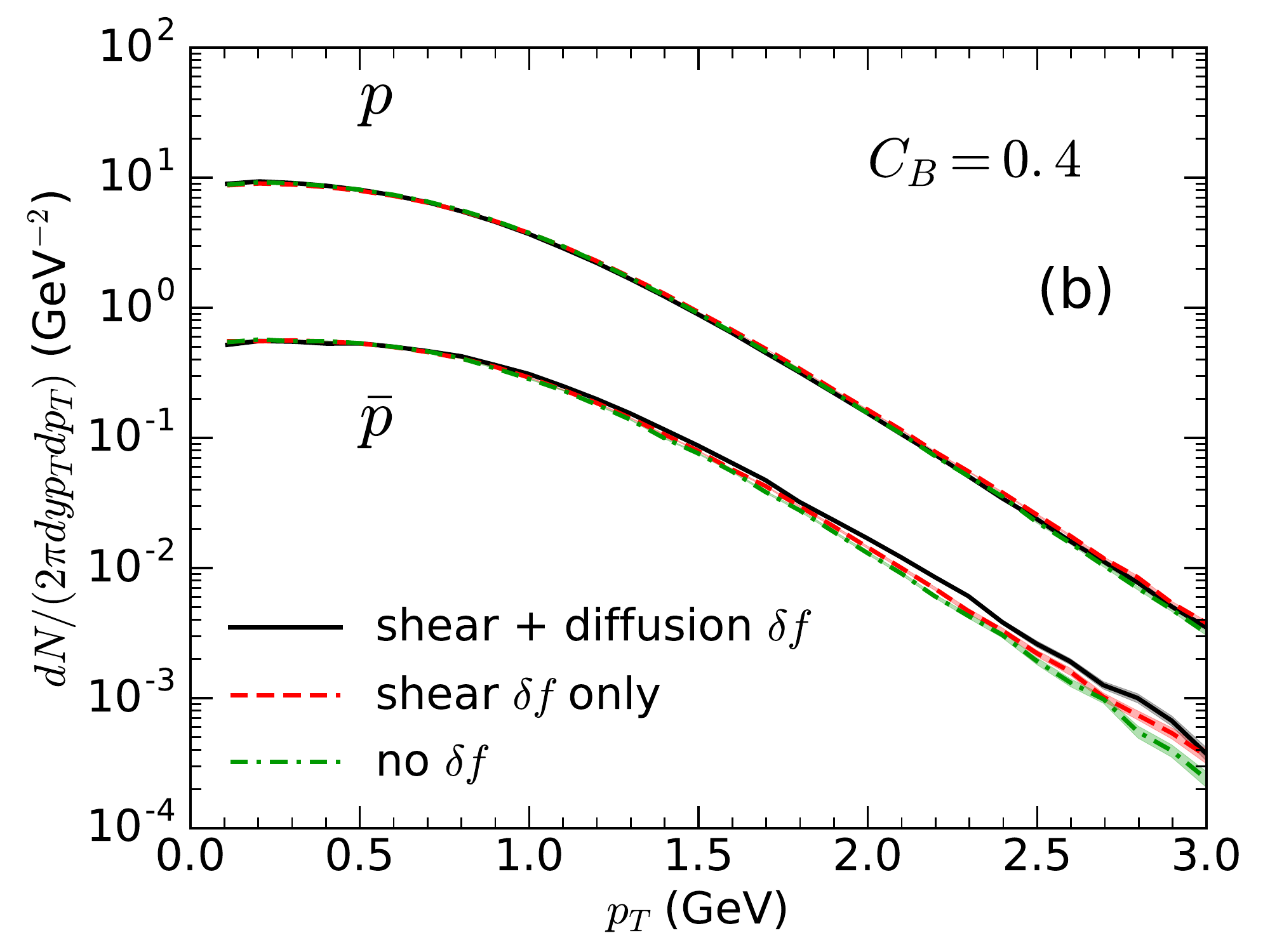}
   \caption{The correction of shear and net baryon diffusion $\delta f$ to the transverse momentum spectra of $\pi^+$, $K^+$, $p$, and $\bar{p}$ at the mid-rapidity in the hybrid simulations. }
  \label{fig:deltaf-pt}
\end{figure}
%

\begin{figure}[ht!]
  \centering
  \includegraphics[width=0.9\linewidth]{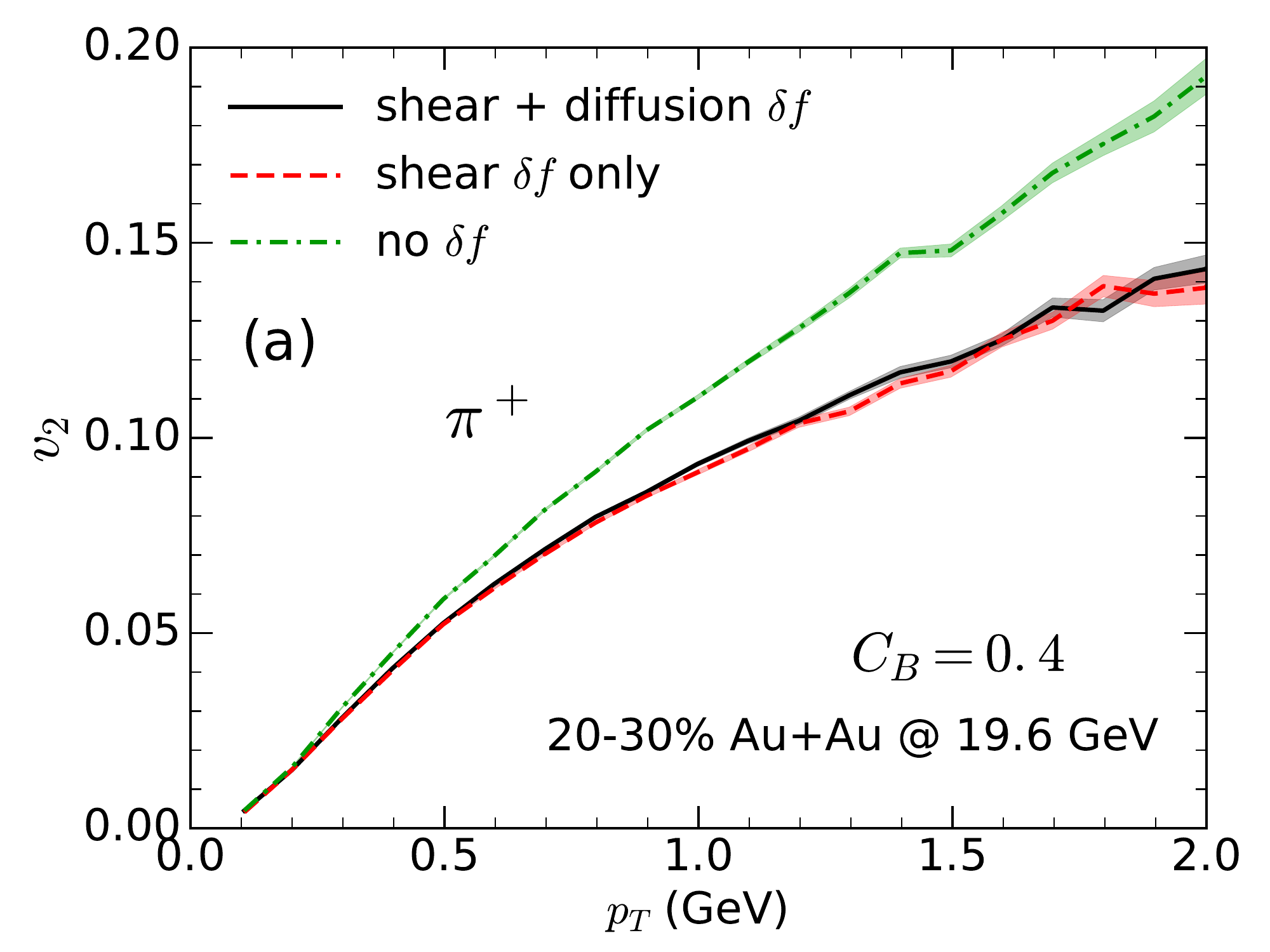}
  \includegraphics[width=0.9\linewidth]{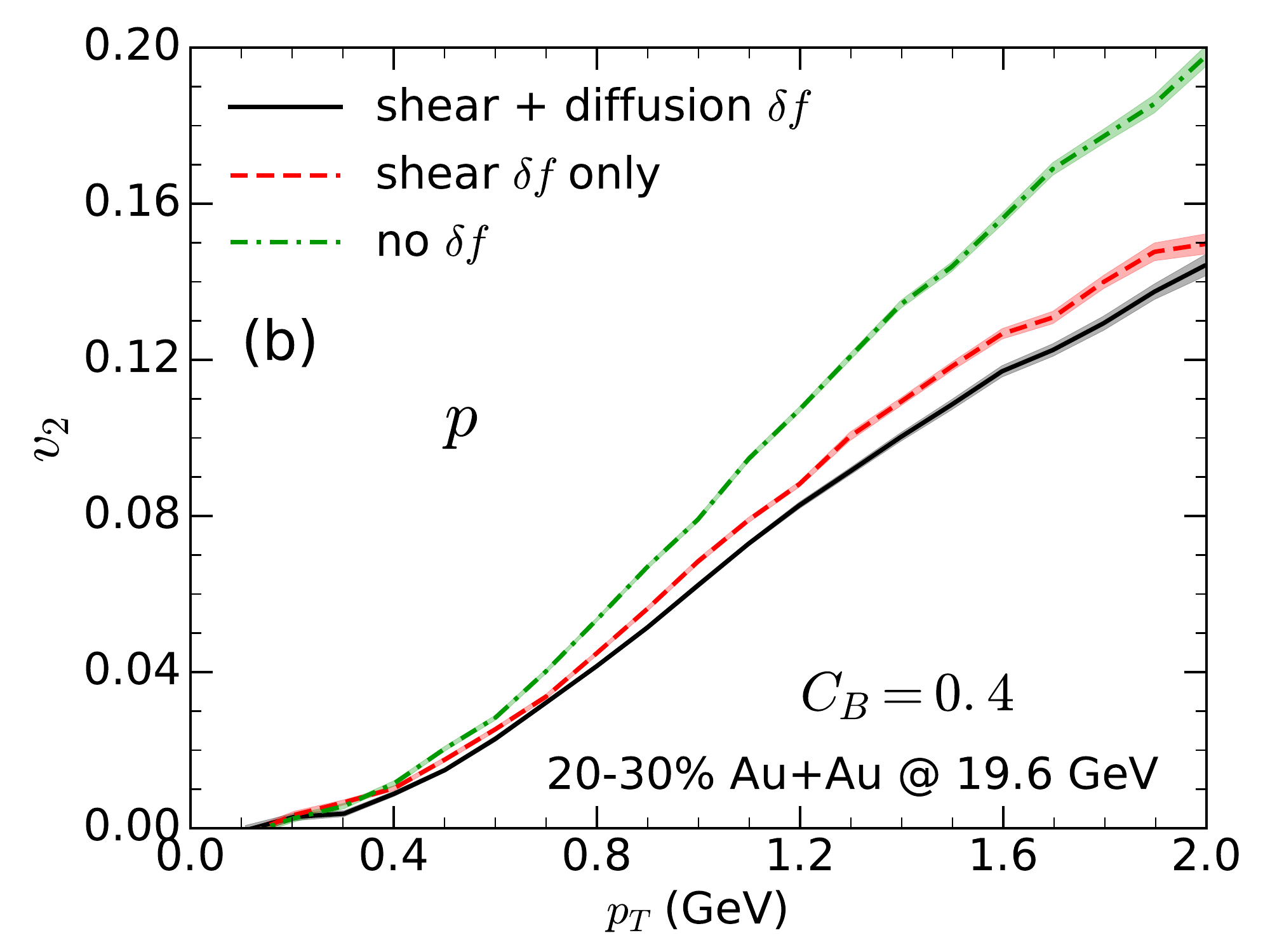}
  \includegraphics[width=0.9\linewidth]{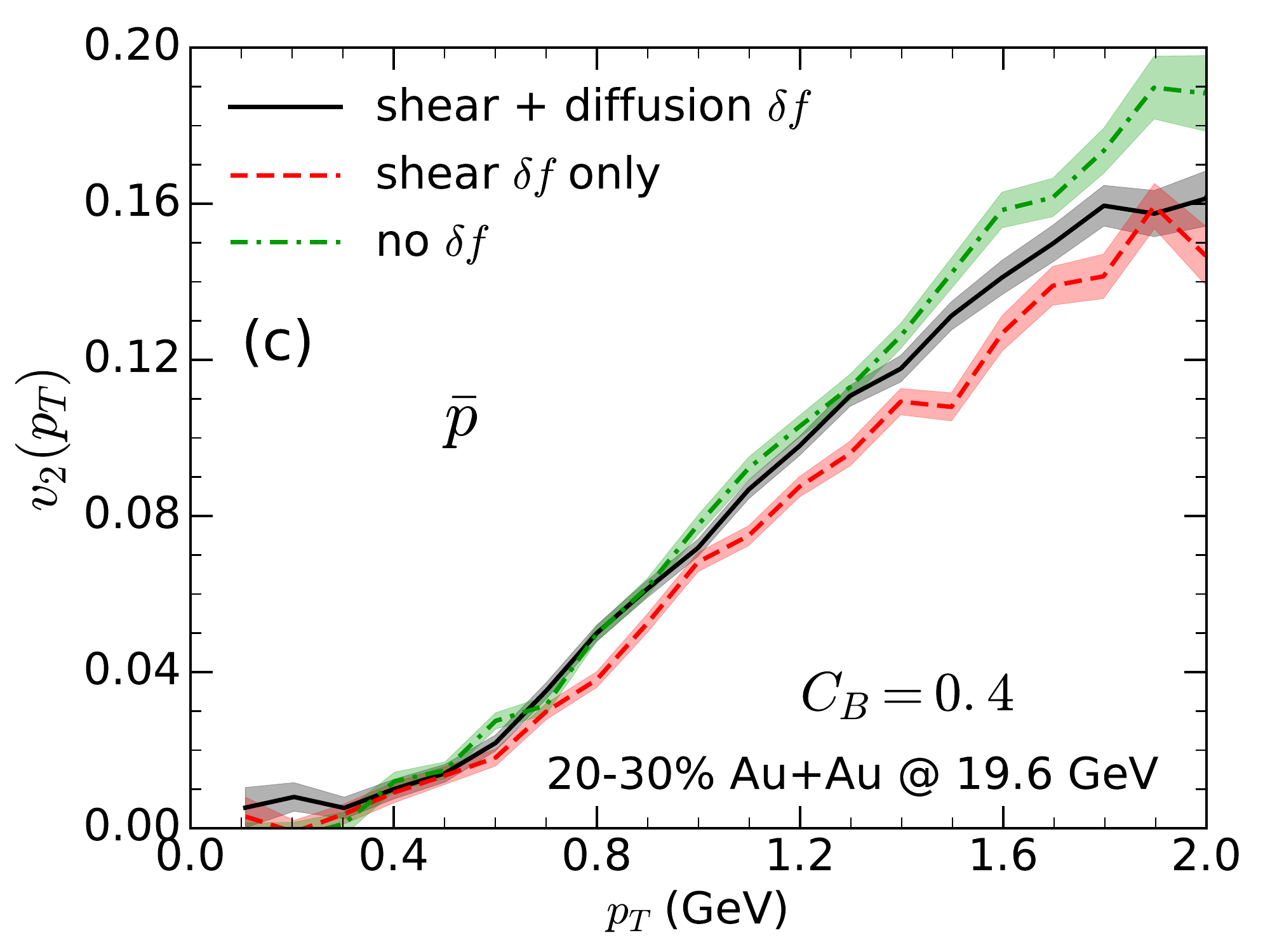}
   \caption{The correction of shear and net baryon diffusion $\delta f$ to the $p_T$-differential $v_2$ of $\pi^+$, $p$, and $\bar{p}$ in hybrid simulations. The shaded bands indicate statistical errors.}
  \label{fig:deltaf-v2}
\end{figure}
%

In Fig.\,\ref{fig:deltaf-pt} we show identified particle spectra and their dependence on shear and baryon diffusion $\delta f$ corrections. For all species the effects are small, with the largest difference visible for anti-protons at $p_T>2$ GeV.

Figure \ref{fig:deltaf-v2} shows the effect of both $\delta f$ corrections on the elliptic flow of pions (a), protons (b), and anti-protons (c). The shear $\delta f$ leads to the typical reduction of $v_2$ for all particle species. Its effects on particle $p_T$-differential $v_2$ is larger than the one from the baryon diffusion. Because the baryon diffusion $\delta f$ depends on the baryon charge of the particle species, it reduces proton $v_2$ more but increases anti-proton $v_2$. It enhanced the difference between proton and antiproton $v_2(p_T)$.

\subsection{Effects of hadronic afterburner at BES energies}

\begin{figure}[ht!]
  \centering
  \includegraphics[width=0.9\linewidth]{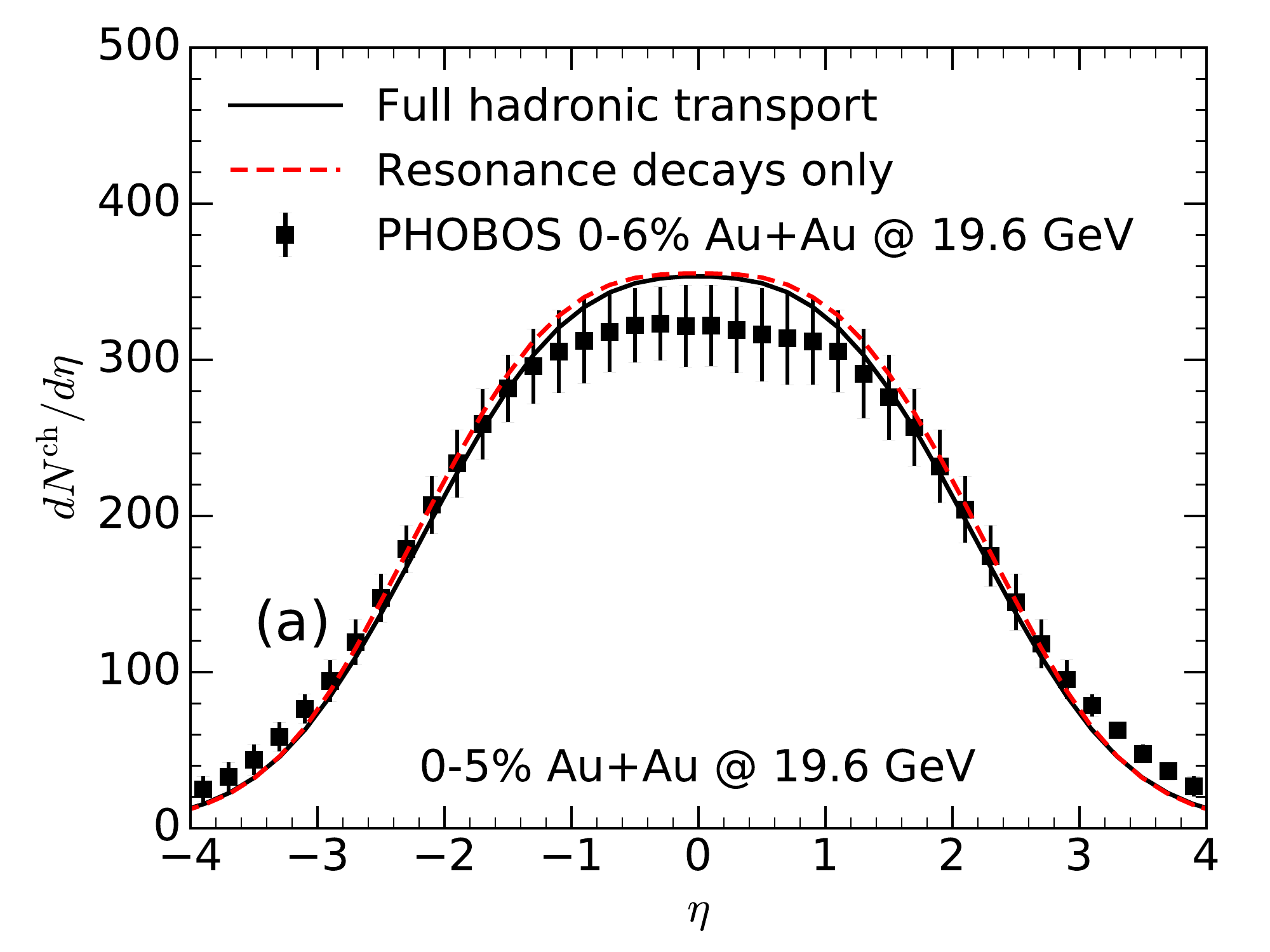}
  \includegraphics[width=0.9\linewidth]{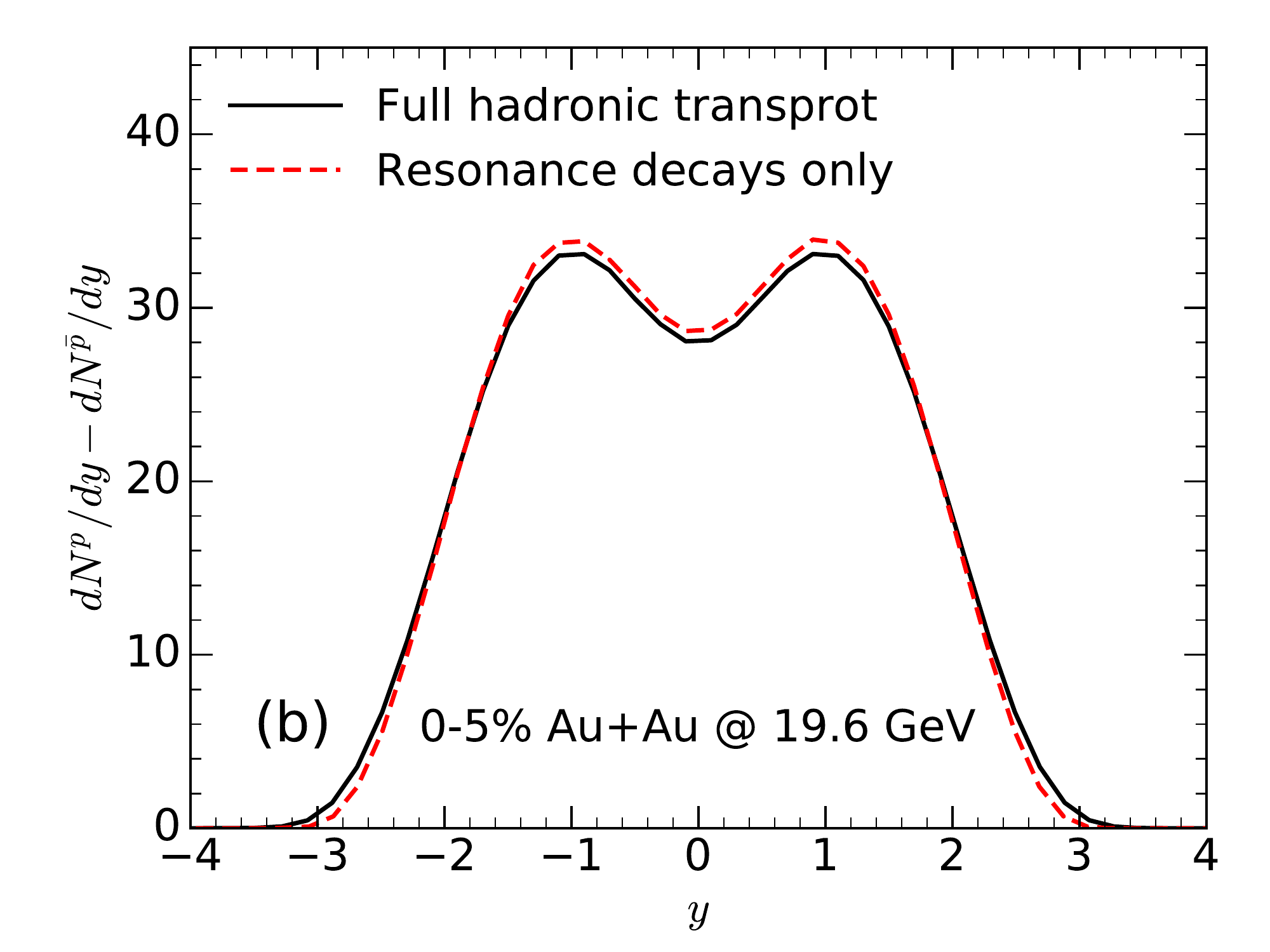}
   \caption{The effects of hadronic rescatterings on charged hadron (a) and net proton (b) rapidity distribution.}
  \label{fig:rescatter}
\end{figure}
%

Figure \ref{fig:rescatter} shows the effect of hadronic rescatterings on the rapidity distributions of identified particles. The late hadronic rescattering phase has a small effect on the shape of the rapidity distribution of charged hadrons. The net protons rapidity distribution is slightly widened by scatterings with other hadrons.

\begin{table*}[ht!]
  \centering
  \begin{tabular}{c|c|c|c}
  \hline \hline
    & $\langle p_T \rangle (p)$ (GeV) & $\langle p_T \rangle (\bar{p})$ (GeV) & $\langle p_T \rangle(\bar{p}) - \langle p_T \rangle(p) $ (GeV) \\ \hline
   Thermal & 0.758 & 0.769 & 0.011  \\ \hline
   Thermal + Corona & 0.753 & 0.766 & 0.013  \\ \hline
   Thermal + Corona + resonances feed down & 0.712 & 0.722 & 0.010  \\ \hline
   Thermal + Corona + full UrQMD & 0.875 & 0.924 & 0.049  \\
   \hline \hline
   \end{tabular}
  \caption{The averaged transverse momentum of protons and anti-protons and their difference at different values from different effects in the hadronic phase.}
  \label{table4}
\end{table*}
%

As in Table \ref{table3}, the mean transverse momentum of anti-protons is slightly larger than the proton mean $p_T$ even without diffusion at $C_B$ = 0. Table~\ref{table4} studies the origin of this difference in detail. Starting with thermally emitted protons and anti-protons from the Cooper-Frye conversion surface, the anti-proton $\langle p_T \rangle$ is only 7 MeV larger than that of the protons. This small difference can be understood by studying the time dependence of $\mu_B/T$ and the flow velocity $u^\tau$ on the hypersurface. The value of $\mu_B/T$ decreases by $\sim 10$\% during the first 4 fm of the hydrodynamic evolution while the radial flow is building up. This anti-correlation between the time evolution of $\mu_B/T$ and $u^\tau$ at early times results in relatively more protons produced when the radial flow is small. The thermal production yields of both protons and anti-protons are small during the first 4 fm of the evolution. Thus, the difference in mean $p_T$ is merely 7 MeV. 

The hadronic corona (see Section \ref{sec:particlization}) produces more protons than anti-protons near the edge of the fireball at the beginning of the hydrodynamic evolution. Because there is no hydrodynamic flow yet and the temperatures of the fluid cells are low, including these particles is expected to reduce the mean $p_T$. Indeed, we found that the proton mean $p_T$ in Table \ref{table4} is reduced twice as much as that of the antiproton when including this contribution.

The resonance feed down contribution from heavy excited baryon states reduces both proton and anti-proton mean $p_T$ similarly. The slight reduction could be attributed to the fact that the shape of heavier particle spectra are less affected by the chemical potential and thus the particle-anti-particle difference in the mean $p_T$ is smaller. Finally, the hadronic rescatterings among light mesons and baryons largely blue-shift proton and anti-proton mean $p_T$. Overall, hadronic rescatterings affect anti-protons more compared to protons. This is because a larger fraction of protons is produced at early times and in the dilute region (compared to anti-protons) and these protons scatter less. In summary, Table~\ref{table4} shows that the mean $p_T$ difference between protons and anti-protons mainly originates from the late stage hadronic rescatterings. Note however, that the other differences in their production, discussed above, are necessary for the rescatterings to have this effect.

\begin{figure}[ht!]
  \centering
  \includegraphics[width=0.9\linewidth]{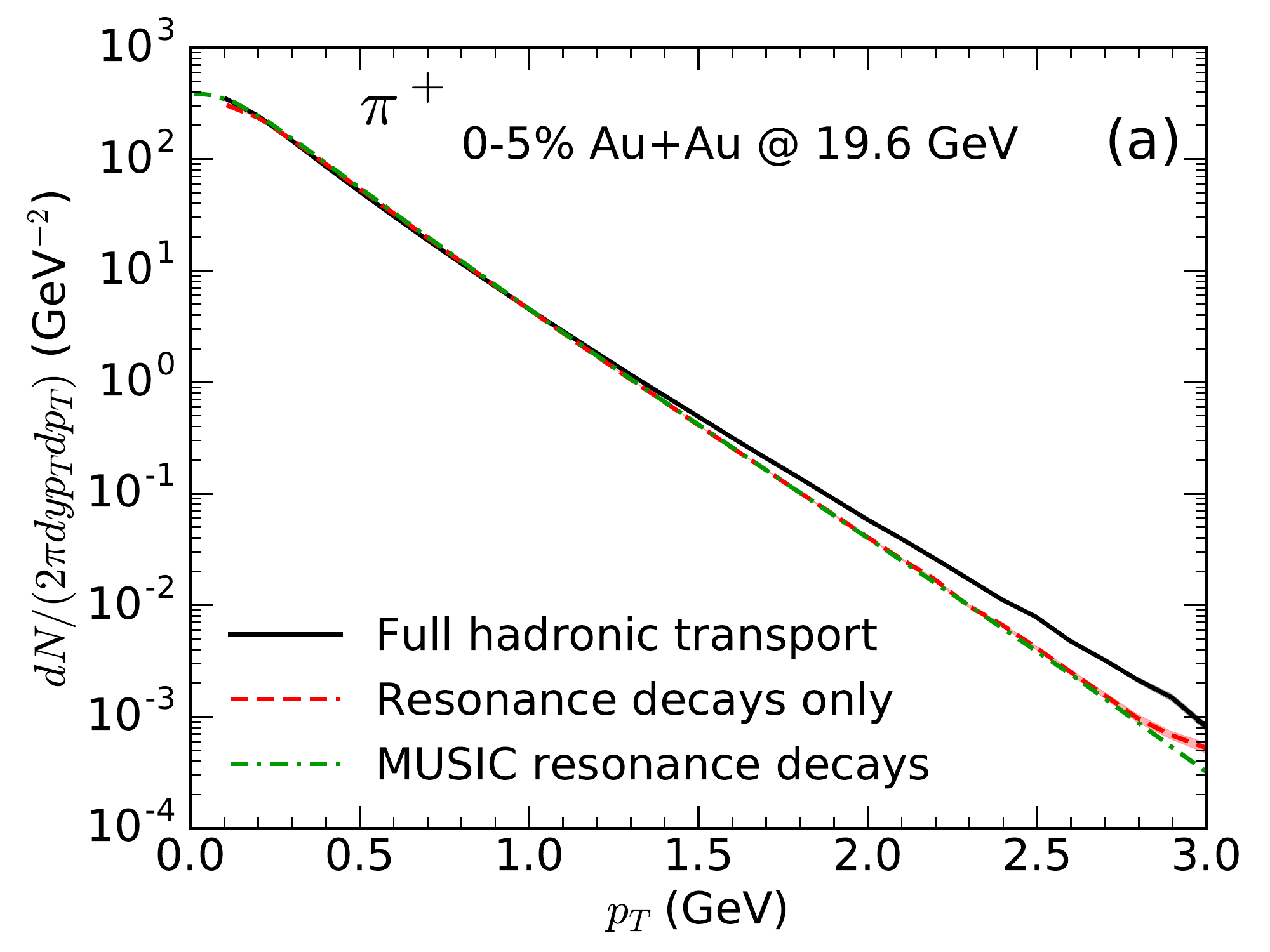}
  \includegraphics[width=0.9\linewidth]{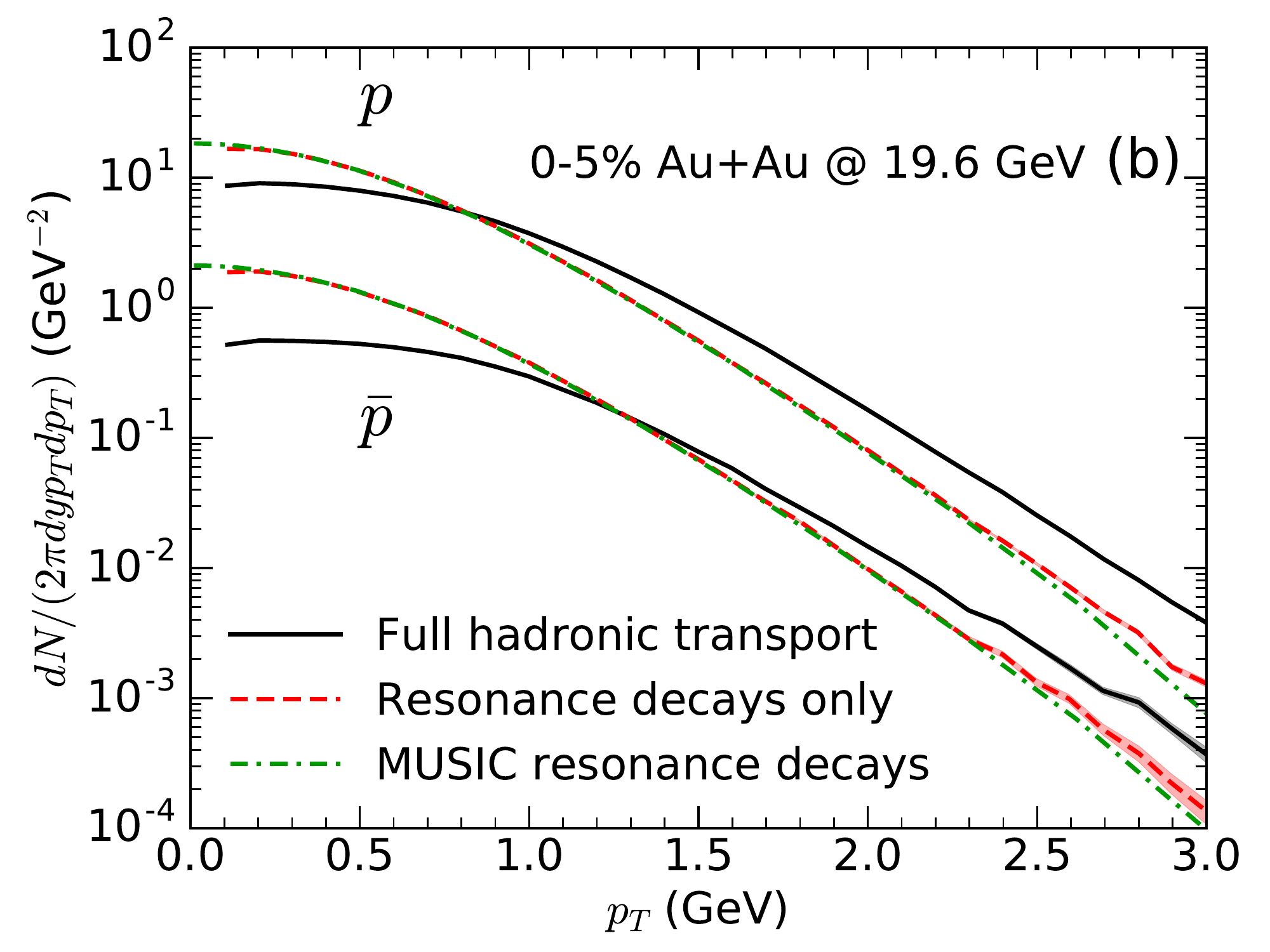}
   \caption{The effects of hadronic transport on the transverse momentum spectra of final $\pi^+$, $K^+$ (a), $p$, and $\bar{p}$ (b).}
  \label{fig3B.2}
\end{figure}
%

Figure \ref{fig3B.2} studies the effect of hadronic transport on particle $p_T$ spectra. The Monte-Carlo results without hadronic rescatterings from the hadronic transport approach are also cross checked with the direct numerical calculations of Cooper-Frye freeze-out and resonance decays. Consistent results are found from the two independent approaches which validates the Monte-Carlo simulations. By comparing feed-down only pion spectra with the results from the full UrQMD simulation, we find that the additional scatterings in the hadronic phase flatten the pion spectra at high $p_T$.

Significant modifications on the shape of proton and anti-proton $p_T$-spectra are found in Fig.~\ref{fig3B.2}b. Both proton and anti-proton spectra get large blue-shifts because of scatterings with light mesons in the hadronic phase. We checked that in this baryon-rich environment the $B\bar{B}$ annihilation processes in the hadronic phase do not have an effect. 

We conclude that at BES collision energies the hadronic transport phase is critical for baryon and anti-baryon spectra.

\begin{figure}[ht!]
  \centering
  \includegraphics[width=0.9\linewidth]{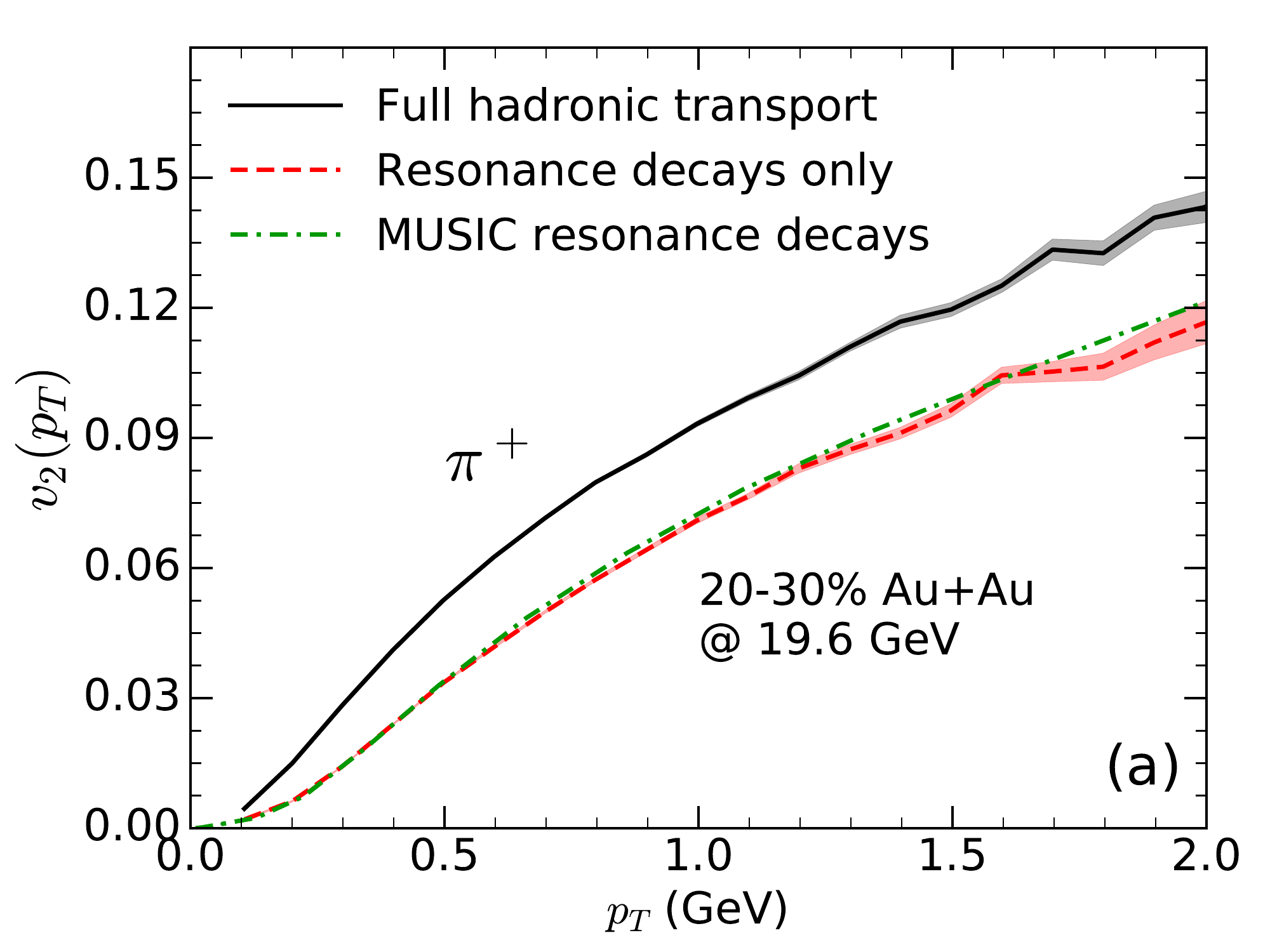}
  \includegraphics[width=0.9\linewidth]{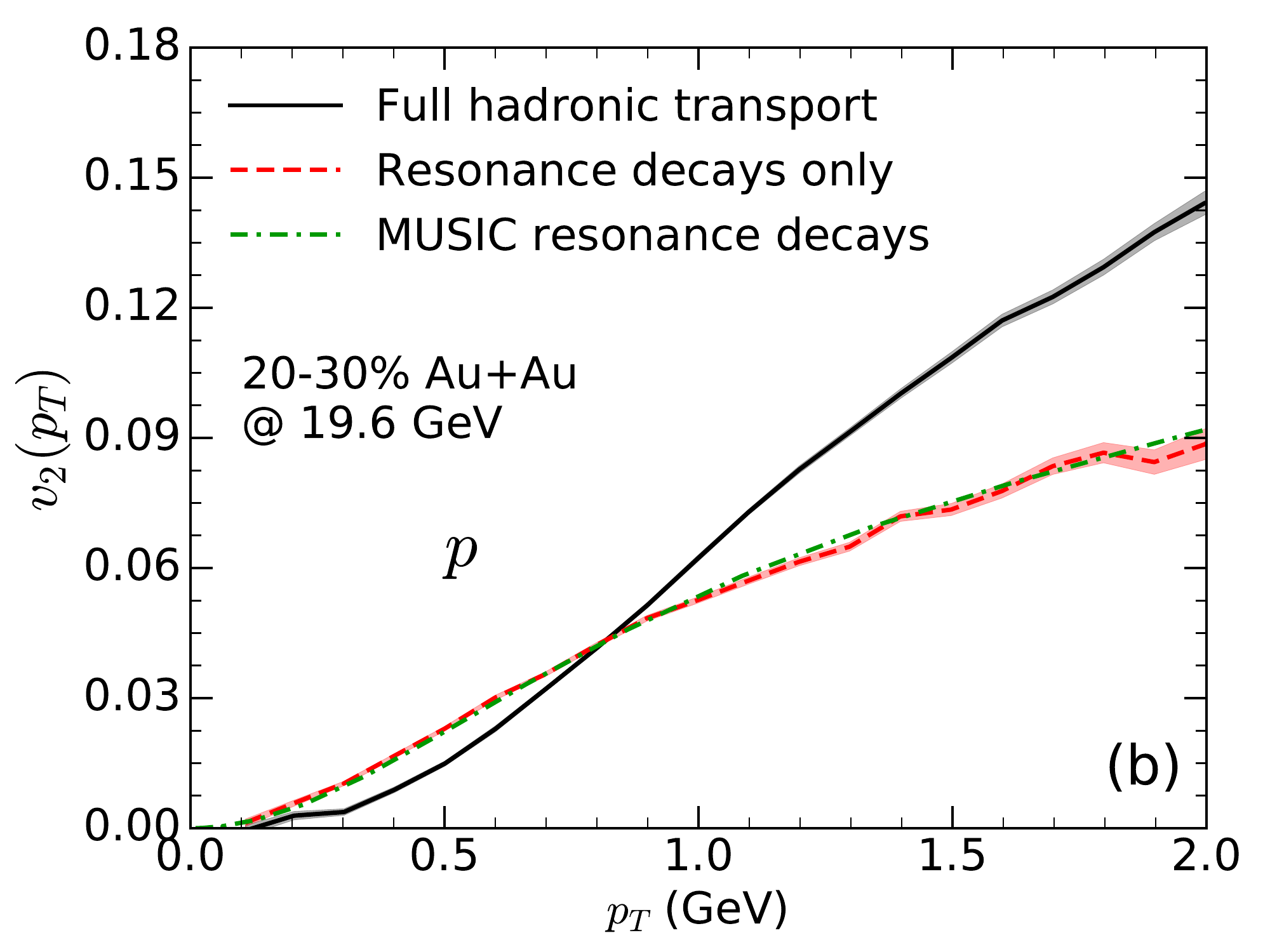}
  \includegraphics[width=0.9\linewidth]{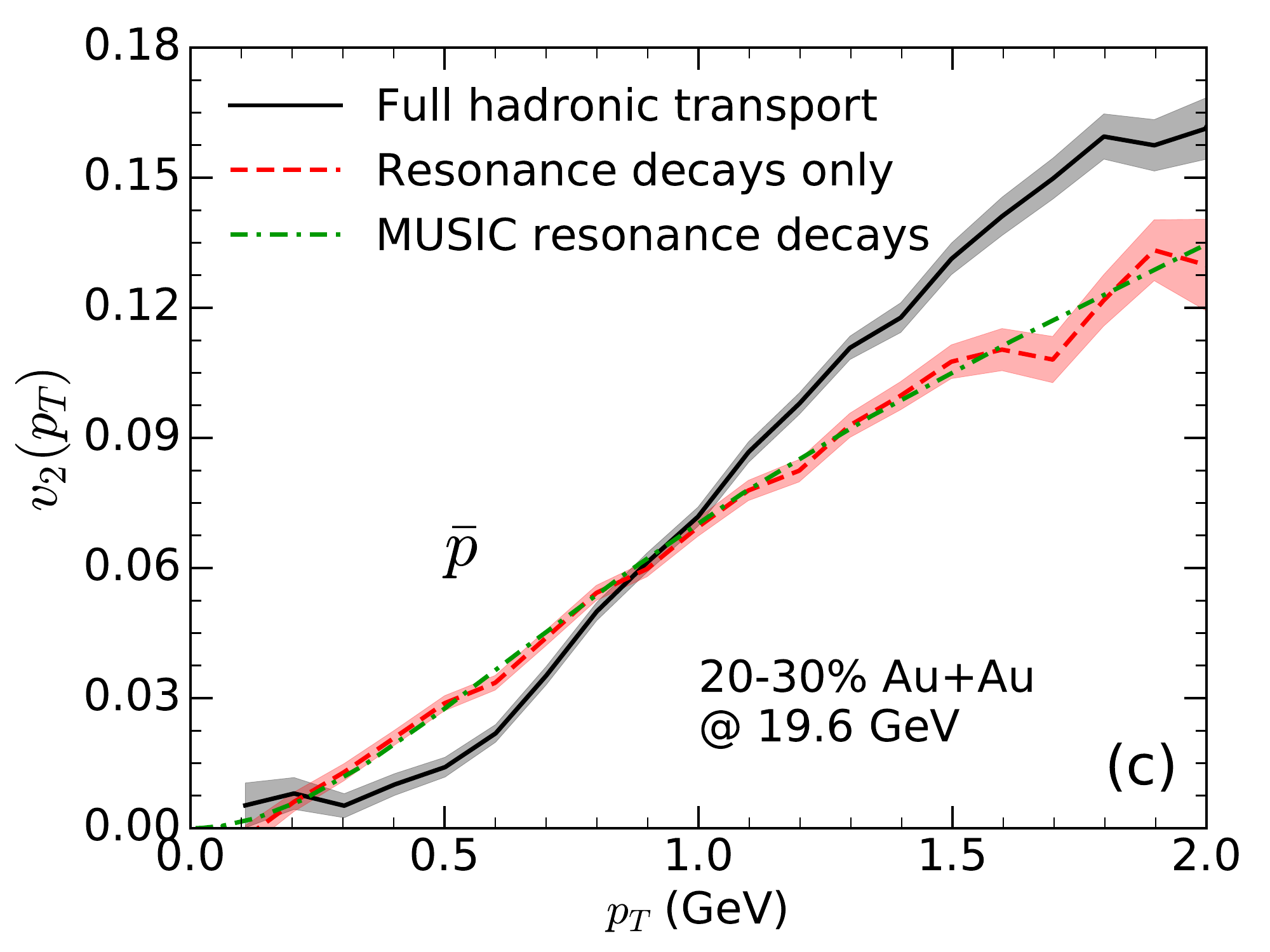}
   \caption{The effects of hadronic transport on the differential $v_2$ of final $\pi^+$ (a), $p$ (b), and $\bar{p}$ (c). The shaded bands indicate statistical errors.}
  \label{fig3B.3}
\end{figure}
%

In Fig.~\ref{fig3B.3} we investigate the effect of the hadronic transport phase on the identified particle $p_T$-differential elliptic flow coefficient. Firstly, consistent results are found between the Monte-Carlo approach (without rescatterings) and the direct numerical calculations of resonance decays. Unlike the minor modifications on the pion $p_T$ spectra, the elliptic flow $v_2(p_T)$ of pions receives a sizable increase from the hadronic rescattering. This can be explained by the hadronic transport converting the remaining spatial eccentricity to particles' momentum anisotropy. Similar to pions, the high $p_T$ (anti-)proton $v_2$ is increased owing to the additional lifetime of the system which converts more spatial eccentricity to momentum anisotropy. Meanwhile, the low $p_T$ proton $v_2$ is reduced. This can be understood as a blue-shift effect, consistent with the modification of the proton spectra.

\section{Conclusions}\label{sec:conclusion}
The theoretical description of heavy ion collisions over a wide range of collision energies requires detailed fluid dynamic simulations with various complications appearing with lower energies. In this work we have introduced and studied the effects of net baryon diffusion that is expected to be present whenever the net baryon density is non-negligible. We have extended the 3+1 dimensional hydrodynamic simulation \textsc{Music} to include baryon diffusion currents and analyzed its effects on a variety of observables in a simplified setup using smooth initial conditions.

Employing an equation of state at finite $\mu_B$ constructed from lattice QCD data and a hadron resonance gas, we were able to evolve systems with non-negligible net baryon density. We found that baryon diffusion, following the gradients of $\mu_B/T$ in the system, tends to transport net baryon number towards mid-rapidity.

While pions and kaons are not affected within the accuracy of the simulation, measurable effects on proton and anti-proton spectra and elliptic flow coefficients are present. In particular, the difference between proton mean transverse momentum and anti-proton mean transverse momentum increases with increasing baryon diffusion. Furthermore, baryon diffusion decreases proton elliptic flow while increasing anti-proton elliptic flow. 

We have also shown that the hadronic microscopic transport stage is very important for baryon spectra and differential elliptic flow coefficients, primarily because of the additional blue-shift given to protons and anti-protons. Apart from this effect, it also continues the conversion from spatial to momentum anisotropy, an effect relevant also for pion elliptic flow.

Finally, we have identified the contributions to the difference in proton- and anti-proton $\langle p_T\rangle$, with the main effect coming from the hadronic afterburner. For the afterburner to have an effect it is also important where and when protons are produced relative to anti-protons, which depends on the distribution of $\mu_B$ on the freeze-out surface.

We have presented one important step towards the development of a comprehensive simulation of heavy ion collision dynamics relevant for collisions in the RHIC BES and BES II as well as the NA61/SHINE program. In the future it will be combined with other important developments in this direction, including a dynamical fluctuating initial state, hydrodynamic fluctuations, and multiple conserved currents with coupled diffusion coefficients \cite{Greif:2017byw}, to result in a powerful theoretical tool that is needed to extract important information on the QCD phase diagram from experimental data.

\begin{acknowledgements}
BPS and CS are supported under DOE Contract No. DE-SC0012704. This research used resources of the National Energy Research Scientific Computing Center, which is supported by the Office of Science of the U.S. Department of Energy under Contract No. DE-AC02-05CH11231. BPS acknowledges a DOE Office of Science Early Career Award. CS thanks a Goldhaber Distinguished Fellowship from Brookhaven Science Associates.
AM is supported by JSPS Overseas Research Fellowship. GSD thanks Conselho Nacional de
Desenvolvimento Cient\'{\i}fico e Tecnol\'{o}gico (CNPq) for financial
support. CG gratefully acknowledges support from the Canada Council for the Arts through its Killam Research Fellowship program.
This work was supported in part by the Natural Sciences and Engineering Research Council of Canada. 
Computations were made in part on the supercomputer Guillimin from McGill University, managed by Calcul Qu\'ebec and Compute Canada. The operation of this supercomputer is funded by the Canada Foundation for Innovation (CFI), NanoQu\'ebec, RMGA and the Fonds de recherche du Qu\'ebec - Nature et technologies (FRQ-NT). This work is supported in part by the U.S. Department of Energy, Office of Science, Office of Nuclear Physics, within the framework of the Beam Energy Scan Theory (BEST) Topical Collaboration.
\end{acknowledgements}

\appendix

\section{Off-equilibrium corrections to the Cooper-Frye Freeze-out}
\label{appendix_A}

In this appendix, we derive the out-of-equilibrium correction for the Cooper-Frye freeze-out procedure from particle diffusion effects. 
The system's energy-momentum tensor and conserved currents can be expressed as,
\begin{equation}
  T^{\mu \nu} = \sum^N_{i = 1} g_i \int d K_i k_i^{\mu} k_i^{\nu} f_i,
\end{equation}
\begin{equation}
  N_Q^{\mu} = \sum_{i = 1}^N g_i Q_i \int d K_i k_i^{\mu} f_i,
\end{equation}
where $N_Q$ is the conserved charge in the system, $g_i$ is the degeneracy factor, $Q_i =
b_i, s_i, c_i$ is the charge, and $\int d K_i = \int \frac{d^3 K_i}{(2 \pi)^3 E_i}$. The Landau matching condition can be written as,
\begin{equation}
  u_\mu u_\nu \delta T^{\mu\nu} = \sum_{i = 1}^N g_i \int d K_i E_i^2 \delta f_{i} = 0,
\end{equation}
\begin{equation}
  \Delta^{\mu\alpha} u^\beta \delta T_{\alpha \beta} = \sum_{i = 1}^N g_i \int d K_i E_i k_i^{\langle \mu \rangle} \delta f_{i} =
  0,
\end{equation}
\begin{equation}
  u_\mu \delta N^\mu_Q = \sum_{i = 1}^N g_i Q_i \int d K_i E_i \delta f_{i} = 0,
\end{equation}
where $k^{\langle \mu \rangle} = \Delta^{\mu\nu} k_\nu$.
The dissipative quantities of the system can be computed as,
\begin{equation}
  \Pi = - \frac{1}{3} \sum_{i = 1}^N g_i m_i^2 \int d K_i \delta f_{i},
  \label{eq6}
\end{equation}
\begin{equation}
  q_Q^{\mu} = \sum_{i = 1}^N g_i Q_i \int d K_i k_i^{\langle \mu \rangle}
  \delta f_{i}, \label{eq7}
\end{equation}
and
\begin{equation}
  \pi^{\mu \nu} = \sum_{i = 1}^N g_i  \int d K_i k_i^{\langle \mu \nobracket}
  k_i^{\nobracket \nu \rangle} \delta f_{i} . \label{eq8}
\end{equation}

The first-order Chapman-Enskog approximation \cite{Chapman:1952} of the Boltzmann equation leads to
\begin{equation}
  D f^{(0)}_{i} + \frac{1}{E_k^i} k_i^{\mu} \nabla_{\mu} f^{(0)}_{i} =
  \frac{1}{E_k^i} C [f^{(1)}_{i}], \label{eq9}
\end{equation}
where $C [f^{(1)}_{i k}]$ is the linearized collision term. In the relaxation
time approximation,
\begin{equation}
  C [f^{(1)}_{i}] = - \frac{E_k^i}{\tau_R} \delta f_{i},
\end{equation}
where $\tau_R$ is the relaxation time. The right hand side of Eq. (\ref{eq9})
becomes $- \frac{1}{\tau_R} \delta f_{i}$. On the left hand side, we have
the equilibrium part of the distribution function,
\begin{equation}
  f^{(0)}_{i} = \frac{1}{\exp [\beta_0 (u_{\mu} k_i^{\mu} - Q_i \alpha_Q)] + \theta_i},
\end{equation}
where $\beta_0 = 1/T$, $\alpha_Q = \mu_Q/T$, and $\theta_i = 1$ for fermions and $\theta_i = -1$ for bosons.
Then
\begin{eqnarray}
  \delta f^{(0)}_{i} &=& - f_{i}^{(0)} \tilde{f}^{(0)}_{i} [\delta \beta_0
  u_{\mu} k_i^{\mu} + \beta_0 \delta u_{\mu} k_i^{\mu} \notag \\
  && \qquad \qquad - Q_i \delta \alpha_Q]
\end{eqnarray}
with
\begin{equation}
\tilde{f}^{(0)}_{i} = (1 - \theta_i f^{(0)}_{i}).
\end{equation}
Next, insert them into Eq. (\ref{eq9}) and organize the terms into bulk, particle diffusion, and shear corrections
according to the rank of the tensor (and the requirement of tracelessness for the shear correction).
For particle diffusion, collecting all rank 1 terms leads to
\begin{eqnarray}
  \delta f^{\tmop{diffusion}}_{i} & = & f_{i}^{(0)} \tilde{f}^{(0)}_{i}
  \tau_R \bigg[ \beta_0 k_i^{\mu} D u_{\mu} + k_i^{\mu} (\nabla_{\mu} \beta_0) \notag \\
  && \qquad \qquad \qquad - \frac{Q_i}{E_{i}} k_i^{\mu} \nabla_{\mu} \alpha_Q \bigg] .  \label{eq18}
\end{eqnarray}
Now, we need to replace all the time derivatives in $\delta f$ using
thermodynamic relations and first order conservation laws. The thermodynamic
relations that we will use are,
\begin{equation}
  d P_0 = s_0 d T + n_Q d \mu_Q
\end{equation}
\begin{equation}
  \varepsilon_0 + P_0 = T s_0 + \mu_Q n_Q.
\end{equation}
\begin{equation}
  d \varepsilon_0 = T d s_0 + \mu_Q d n_Q
\end{equation}
Using the chain rule,
\begin{equation}
  d T = - \frac{1}{\beta_0^2} d \beta_0
\end{equation}
\begin{equation}
  \beta_0 d \mu_Q = d \alpha_Q - \alpha_Q \frac{d \beta_0}{\beta_0}
\end{equation}
The conservation laws at first order in Knudsen number are,
\begin{equation}
  D \varepsilon_0 = - (\varepsilon_0 + P_0) \theta
\end{equation}
\begin{equation}
  (\varepsilon_0 + P_0) D u^{\mu} = \nabla^{\mu} P_0
\end{equation}
\begin{equation}
  D n_Q = - n_Q \theta
\end{equation}
For diffusion in Eq. (\ref{eq18}),
\begin{eqnarray}
  \delta f^{\tmop{diffusion}}_{i} 
  & = & f_{i}^{(0)} \tilde{f}^{(0)}_{i} \tau_R \bigg[
  \frac{n_Q}{\varepsilon_0 + P_0} k_i^{\mu} \nabla_{\mu} \alpha_Q \notag\\
  && \qquad\qquad\qquad - \frac{Q_i}{E_{i k}} k_i^{\mu} \nabla_{\mu} \alpha_Q \bigg]
  \label{eq33}
\end{eqnarray}

All the $\delta f$s are proportional to the relaxation time
$\tau_R$. In the following we derive expressions for the relaxation time in terms of the
system's heat conductivity. In the Navier-Stokes limit, we have,
\begin{equation}
  q^{\mu}_Q = \kappa \nabla^{\mu} \alpha_Q .
\end{equation}
Here we define useful thermodynamic integrals
\begin{equation}
  J_{n q}^i = \frac{1}{(2 q + 1) !!} \int d K_i (u_{\mu} k_i^{\mu})^{n - 2 q} (-
  \Delta_{\mu \nu} k_i^{\mu} k_i^{\nu})^q f_{i k}^{(0)} \tilde{f}_{i k}^{(0)} \,.
\end{equation}

For charge diffusion, the situation for net baryon, net strangness, and net
charge are very similar. So here we only consider net baryon charge as an example. Inserting
Eq. (\ref{eq33}) into Eq. (\ref{eq7}) we find,
\begin{eqnarray*}
  q_B^{\mu} & = & \sum_{i = 1}^N g_i b_i \int d K_i k_i^{\langle \mu \rangle}
  \delta f_k^i\\
  & = & \tau_R \Delta^{\mu} \,_{\nu} \nabla_{\alpha} \alpha_B \sum_{i = 1}^N
  g_i b_i \notag \\
  && \times \int d K_i k_i^{\nu} k_i^{\alpha} f_{i k}^{(0)} \tilde{f}^{(0)}_{i
  k} \tau_R \left[ \frac{n_B}{\varepsilon_0 + P_0} - \frac{b_i}{E_{i k}}
  \right]
\end{eqnarray*}
We define
\[ I^{\nu \alpha}_B = \sum_{i = 1}^N g_i b_i \int d K_i k_i^{\nu} k_i^{\alpha}
   f_{i k}^{(0)} \tilde{f}^{(0)}_{i k} \left[ \frac{n_B}{\varepsilon_0 + P_0}
   - \frac{b_i}{E_{i k}} \right] \]
and
\[ q_B^{\mu}  = \kappa_B \nabla^{\mu} \alpha_B = \tau_R \hat{\kappa}_B \nabla^{\mu} \alpha_B, \]
where
\begin{eqnarray*}
  \hat{\kappa}_B & = & \frac{1}{3} \Delta_{\mu \nu} I^{\mu \nu}_B\\
   & = & - \sum_{i = 1}^{N_B} \left[ \frac{n_B}{\varepsilon_0 + P_0} b_i
  J_{21}^i - J_{11}^i \right].
\end{eqnarray*}
Here $N_B$ runs over all the baryons. In the massless limit,
\begin{equation}\label{eq:kappaB}
\kappa_B = \tau_R n_B \left( \frac{1}{3}\coth \left(\alpha_B \right) - \frac{n_B T}{\varepsilon_0 + P_0} \right) .
\end{equation}
In the small $\alpha_B$ limit
\begin{equation}
  \kappa_B = \tau_R n_B \frac{1}{3 \alpha_B} = \frac{\tau_R T}{3} \frac{n_B}{\mu_B}.
\end{equation}
In the relaxation time approximation, we can relate the net baryon diffusion
constant to the specific shear viscosity through the relaxation time,
\begin{equation}
    \tau_R = \frac{5 \eta}{e + P}
\end{equation}
then
\begin{equation}
  \kappa_B = \frac{5\eta}{e + \mathcal{P}} \frac{T}{3} \frac{n_B}{\mu_B} = \frac{5}{3} \frac{\eta T}{e + \mathcal{P}}
  \frac{n_B}{\mu_B}.
\end{equation}
For a small specific shear viscosity $\frac{\eta T}{e + \mathcal{P}} = \frac{1}{4 \pi}$,
\begin{equation}
    \kappa_B = \frac{5}{12 \pi} \frac{n_B}{\mu_B}.
\end{equation}

The net baryon diffusion out-of-equilibrium correction to the Cooper-Frye formula
$\delta f$ is 
\begin{eqnarray}
  \delta f^{B - \tmop{diffusion}}_{i k} 
   \hspace{-0.2cm}& = &  \hspace{-0.2cm}f_{i k}^{(0)} \tilde{f}^{(0)}_{i k} \left[ \frac{n_B}{\varepsilon_0 +
  P_0} - \frac{b_i}{E_{i k}} \right] \frac{k_i^{\mu} q_{\mu}}{\hat{\kappa}_B}. ~\label{eq:deltafbaryon}
\end{eqnarray}

\section{Particlization at finite baryon density}
\label{appendix_B}

In this appendix, we discuss about how to generate Monte-Carlo samples of
particles from a hydrodynamic hypersurface with finite baryon chemical
potential and baryon diffusion current.

Different species of particles and unstable resnonances are sampled one by one.
For a given particle species, we first need to compute its particle yield
in every freeze-out fluid cell.
The space-time positions of the emitted particles are sampled from different
freeze-out fluid cells with particle yields as relative weights.
Sec.~\ref{appB:sec1} discuss how to compute the particle yield at finite
baryon chemical potential and net baryon diffusion current.
Once the particle's spatial position is determined, its momentum information
is sampled using the acceptance-rejection method. This sampling method requires
us to estimate the maximum of the particle momentum distribution.
This is derived in Sec.~\ref{appB:sec2} with the presence of the net baryon
diffusion out-of-equilibrium correction for the first time.
All the numerics were implemented in the open-source iSpectraSampler
(\textsc{iSS}) code package. The numerical validation
is presented in Sec.~\ref{appB:sec3}.

\subsection{Estimation of particle yields} \label{appB:sec1}
Particle momentum distributions from a fluid cell can
be calculated using the Cooper-Frye formula,
\begin{equation}
  E \frac{d N_i}{d^3 p} = \frac{g_i}{(2 \pi)^3} \Delta^3 \sigma_{\mu} p^{\mu} 
  (f_{i}^{(0)} + \delta f_{i}) .
\end{equation}
So the particle yield is given by
\begin{equation}
  N_i = \frac{g_i}{(2 \pi)^3} \Delta^3 \sigma_{\mu} \int \frac{d^3 p}{E}
  p^{\mu}  (f_{i}^{(0)} + \delta f_{i}) .
\end{equation}
For the thermal equilibrium part,
\begin{equation}
  f_{i}^{(0)} = \frac{1}{e^{\beta (E - \mu_i)} - \theta}
\end{equation}
with $\theta = - 1$ for fermions and $\theta = + 1$ for bosons. The
equilibrium part of the integral can be evaluated as follows,
\begin{eqnarray}
  N_i^{\tmop{eq}} & = & \frac{g_i}{(2 \pi)^3} \Delta^3 \sigma_{\mu} \int
  \frac{d^3 p}{E} p^{\mu} \frac{1}{e^{\beta (E - \mu_i)} - \theta} \nonumber\\
  & = & \frac{g_i}{2 \pi^2} \Delta^3 \sigma_{\mu} u^{\mu} \int d E E
   \frac{\sqrt{E^2 - m_i^2} }{e^{\beta (E - \mu_i)} - \theta}.
\end{eqnarray}
Now, we can expand the thermal equilibrium distribution function,
\begin{eqnarray}
  \frac{1}{e^{\beta (E - \mu_i)} - \theta} 
  & = & \sum_{n = 1}^{\infty} \theta^{n - 1} e^{n \beta \mu_i} e^{- n \beta
  E} .
\end{eqnarray}
In practice, we compute the summation up to $n = 10$.
Thus,
\begin{eqnarray}
  N_i^{\tmop{eq}}
  \hspace{-0.2cm}& = & \hspace{-0.2cm} \frac{g_i}{2 \pi^2} \Delta^3 \sigma_{\mu} u^{\mu} \frac{m_i^2}{\beta}
  \sum_{n = 1}^{\infty} \frac{\theta^{n - 1}}{n} e^{n \beta \mu_i} K_2 (n m_i
  \beta).
\end{eqnarray}

Now, we focus on the contribution from baryon diffusion in the following
because shear viscous effects do not modify the particle number and
bulk viscosity is not considered in this work.
The form of the off-equilibrium correction to the distribution function from net
baryon diffusion is computed in the relaxation time approximation and
given by (\ref{eq:deltafbaryon}).
This leads to a correction to the particle yield of the form
\begin{eqnarray}
  \delta N_i^q & = & \frac{g_i \Delta^3 \sigma_{\mu}}{(2 \pi)^3} \int
  \frac{d^3 p}{E} \frac{ p^{\mu} p^{\nu} q_{\nu}}{\hat{\kappa}
  (T \nocomma, \mu_B)} f_{i}^{(0)} \tilde{f}^{(0)}_{i}  \left( \frac{n_B}{\varepsilon +
  \mathcal{P}} - \frac{b_i}{E} \right) \nonumber\\
  & = & \frac{g_i}{(2 \pi)^3} \frac{\Delta^3 \sigma_{\mu}
  q^{\mu}}{\hat{\kappa} (T \nocomma, \mu_B)} I (T, m_i) 
\end{eqnarray}
with
\begin{eqnarray*}
  I (T, m_i) & = & \frac{\Delta_{\mu \nu}}{3}  \int \frac{d^3 p}{E} p^{\mu}
  p^{\nu} f_{i}^{(0)} \tilde{f}^{(0)}_{i} \left( \frac{n_B}{\varepsilon +
  \mathcal{P}} - \frac{b_i}{E} \right)\notag \\
& = & 4 \pi \bigg[ - \frac{n_B}{\varepsilon + \mathcal{P}} \left(
  \frac{m^2_i}{\beta^2} \sum_{n = 1}^{\infty} \frac{\theta^{n - 1}}{n} e^{n
  \beta \mu_i} K_2 (n m_i \beta) \right) \notag \\
  && \qquad - b_i \left( \frac{1}{3 \beta^3}
  \sum_{n = 1}^{\infty} n \theta^{n - 1} e^{n \beta \mu_i} I_1 (m_i \beta
  \nocomma, n) \right) \bigg]
\end{eqnarray*}
The second integral 
\begin{equation}
  I_1 (m_i \beta, n) \equiv \int_{m_i \beta}^{\infty} \frac{d \xi}{\xi} \left(
  - \sqrt{\xi^2 - (m_i \beta)^2} \right)^3 e^{- n \xi}
\end{equation}
does not have a closed form, but can be expanded as follows
\begin{eqnarray}
  I_1 (m_i \beta, n) 
  & = & - (m_i \beta)^3 \left[ \frac{e^{- n m_i \beta}}{n m_i \beta} \left(
  \frac{2}{(n m_i \beta)^2} + \frac{2}{n m_i \beta} - \frac{1}{2} \right) \right. \notag \\
  &  & \hspace{4em} \left. + \sum_{k = 3}^{\infty} \frac{3}{k!} \frac{(2 k -
  5) !!}{2^k} E_{2 k - 2} (n m_i \beta) \right. \nonumber\\
  && \hspace{4em} \left. + \frac{3}{8} E_2 (n m_i \beta) \right] \,,
    \label{B9}
\end{eqnarray}
where $E_k (n m_i \beta)$ is the exponential integral function.
Finally, the correction to the particle yield from the baryon diffusion term becomes
\begin{eqnarray}
  \delta N_i^q & = & \frac{g_i}{2 \pi^2} \frac{\Delta^3 \sigma_{\mu}
  q^{\mu}}{\hat{\kappa} (T \nocomma, \mu_B)} \notag \\
  &  & \times \bigg\{ - \frac{n_B}{\varepsilon + \mathcal{P}} \left(
  \frac{m^2_i}{\beta^2} \left[ \sum_{n = 1}^{\infty} \frac{\theta^{n - 1} }{n}
  e^{n \beta \mu_i} K_2 (n m_i \beta) \right] \right) \notag \\
  && - b_i \left( \frac{1}{3
  \beta^3} \left[ \sum_{n = 1}^{\infty} n \theta^{n - 1} e^{n \beta \mu_i} I_1
  (m_i \beta \nocomma, n) \right] \right) \bigg\}
  \label{B10}
\end{eqnarray}
In the numerical implementation, we truncate the summations in
Eqs.~(\ref{B9}) and (\ref{B10}) at $n = 10$.

\subsection{Estimation of the maximum in the particle distribution} \label{appB:sec2}

In the processes of sampling, we need to estimated the maximum of the particle
momentum distribution, $E \frac{d N}{d^3 p}$.
Using the H\"{o}lder inequality \cite{Shen:2014vra}, we have
\begin{equation}
  p^{\mu} \Delta^3 \sigma_{\mu} < (p \cdot u) \left( | \Delta^3 \sigma_{\mu}
  u^{\mu} | + \sqrt{| \Delta^3 \sigma_{\mu} \Delta^3 \sigma_{\nu} \Delta^{\mu
  \nu} |} \right) .
\end{equation}

In order to estimate the maximum of the particle distribution, it is useful to define the following function \cite{Shen:2014vra},
\begin{equation}
G(E; A, \theta) = \frac{E^A}{e^{\beta (E - \mu)} - \theta}.
\end{equation}
By setting its derivatives to zero, the extrema can be computed and denoted as $G^{(A,\theta)}_{\max}$. The detailed expression of $G^{(A,\theta)}_{\max}$ was discussed in Ref. \cite{Shen:2014vra}.
For the equilibrium part we need to calculate the maximum of the function
\begin{equation}
  E f_0 = \frac{E}{e^{\beta (E - \mu_i)} - \theta} = G (E ; 1, \theta)
\end{equation}
The solution is $G^{(1,\theta)}_{\max}$.

For baryon diffusion we have
\begin{eqnarray}
  E \delta f_q   \hspace{-0.2cm}& < & \hspace{-0.2cm}\frac{\sqrt{- q^{\mu} q_{\mu}}}{\hat{\kappa} (T \nocomma, \mu_B)}
  \left( \frac{n_B}{\varepsilon + \mathcal{P}} \lambda G^{(2,\theta)}_{\max} + | b_i
  | \lambda G^{(1,\theta)}_{\max} \right)
\end{eqnarray}
and the shear viscous correction yields
\begin{eqnarray}
  E \delta f_{\pi} 
  & < & \frac{\sqrt{\pi^{\mu \nu} \pi_{\mu \nu}}}{2 (\varepsilon +
  \mathcal{P}) T^{}} \lambda G^{(2,\theta)}_{\max} \,.
\end{eqnarray}
Here $\lambda = 1$ for fermions and $\lambda = 2$ for bosons.

So the maximum of the particle momentum spectra can be estimated as,
\begin{eqnarray}
  P_{\max} & = & \frac{g_a}{(2 \pi)^3} \left( | \Delta^3 \sigma_{\mu} u^{\mu}
  | + \sqrt{| \Delta^3 \sigma_{\mu} \Delta^3 \sigma_{\nu} \Delta^{\mu \nu} |}
  \right) \nonumber\\
  &  & \times \bigg( G^{(1,\theta)}_{\max} \nonumber \\
  & & \qquad  + \frac{\sqrt{- q^{\mu}
  q_{\mu}}}{\hat{\kappa} (T \nocomma, \mu_B)} \left( \frac{n_B}{\varepsilon +
  \mathcal{P}} \lambda G^{(2,\theta)}_{\max} + | b_i | \lambda G^{(1,\theta)}_{\max} \right)
  \nonumber\\
  &  & \qquad  + \frac{\sqrt{\pi^{\mu \nu} \pi_{\mu \nu}}}{2
  (\varepsilon + \mathcal{P}) T^{}} \lambda G^{(2,\theta)}_{\max} \bigg) 
\end{eqnarray}

\subsection{Validation of the particle sampling procedure} \label{appB:sec3}

In this section we present a validation of the numerical particle sampler by comparing to results computed directly from the Cooper-Frye formula. 
Individual particles are sampled from a (3+1)D hydrodynamic hyper-surface for 10-40\% Au+Au collisions at 19.6 $A$\,GeV. We include effects from finite net baryon density and net baryon diffusion to the particle momentum distribution. The net baryon diffusion coefficient is set to $\kappa = 0.2 n_B/\mu_B$ in the hydrodynamic simulation.

Please note that the Cooper-Frye results shown in this section are produced using \textsc{Music}, which is completely independent of the particle sampler \textsc{iSS}. Hence, the following numerical comparisons also provide a cross check of computing particle momentum distributions using the Cooper-Frye procedure in the two numerical codes.
 
\begin{figure*}[ht!]
  \centering
  \begin{tabular}{cc}
  \includegraphics[width=0.5\linewidth]{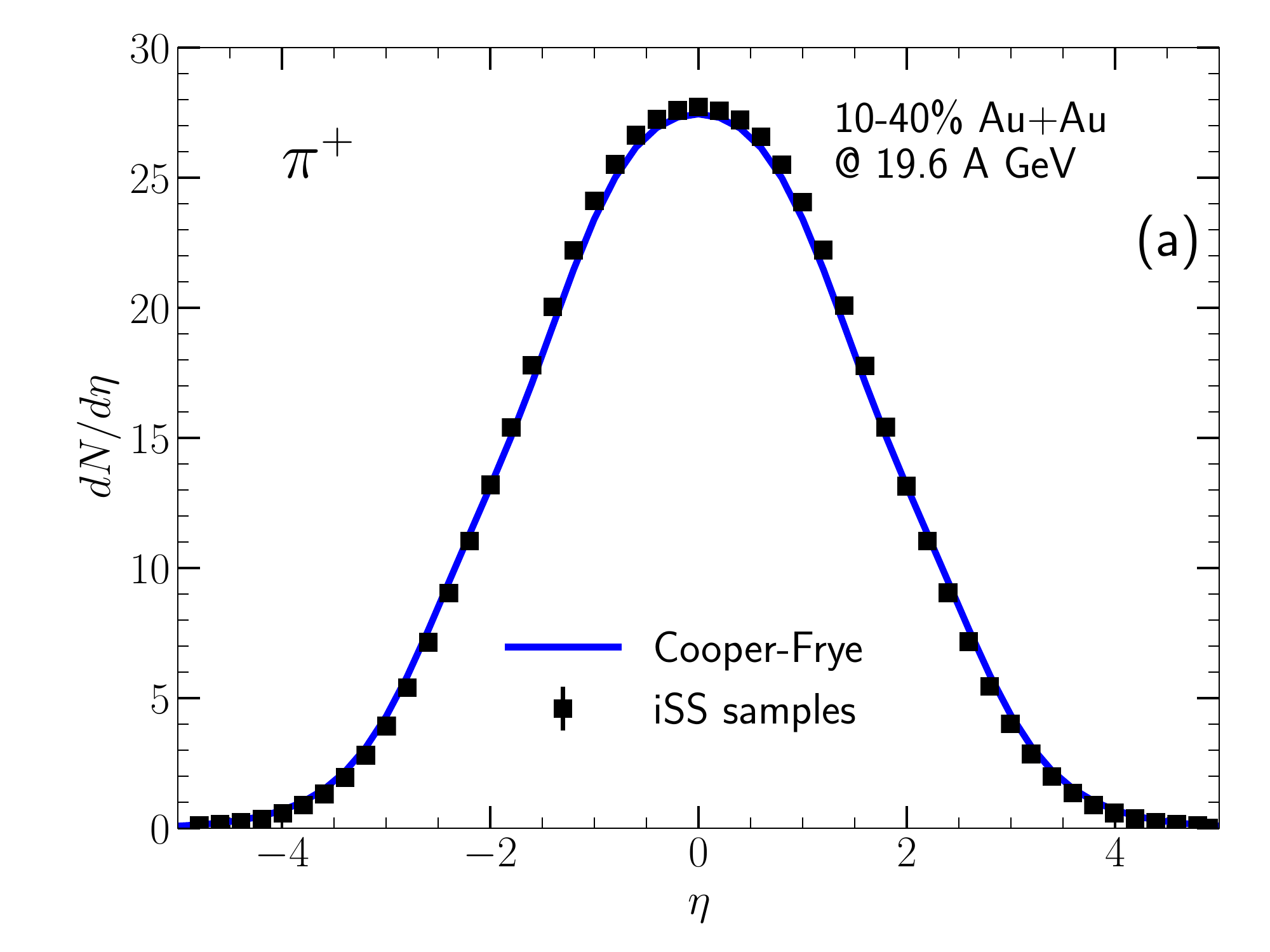} &
  \includegraphics[width=0.5\linewidth]{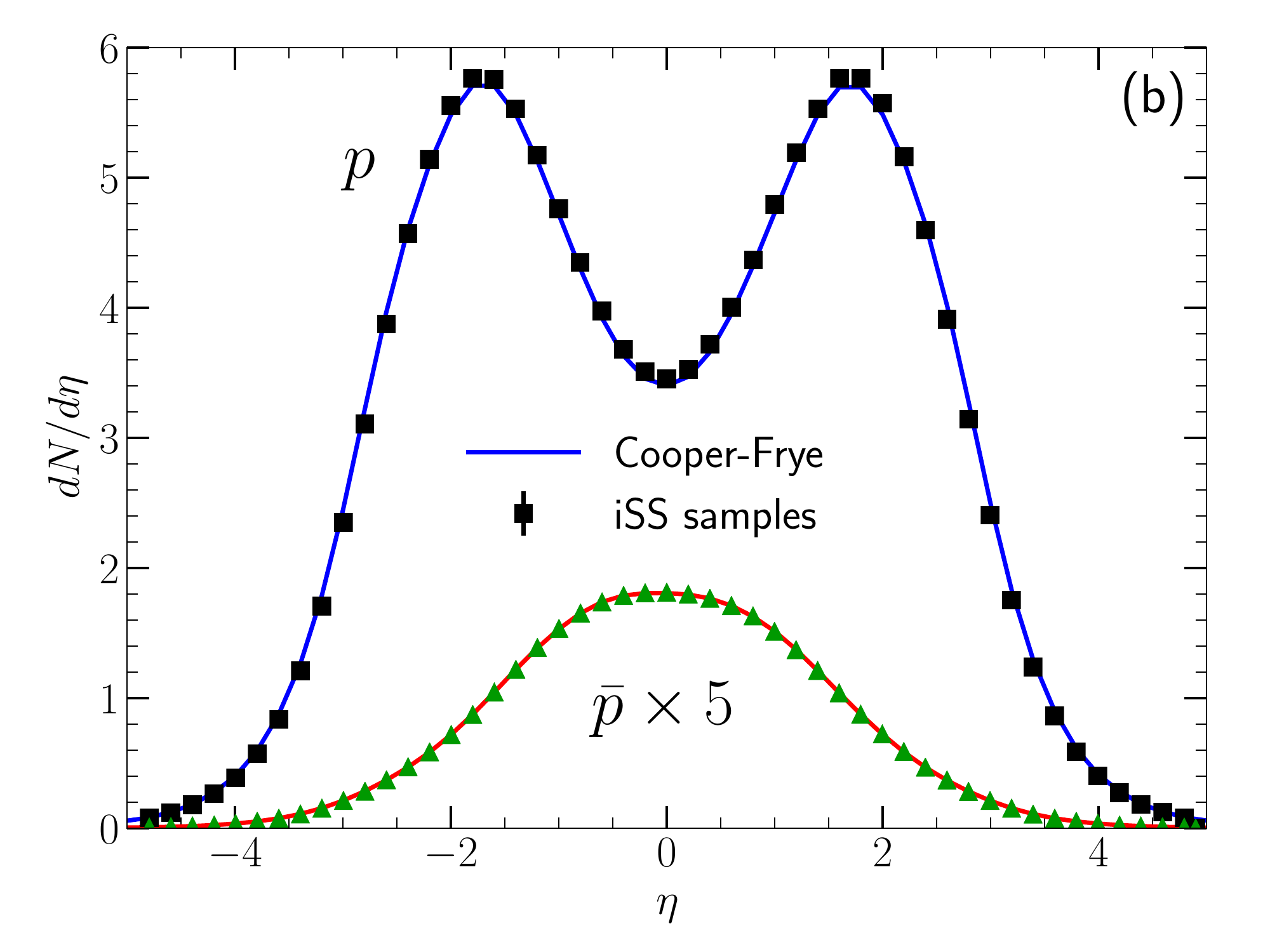}
  \end{tabular}
  \caption{Cross check the pseudo-rapidity distribution of thermally emitted $\pi^+$, $p$, and $\bar{p}$ between the numerical results from Cooper-Frye formula (lines) and the statistical results from particle sampler {\tt iSS} (markers). }
  \label{appendix2.fig1}
\end{figure*}
%
\begin{figure*}[ht!]
  \centering
  \begin{tabular}{cc}
  \includegraphics[width=0.5\linewidth]{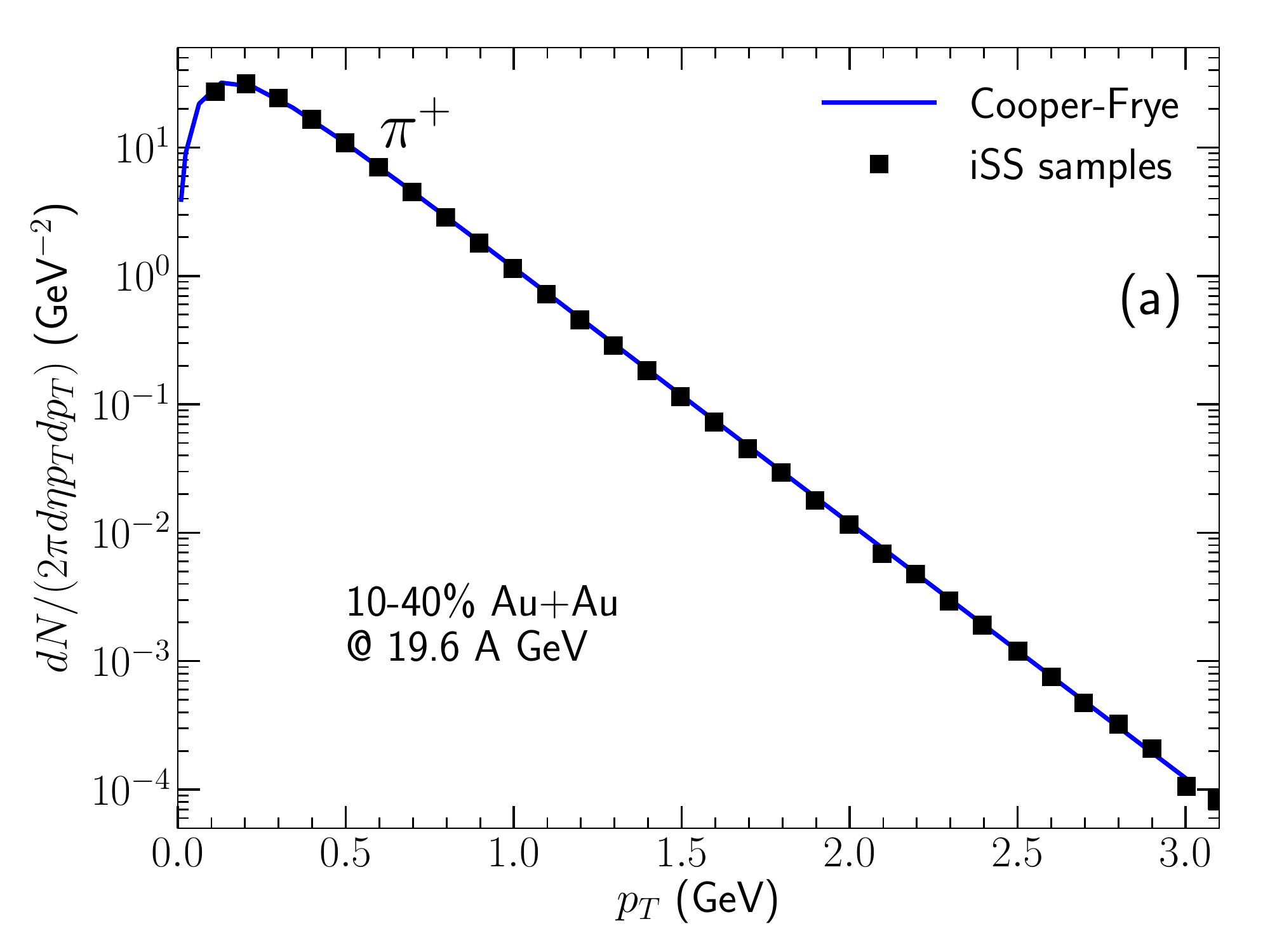} &
  \includegraphics[width=0.5\linewidth]{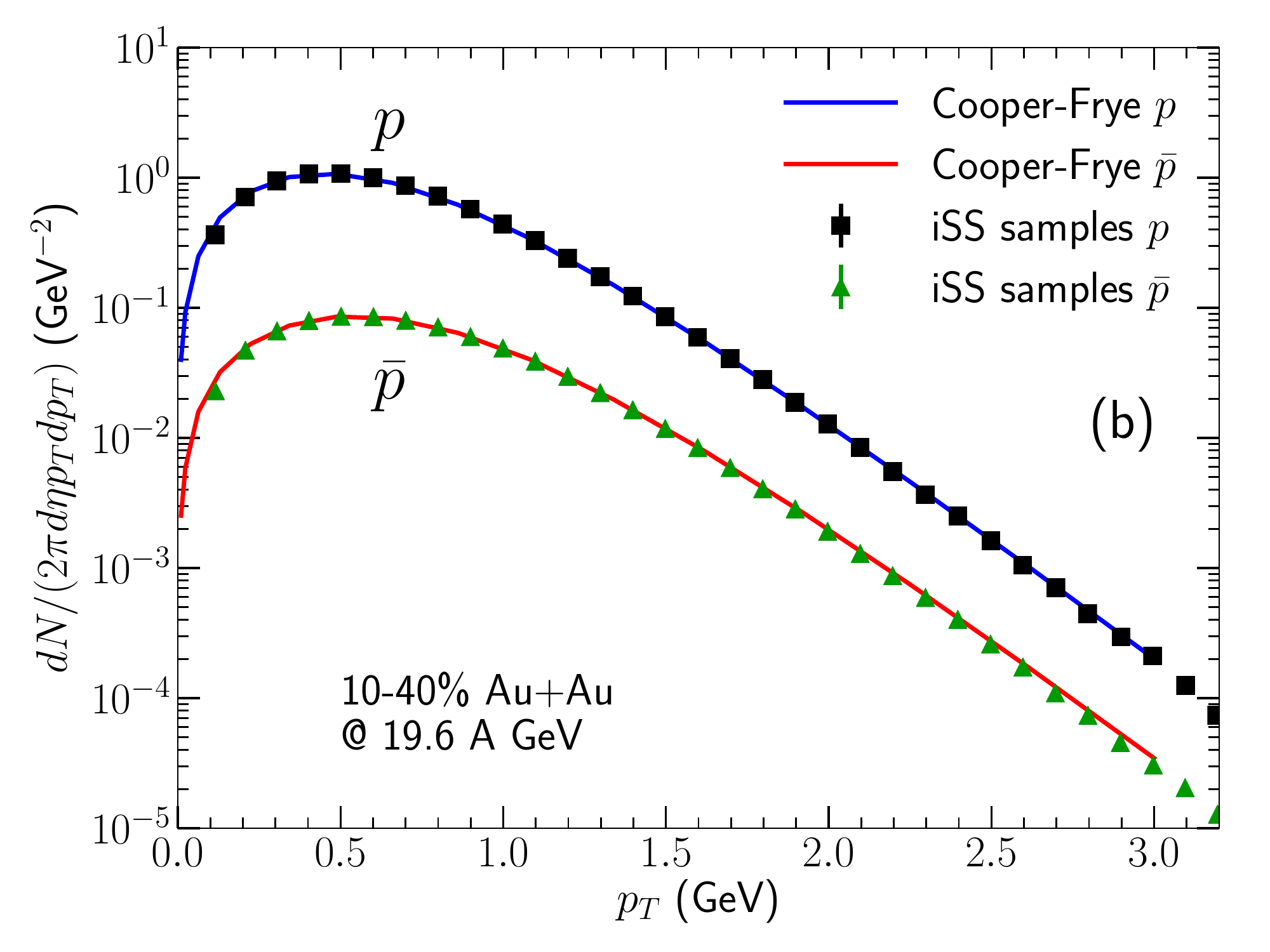}
  \end{tabular}
  \caption{Cross check the $p_T$-spectra of thermally emitted $\pi^+$, $p$, and $\bar{p}$ between  the numerical results from Cooper-Frye formula (lines) and the statistical results from particle sampler {\tt iSS} (markers).}
  \label{appendix2.fig2}
\end{figure*}
%
\begin{figure*}[ht!]
  \centering
  \begin{tabular}{cc}
  \includegraphics[width=0.5\linewidth]{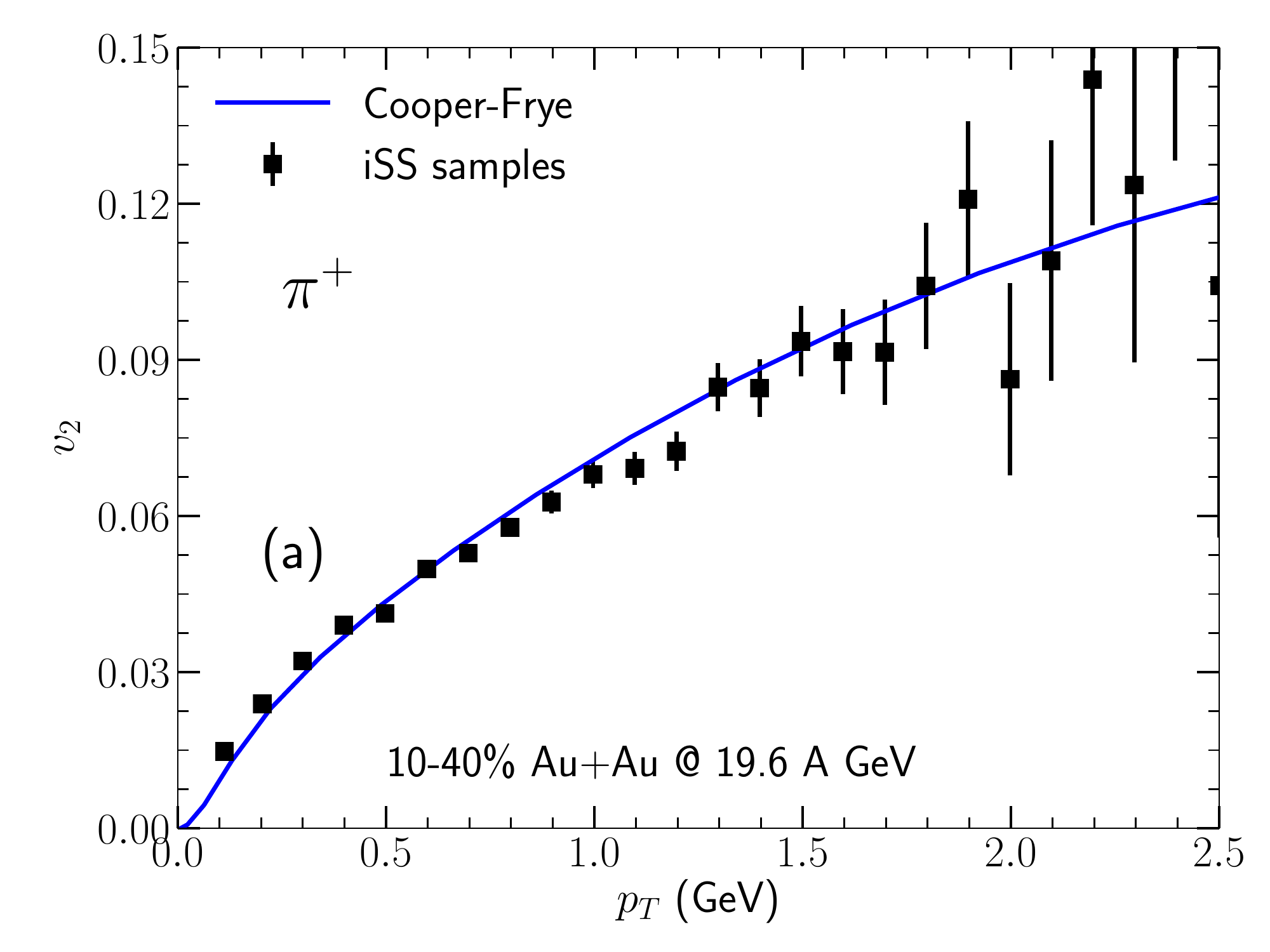} &
  \includegraphics[width=0.5\linewidth]{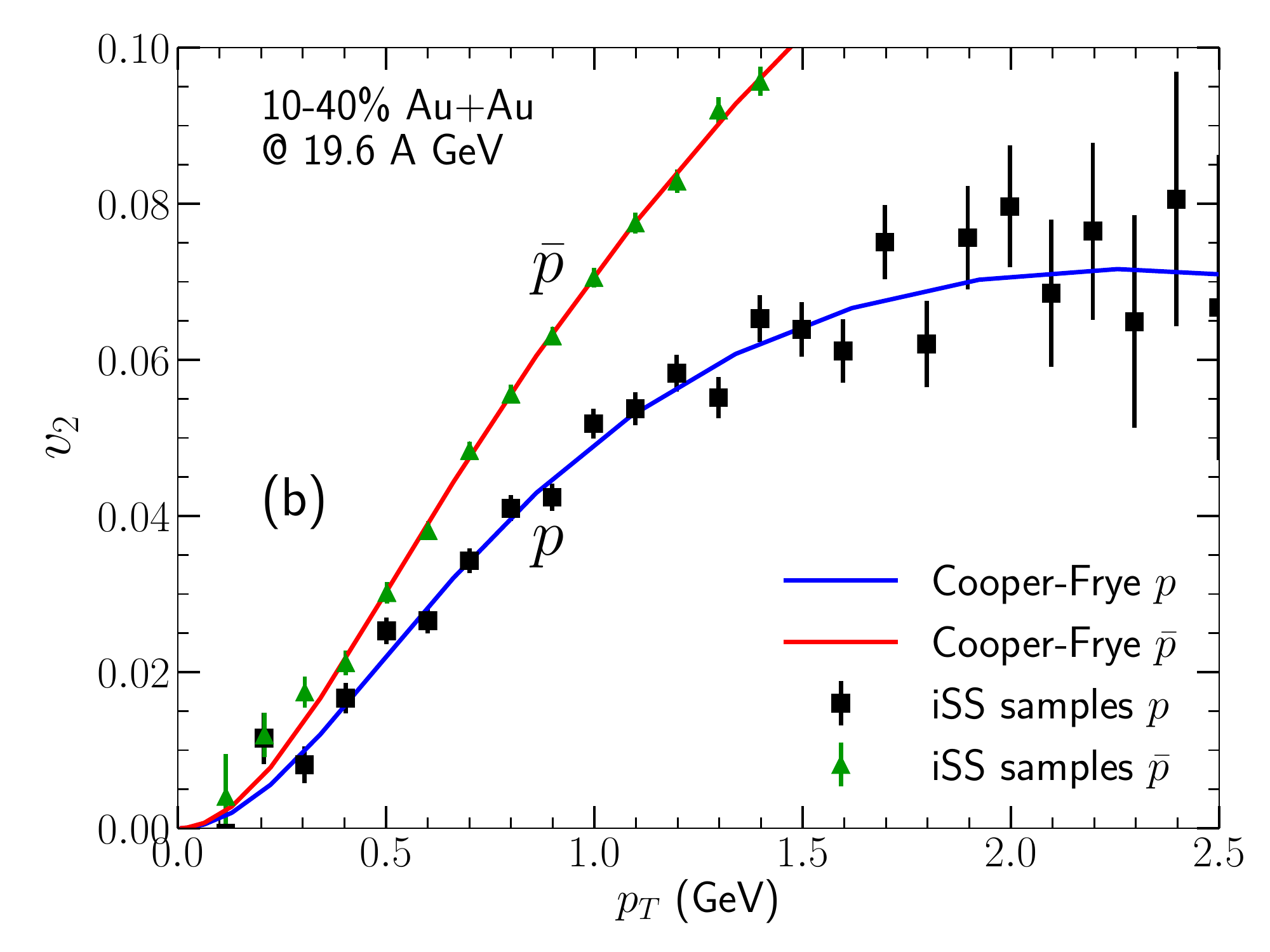}
  \end{tabular}
  \caption{Cross check the $p_T$-differential $v_2$ of thermally emitted $\pi^+$, $p$, and $\bar{p}$ between  the numerical results from Cooper-Frye formula (lines) and the statistical results from particle sampler {\tt iSS} (markers).}
  \label{appendix2.fig3}
\end{figure*}
%

As presented in Figs.~\ref{appendix2.fig1}, \ref{appendix2.fig2}, and \ref{appendix2.fig3}, the particle sampler \textsc{iSS} reproduces the numerical results from the Cooper-Frye formula as desired. Particle pseudo-rapidity distributions, $p_T$-spectra, and $p_T$-differential elliptic flow coefficients $v_2(p_T)$ are shown to be in good agreement between the two numerical codes. Out-of-equilibrium corrections from shear viscosity and net baryon diffusion are included in the calculations.

\section{Validation of hydrodynamic evolution at finite baryon density}
\label{appendix_C}

In this appendix, we provide two numerical checks for the hydrodynamic evolution at finite baryon density. 

\subsection{Gubser Solution at finite $\mu_B$}

Similar to the analyses performed in Ref.\ \cite{Marrochio:2013wla}, we use
hydrodynamic solutions obtained assuming a Gubser flow background \cite%
{Gubser:2010ze,Gubser:2010ui} to test our simulation code. These types of
solutions display a strong radial expansion rate and allow us to check the
performance of our code in the transverse plane. We note that the Gubser
solutions are valid only for conformal fluids and, in this sense, they
assume an ideal equation of state, $\varepsilon \sim T^{4}$, $P\sim T^{4}$,
and $\varepsilon =3P$. Furthermore, such solutions cannot display any
effects of diffusion, since the only scale present in the problem is the
temperature and, consequently, the chemical potential must behave as $\mu
_{B}\sim T$. This leads to a thermal potential $\mu _{B}/T$ that is constant
and to a vanishing Navier-Stokes term $\nabla ^{\mu }\left( \mu
_{B}/T\right) =0$. Nevertheless, we can still test the code in the ideal
fluid limit, in the presence of a finite net-charge (here, the baryon
number) density (this will keep $\varepsilon \sim P\sim T^{4}$, and $%
\varepsilon =3P$, only changing the proportionality factors between $%
\varepsilon $, $P$ and $T^{4}$).

In the ideal fluid limit, an analytic solution of the velocity, energy
density and net-charge density were obtained in Refs. \cite%
{Gubser:2010ze,Gubser:2010ui}. Such solutions provides us with a non-trivial
numerical check for the dynamical evolution of $\rho _{B}$ in the transverse
plane. The analytic solutions for the hydrodynamic fields are, 
\begin{eqnarray}
e(\tau ,r) \hspace{-0.1cm}&=&\hspace{-0.1cm}\frac{e_{0}}{\tau ^{4}}\frac{(2q\tau )^{8/3}}{\left[
1+2q^{2}(\tau ^{2}+r^{2})+q^{4}(\tau ^{2}-r^{2})^{2}\right] ^{4/3}}\hspace{0.4cm}\\
n_{B}(\tau ,r) \hspace{-0.1cm}&=&\hspace{-0.1cm}\frac{n _{B0}}{\tau ^{3}}\frac{4q^{2}\tau ^{2}}{%
\left[ 1+2q^{2}(\tau ^{2}+r^{2})+q^{4}(\tau ^{2}-r^{2})\right] ^{2}}
\end{eqnarray}%
with the radial velocity being given by (this radial velocity is unaffected
by the inclusion of viscosity or of a finite baryon density) 
\begin{equation}
v^{r}(\tau ,r)=\frac{2q^{2}\tau r}{1+(q\tau )^{2}+(qr)^{2}},
\end{equation}%
where $r=\sqrt{x^{2}+y^{2}}$ is the transverse radius. The constant $q$,
introduced in the expressions above, is a free parameter of the solution and
determines the system size, with larger values of $q$ corresponding to
smaller systems. In our simulations, we fix $q=1$ fm$^{-1}$ -- the same value as
used in Ref.\,\cite{Marrochio:2013wla}. Also, $e_{0} = 1$ fm$^{-4}$ and $\rho _{B0} = 0.5$ fm$^{-3}$
are free parameters that determine the initial value at, $r=0$, of the
energy and net-charge densities.
\begin{figure*}[th]
\centering
\begin{tabular}{cc}
\includegraphics[width=0.5\linewidth]{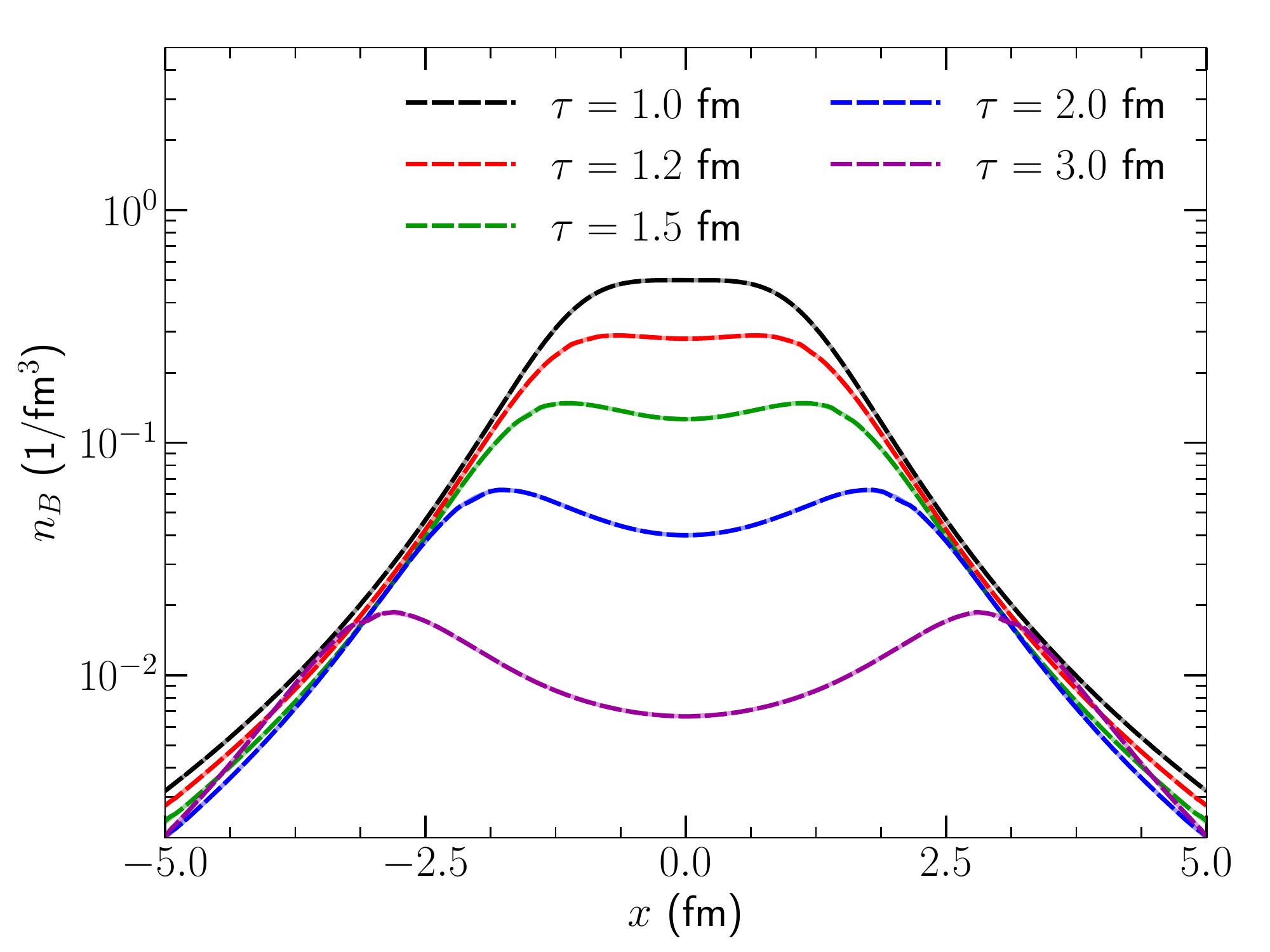} & %
\includegraphics[width=0.5\linewidth]{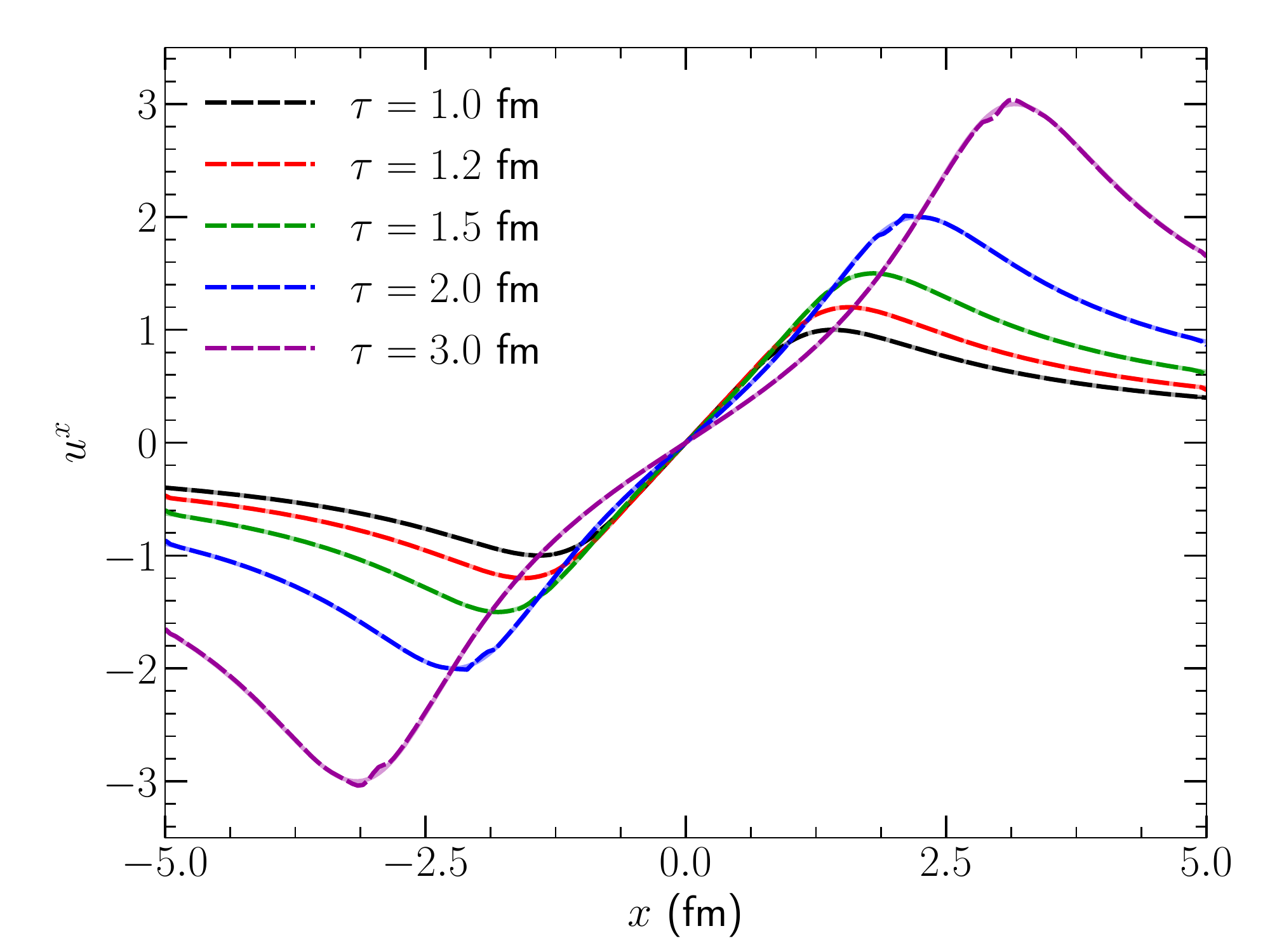}%
\end{tabular}
\caption{The evolution of $n_{B}$ and flow velocity $u^{x}$
compared with analytic Gubser solution (light solid lines in the background).
}
\label{appendix3.fig1}
\end{figure*}

Figure~\ref{appendix3.fig1} shows the comparison between numerical solutions
found using \textsc{Music} (dark dashed lines) and the analytic solution
(light solid lines). An excellent agreement is obtained over the entire
space ($r<5$ fm) for all times between $1$ fm$<\tau <3.0$ fm. We note that,
even though we only considered a time evolution of $2$ fm, this was enough
for the charge density of the fluid to decrease by a factor $\sim $100 and
that \textsc{Music} was able to describe the solution over such a wide range
of densities.

\subsection{Cross check hydrodynamic evolution with net baryon diffusion in 1+1D}

\begin{figure*}[ht!]
  \centering
  \begin{tabular}{cc}
  \includegraphics[width=0.5\linewidth]{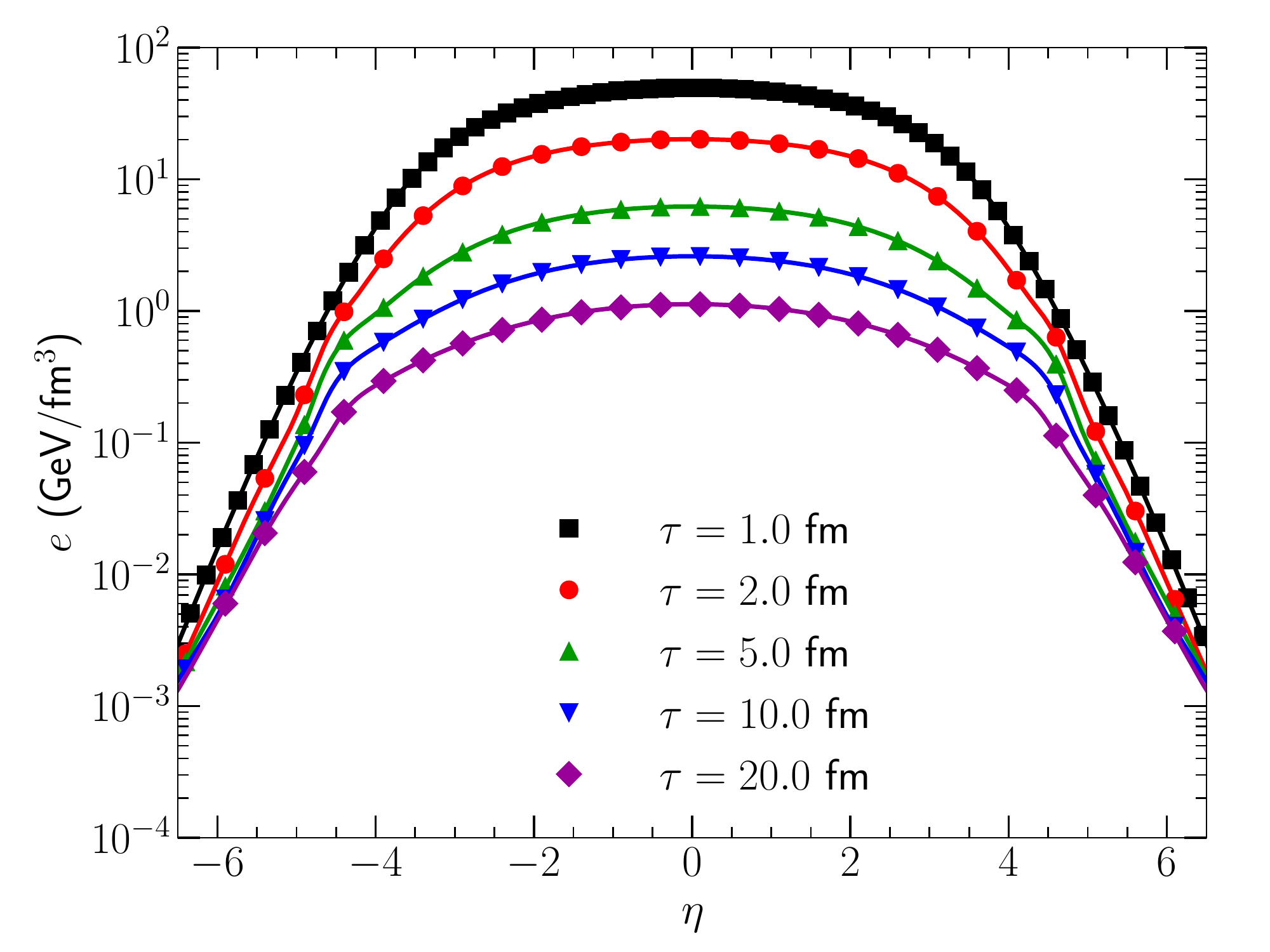} &
  \includegraphics[width=0.5\linewidth]{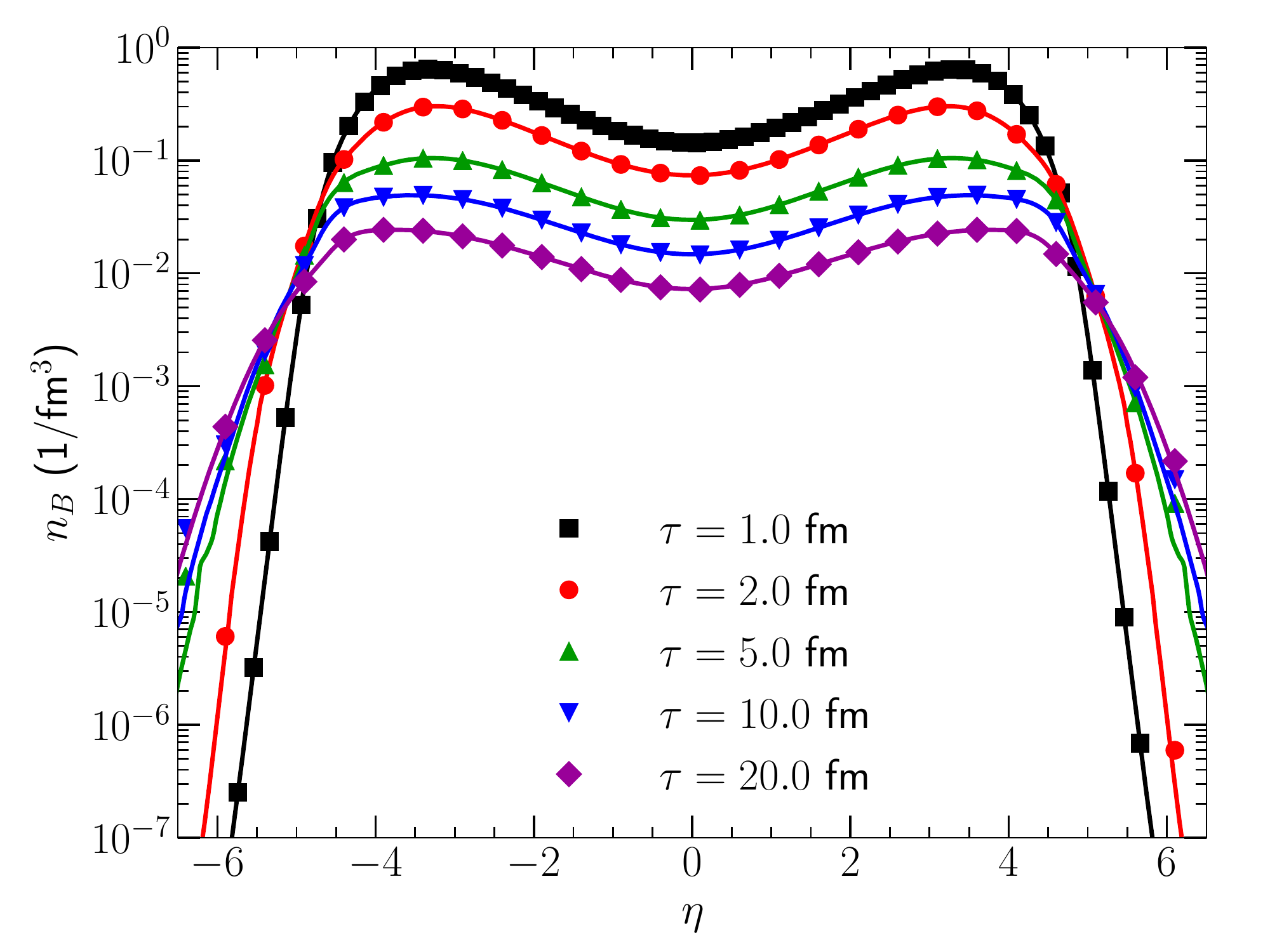}
  \end{tabular}
  \caption{The evolution of local energy density $e$ and net baryon density $n_B$ compared with (1+1)-D numerical solutions (markers).}
  \label{appendix3.fig2}
\end{figure*}

We next present a comparison of the evolution including net baryon diffusion with numerical solutions in longitudinal 1+1D \cite{Monnai:2012jc}. The equations of motion are as follows,
\begin{equation}
\partial_\mu T^{\mu \nu} = 0; \quad \partial_\mu J_B^\mu = 0
\end{equation}
with $J_B^\mu = n_B u^\mu + q^\mu$. The baryon diffusion current $q^\mu$ evolves with
\begin{equation}
\Delta^{\mu\nu} D q_\nu = -\frac{1}{\tau_q} \left( q^\mu - \kappa \nabla^\mu \frac{\mu_B}{T} \right).
\end{equation}
In the numerical test shown in Fig.~\ref{appendix3.fig2}, the net baryon diffusion coefficient is chosen to be $\kappa = 0.2 n_B/\mu_B$ and the diffusion relaxation time, $\tau_q = 0.2/T$. Good agreements are achieved for the rapidity evolution of local energy density and net baryon density between the two numerical algorithms. This validates our numerical solver with baryon diffusion along the longitudinal direction.
%

\subsection{Stabilized numerical evolution with net baryon diffusion}
Viscous hydrodynamics considers the net baryon diffusion current $q^\mu$ as a
perturbative correction to the equilibrium net baryon current.
However, the size of $q^\mu$ can be comparable or even larger than $n_B$
in dilute density regions in realistic event-by-event simulations.
Although these fluid cells are far outside the freeze-out surface
and their dynamical evolution does not affect any physical observables,
they may cause numerical instability problems during the evolution.
To stabilize the simulations, we regulate the ill-behaved $q^\mu$ in fluid cells.

First we need to make sure that the diffusion current is orthogonal
to the flow velocity $u^\mu$.
This is done by evolving the three spatial components of $q^\mu$ as independent fields,
while the $q^\tau$ component is constructed using
$q^\tau = q^i u_i/u^\tau$ where $i$ is summed over $x, y$, and $\eta$.
Secondly, the relative size of $q^\mu$ compared to the net baryon density
$n_B$ can be computed as,
\begin{equation}
    r_q =\frac{1}{f_\mathrm{strength}} \sqrt{\frac{q^\mu q_\mu}{n_B^2}},
    \label{appendix3.eq6}
\end{equation}
where the regulation strength parameter
\begin{equation}
    f_\mathrm{strength} = \chi_0 \bigg[\frac{1}{\exp(-(e - e_0)/\xi_0) + 1} - \frac{1}{\exp(e_0/\xi_0) + 1} \bigg].
    \label{appendix3.eq7}
\end{equation}
The parameter $\chi_0$ controls the overall strength of the regulation for
local energy density $e$ above a critical energy density $e_0$.
For the fluid cell whose local energy density is smaller than $e_0$,
the regulation strength increases exponentially as the local energy density $e$
decreases. In Eq.~(\ref{appendix3.eq7}), we chose $\chi_0 = 10$,
$e_0 = 0.1$\,GeV/fm$^3$, and the width parameter $\xi_0 = 0.01$\,GeV/fm$^3$.

When the ratio $r_q$ is larger than some maximum value $r^\mathrm{max}_q$,
the $q^\mu$ current is regulated as,
\begin{equation}
\tilde{q}^\mu = \frac{r^\mathrm{max}_q}{r_q} q^\mu.
\end{equation}
With this regulation scheme, we found that 100\% of
event-by-event hydrodynamic simulations could be stabilized
with $r^\mathrm{max}_q = 1$.

We have checked that the majority of the regulations are triggered
in the dilute density region outside the freeze-out surface,
leaving the physical region untouched.
The final hadronic flow observables show negligible dependence on $\xi_0$ in
the range of 1 to 30.

\section{Comparison between two microscopic hadronic transport models: UrQMD and JAM}

In this section, we use the same inputs for two publicly available hadronic transport codes and compare the final hadronic observables. The difference from the two hadronic cascade models can help us to estimate theoretical uncertainties in the late dilute phase. 

\begin{figure*}[ht!]
  \centering
  \begin{tabular}{cc}
  \includegraphics[width=0.5\linewidth]{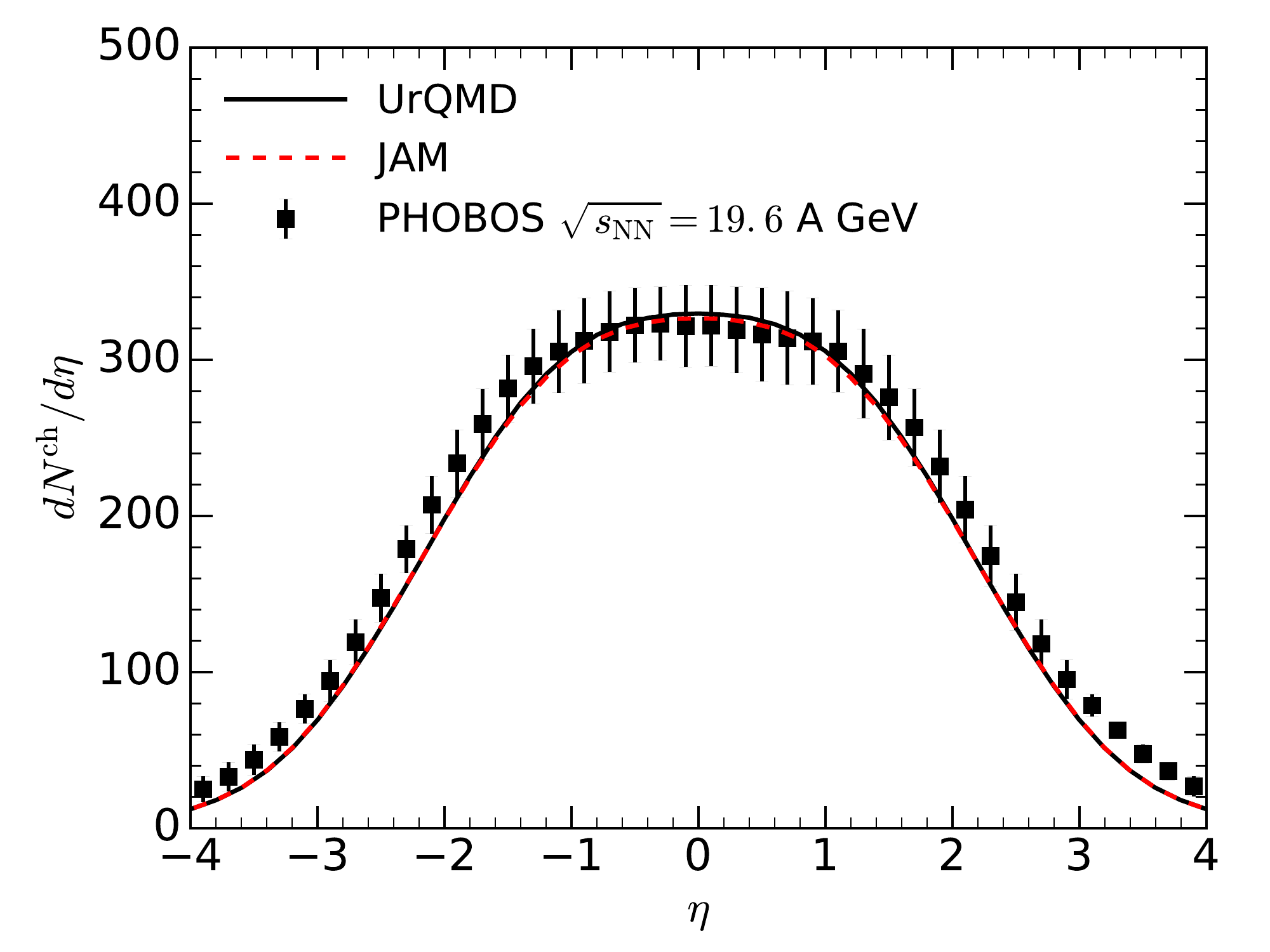} &
  \includegraphics[width=0.5\linewidth]{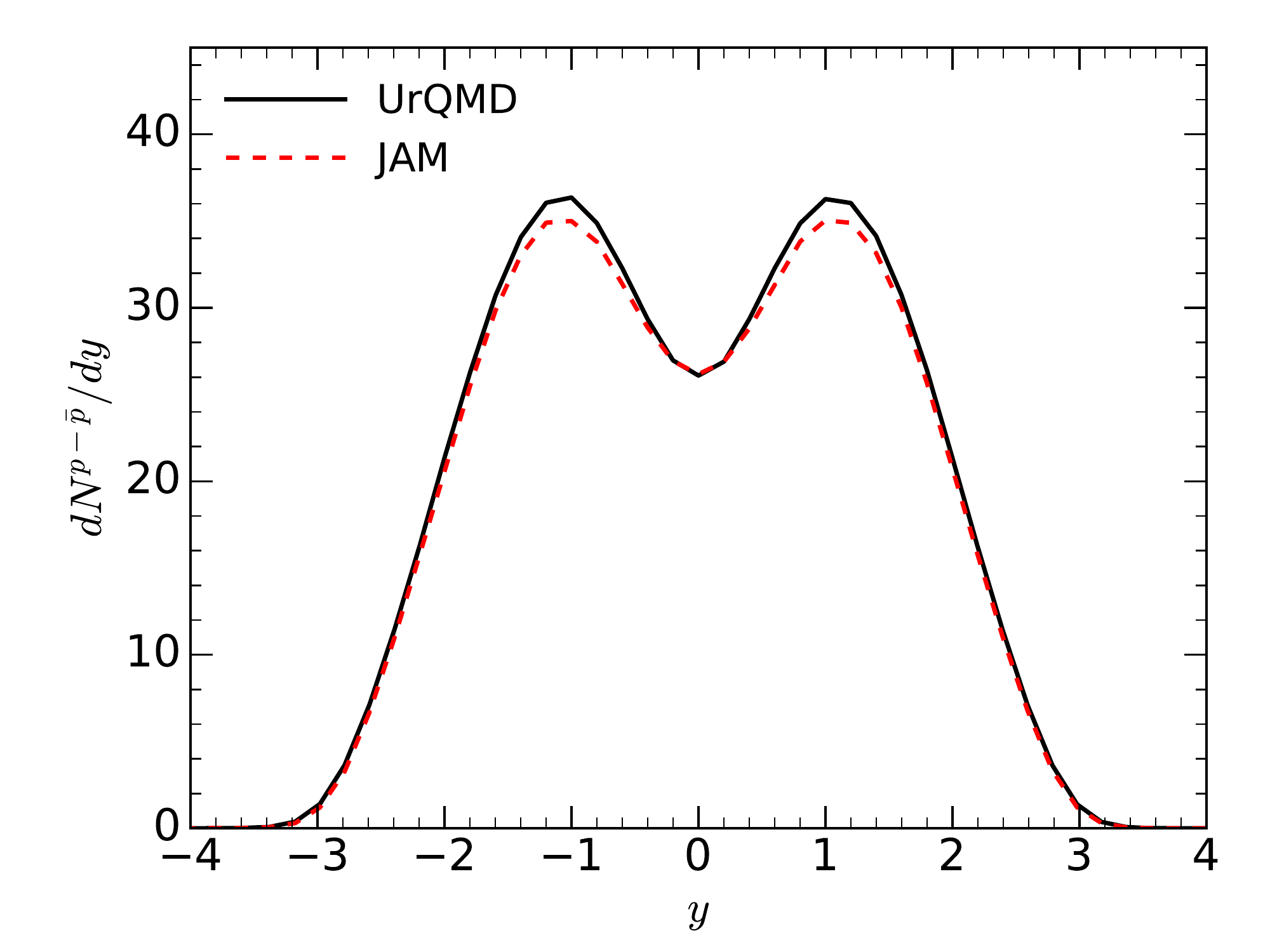}
  \end{tabular}
  \caption{The (pseudo-)rapidity distribution of charged hadrons and net protons from the two hadronic cascade simulations for 0-5\% centrality Au+Au collisions at 19.6 GeV.}
  \label{appendix4.fig1}
\end{figure*}
%

\begin{figure*}[ht!]
  \centering
  \begin{tabular}{cc}
  \includegraphics[width=0.5\linewidth]{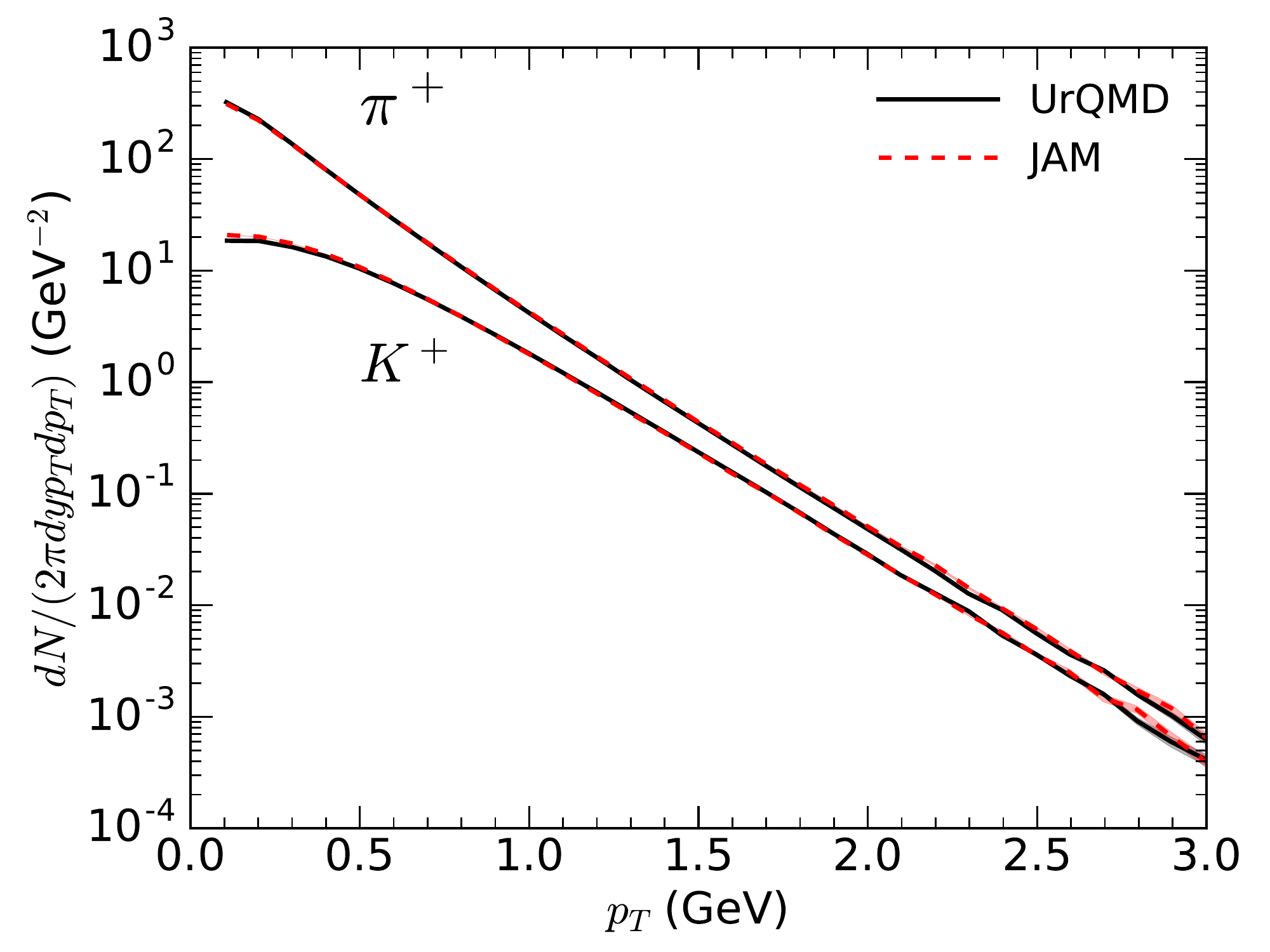} &
  \includegraphics[width=0.5\linewidth]{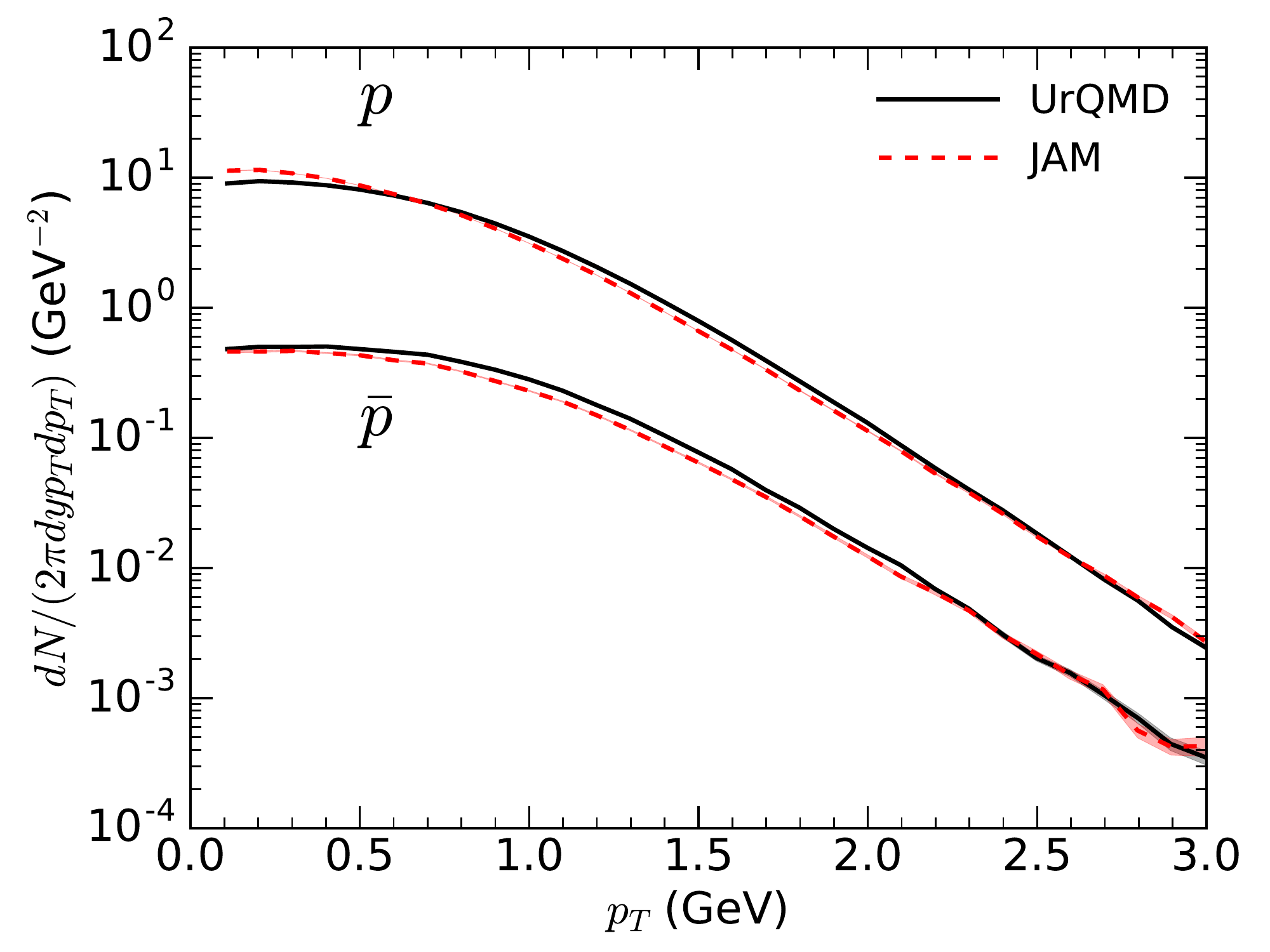}
  \end{tabular}
  \caption{The comparison of $p_T$-spectra of final $\pi^+$, $K^+$, $p$, and $\bar{p}$ from the two hadronic cascade simulations.}
  \label{appendix4.fig2}
\end{figure*}
%

\begin{figure*}[ht!]
  \centering
  \begin{tabular}{cc}
  \includegraphics[width=0.5\linewidth]{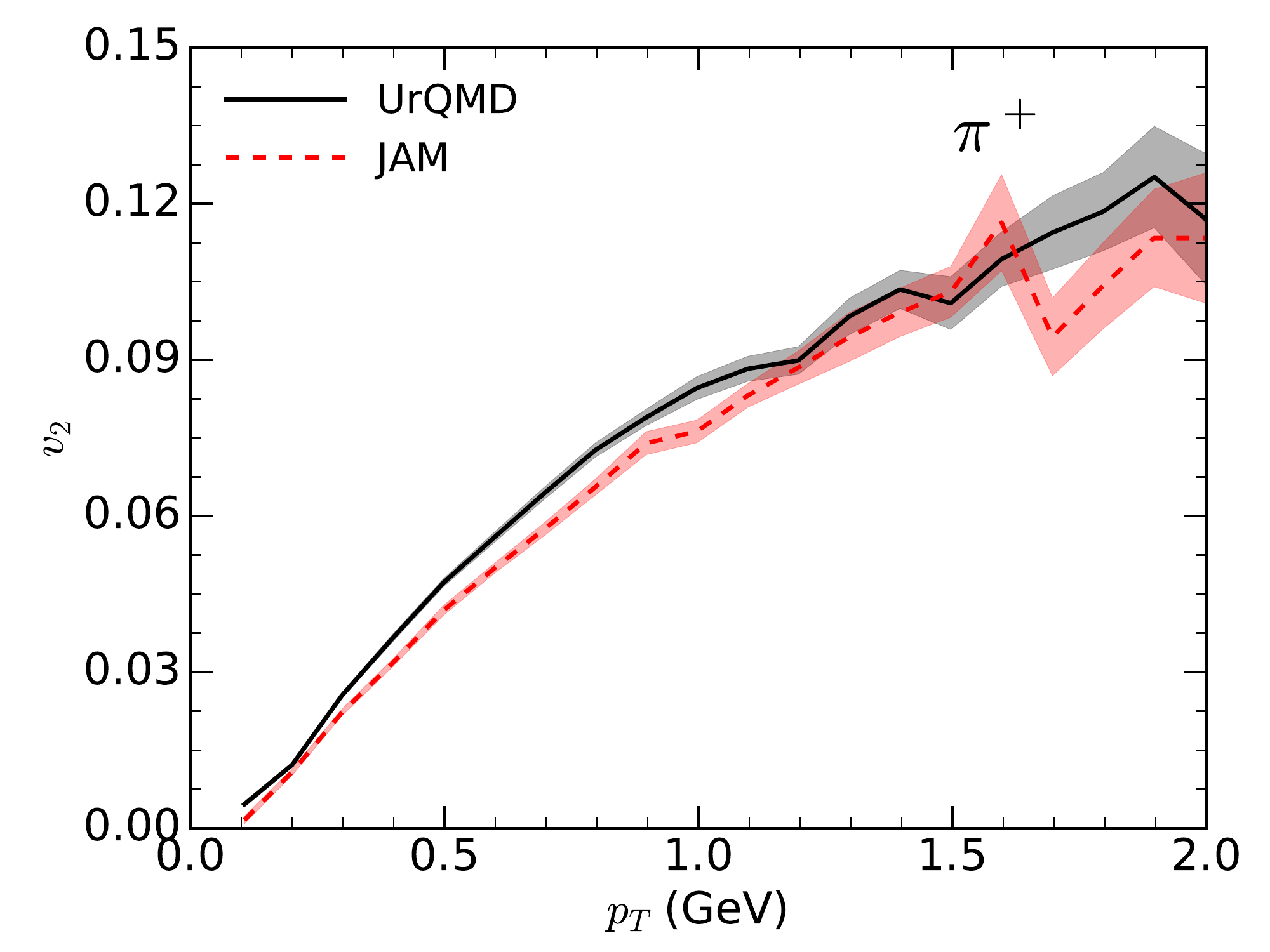} &
  \includegraphics[width=0.5\linewidth]{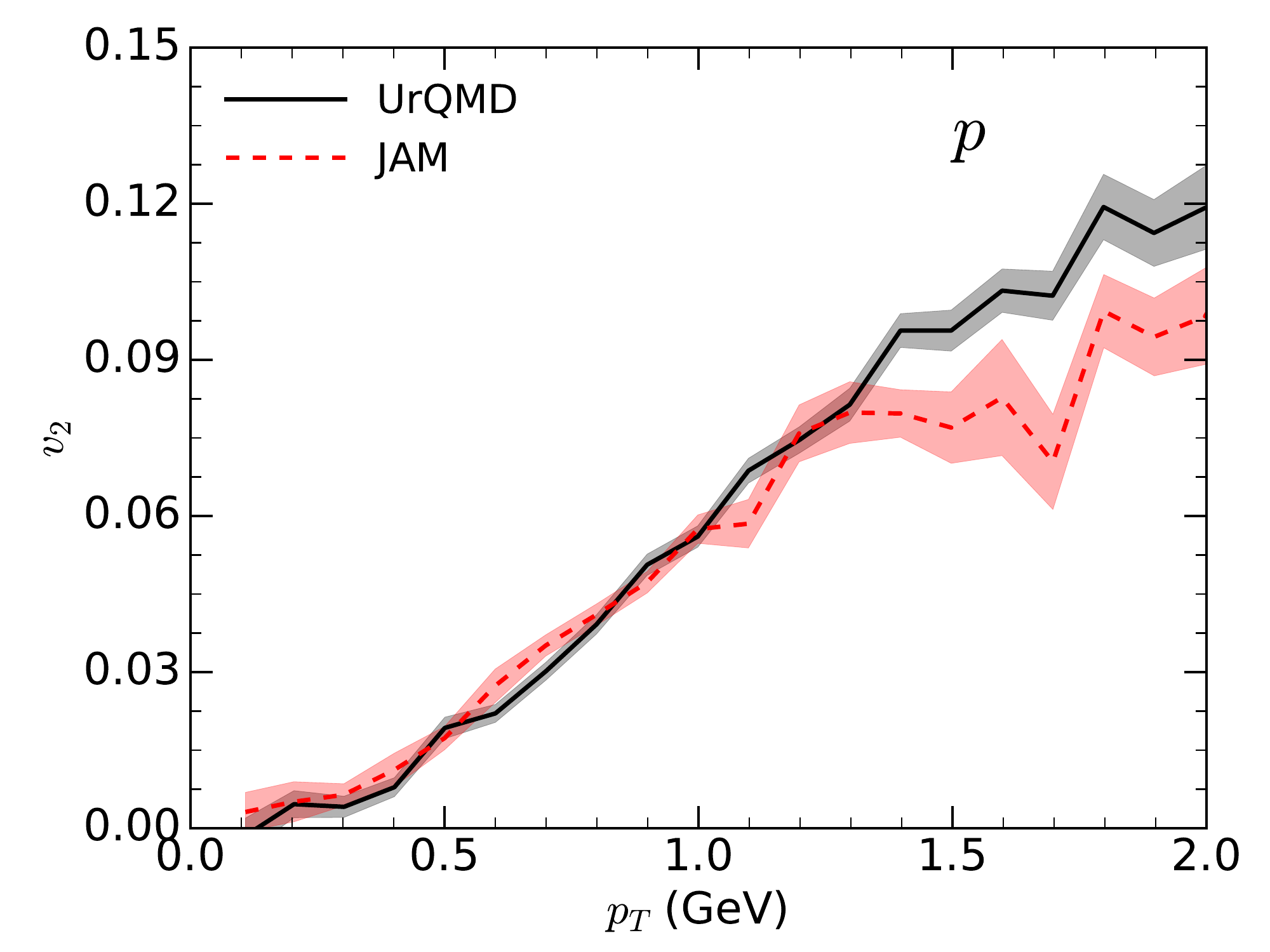}
  \end{tabular}
  \caption{The $p_T$-differential $v_2$ of final $\pi^+$, $K^+$, and $p$ from the two hadronic cascade simulations. The shaded bands indicate statistical errors.}
  \label{appendix4.fig3}
\end{figure*}
%

Figures \ref{appendix4.fig1}, \ref{appendix4.fig2}, and \ref{appendix4.fig3} show the comparisons between the two hadronic transport models. We find identical results for charged hadron pseudorapidity distributions and $p_T$-spectra of $\pi^+$ and $K^+$. This means that the dynamical evolution of the light mesons are very close in the two hadronic cascade models. Meanwhile, small but visible differences are present in proton and antiproton transverse momentum and rapidity distributions. The net proton numbers from the JAM model are slightly smaller than those from UrQMD. This might indicate a slightly stronger baryon-anti-baryon annihilation in the JAM model. The $p_T$ spectrum of protons from the JAM model is slightly steeper than that from UrQMD. The two transport models also produce slightly different results for $\pi^+$ and $p$ $p_T$ differential elliptic flow. The JAM model gives smaller $v_2$ value compared to the UrQMD results, which is statistically significant for $p_T<1\,{\rm GeV}$. 

\bibliography{references}

\end{document}